\documentclass[journal,onecolumn,draftclsnofoot,12pt]{IEEEtran}
\IEEEoverridecommandlockouts

\usepackage{cite}
\usepackage{amsmath,amssymb,amsfonts,bm}
\usepackage{latexsym}
\usepackage{units}
\usepackage{algorithmicx}
\usepackage{algpseudocode}
\usepackage{algorithm} 
\usepackage{mathtools}
\usepackage{graphicx}
\usepackage{textcomp}
\usepackage{xcolor}
\usepackage{cancel}
\usepackage{accents}
\usepackage{leftidx}
\usepackage{nicematrix}
\usepackage{etoolbox}
\usepackage{tikz}
\usetikzlibrary{arrows}
\usetikzlibrary{matrix}
\usepackage{arydshln}
\usepackage{drawmatrix}
\usepackage{diffcoeff}  
\usepackage{fancyvrb}
\usepackage{setspace}
\usepackage{stackengine}
\stackMath
\usepackage{pdflscape}
\usepackage{multirow}
\usepackage{multicol}
\usepackage{fullwidth}

\ifCLASSOPTIONcompsoc
\usepackage[caption=false,font=normalsize,labelfon
t=sf,textfont=sf]{subfig}
\else
\usepackage[caption=false,font=footnotesize]{subfi
g}
\fi

\def\BibTeX{{\rm B\kern-.05em{\sc i\kern-.025em b}\kern-.08em
    T\kern-.1667em\lower.7ex\hbox{E}\kern-.125emX}}

\definecolor{darkgreen}{rgb}{0,0.75,0}
\usepackage{hyperref}
\hypersetup{
  pdfauthor = {Mohammad M. Mansour},
  pdfcreator = {Mohammad M. Mansour},
  pdfnewwindow,
  pdffitwindow=true,
  colorlinks=true,
  linkcolor=blue,       
  citecolor=darkgreen,      
  filecolor=red,        
  urlcolor=cyan         
}
\newtheorem{theorem}{Theorem}
\newtheorem{lemma}{Lemma}
\newtheorem{corollary}{Corollary}
\newcommand{\IEEEQEDright}{\hfill\IEEEQED}

\newcommand{\newthinspace}{%
  \kern\dimexpr0.05em\relax
}

\usetikzlibrary{tikzmark}
\usetikzlibrary{calc}
\usetikzlibrary{matrix,backgrounds,fit,decorations.pathreplacing}
\errorcontextlines\maxdimen

\newcommand{\ALGtikzmarkcolor}{gray!50}
\newcommand{\ALGtikzmarkextraindent}{4pt}
\newcommand{\ALGtikzmarkverticaloffsetstart}{-.5ex}
\newcommand{\ALGtikzmarkverticaloffsetend}{-.5ex}
\makeatletter
\newcounter{ALG@tikzmark@tempcnta}

\newcommand\ALG@tikzmark@start{%
    \global\let\ALG@tikzmark@last\ALG@tikzmark@starttext%
    \expandafter\edef\csname ALG@tikzmark@\theALG@nested\endcsname{\theALG@tikzmark@tempcnta}%
    \tikzmark{ALG@tikzmark@start@\csname ALG@tikzmark@\theALG@nested\endcsname}%
    \addtocounter{ALG@tikzmark@tempcnta}{1}%
}

\def\ALG@tikzmark@starttext{start}
\newcommand\ALG@tikzmark@end{%
    \ifx\ALG@tikzmark@last\ALG@tikzmark@starttext
    \else
        \tikzmark{ALG@tikzmark@end@\csname ALG@tikzmark@\theALG@nested\endcsname}%
        \tikz[overlay,remember picture] \draw[\ALGtikzmarkcolor] let \p{S}=($(pic cs:ALG@tikzmark@start@\csname ALG@tikzmark@\theALG@nested\endcsname)+(\ALGtikzmarkextraindent,\ALGtikzmarkverticaloffsetstart)$), \p{E}=($(pic cs:ALG@tikzmark@end@\csname ALG@tikzmark@\theALG@nested\endcsname)+(\ALGtikzmarkextraindent,\ALGtikzmarkverticaloffsetend)$) in (\x{S},\y{S})--(\x{S},\y{E});%
    \fi
    \gdef\ALG@tikzmark@last{end}%
}

\apptocmd{\ALG@beginblock}{\ALG@tikzmark@start}{}{\errmessage{failed to patch}}
\pretocmd{\ALG@endblock}{\ALG@tikzmark@end}{}{\errmessage{failed to patch}}
\makeatother

\makeatletter
\newdimen{\algindent}
\NewDocumentCommand{\LeftComment}{s m}{%
  \Statex \IfBooleanF{#1}{\hspace*{\ALG@thistlm}}\green{\(\triangleright\) \textit{#2}}}
\NewDocumentCommand{\FirstLeftComment}{s m}{%
  \Statex \IfBooleanF{#1}{\hspace*{\ALG@thistlm}}\green{\hskip\algorithmicindent\(\triangleright\) \textit{#2}}}
\NewDocumentCommand{\LeftCommentNoTriangle}{s m}{%
  \Statex \IfBooleanF{#1}{\hspace*{\ALG@thistlm}}\green{\textit{#2}}}
\makeatother

\newcommand{\algsnameabbr}{Algs.}
\newcommand{\algnameabbr}{Alg.}
\floatname{algorithm}{\algnameabbr}

\newcommand{\smallsim}{\smallsym{\mathrel}{\sim}}
\makeatletter
\newcommand{\smallsym}[2]{#1{\mathpalette\make@small@sym{#2}}}
\newcommand{\make@small@sym}[2]{%
  \vcenter{\hbox{$\m@th\downgrade@style#1#2$}}%
}
\newcommand{\downgrade@style}[1]{%
  \ifx#1\displaystyle\scriptstyle\else
    \ifx#1\textstyle\scriptstyle\else
      \scriptscriptstyle
  \fi\fi
}
\makeatother

\def\onenegskip{\hspace{-0.2pt}}
\def\twonegskip{\onenegskip\onenegskip}
\def\threenegskip{\twonegskip\onenegskip}
\def\fournegskip{\threenegskip\onenegskip}
\def\fivenegskip{\fournegskip\onenegskip}
\def\sixnegskip{\fivenegskip\onenegskip}
\def\sevennegskip{\sixnegskip\onenegskip}
\def\eightnegskip{\sevennegskip\onenegskip}

\newcommand{\blue}[1]{{\color{blue}{#1}}}           
\newcommand{\magenta}[1]{{\color{magenta}{#1}}}     
\definecolor{armygreen}{rgb}{0.0, 0.5, 0.0}
\newcommand{\green}[1]{{\color{armygreen}{#1}}}     

\def\arrowtext#1#2{\hbox to#1{\arrowtextA\ #2 \arrowtextA\kern2pt\llap{$\succ$}}}
\def\arrowtextA{\leaders\vrule height2.7pt depth-2.3pt\hfil}

\newcommand\tsup[2][2]{%
 \def\useanchorwidth{T}%
  \ifnum#1>1%
    \stackon[-1.2ex]{\tsup[\numexpr#1-1\relax]{#2}}{\mathchar"307E\kern-.5pt}%
  \else%
    \stackon[-1ex]{#2}{\mathchar"307E\kern-.5pt}%
  \fi%
}

\newcommand\supsc[2]{#1^{\!\raisebox{+.3ex}{\scalebox{.8}{$\scriptstyle #2$}}}}
\newcommand\subsc[2]{#1_{\raisebox{+.1ex}{\scalebox{.8}{$\scriptstyle  #2$}}}}
\newcommand\subpsc[3]{#1_{\raisebox{+.1ex}{\scalebox{.8}{$\scriptstyle  #2$}}}^{\!\raisebox{+.3ex}{\scalebox{.8}{$\scriptstyle #3$}}}}

\newcommand{\mbf}[1]{\mathbf{#1}}

\newcommand{\nth}[1]{{#1}{\text{th}}}

\newcommand{\herm}{\dag}
\newcommand{\tran}{\mathsf{T}}

\renewcommand{\Re}[1]{\mathfrak{R}\!\left\{ #1 \right\}}
\newcommand{\Ree}[1]{\mathfrak{R}\{ #1 \}}
\renewcommand{\Im}[1]{\mathfrak{I}\!\left\{ #1 \right\}}

\DeclareMathOperator*{\argmax}{\mathrm{argmax}}
\DeclareMathOperator*{\argmin}{\mathrm{argmin}}

\renewcommand{\a}[0]{\alpha}

\DeclareRobustCommand\parent{1} 
\DeclareRobustCommand\child{2}  

\DeclareRobustCommand\mupnoarg{\subsc{\mu}{\mathrm{p}}}
\DeclareRobustCommand\mup[1]{\mupnoarg(#1)}
\DeclareRobustCommand\muapnoarg{\subsc{\mu}{\mathrm{ap}}}
\DeclareRobustCommand\muap[1]{\muapnoarg(#1)}
\DeclareRobustCommand\mupznoarg{\subsc{\mu}{\mathrm{pz}}}

\DeclareRobustCommand\mum[1]{\subsc{\mu}{\mathrm{m}}\!\left(#1\right)}

\DeclareRobustCommand\muparentnoarg{\subpsc{{\mu\twonegskip}}{\parent}{}}
\DeclareRobustCommand\muparent[1]{\muparentnoarg\eightnegskip(\twonegskip#1\sixnegskip)}
\DeclareRobustCommand\muchildnoarg{\subpsc{\mu\onenegskip}{\child}{}}
\DeclareRobustCommand\muchild[1]{\muchildnoarg\fivenegskip(\twonegskip#1\sixnegskip)}

\DeclareRobustCommand\muhatchildnoarg{\subsc{\hat{\mu}}{\child}}
\DeclareRobustCommand\muhatchild[1]{\muhatchildnoarg(\twonegskip#1\sixnegskip)}

\newcommand{\C}[0]{\mathcal{C}}
\newcommand{\X}[0]{\mathcal{X}}

\newcommand{\abs}[1]{\left|{#1}\right|}
\newcommand{\abss}[1]{|{#1}|}

\newcommand{\normm}[1]{\lVert{#1}\rVert}

\newcommand{\fnormm}[1]{\lvert\hspace{-0.15em}\lvert{#1}\rvert\hspace{-0.15em}\rvert}
\newcommand{\fnormmsq}[1]{\supsc{\lvert\hspace{-0.15em}\lvert{{#1}}\rvert\hspace{-0.15em}\rvert}{\,2}}

\newcommand{\slice}[1]{{\left\lfloor{#1}\right\rceil}}

\newcommand{\flr}[1]{\left\lfloor #1\right \rfloor}

\newcommand{\Tr}[1]{\mathrm{Tr}\!\left( #1 \right)}
\newcommand{\Trr}[1]{\mathrm{Tr}( #1 )}

\newcommand{\diagg}[1]{\mbox{\small$\mathrm{diag}$}( #1 )}

\newcommand{\ML}{\mbox{\tiny$\mathrm{ML}$}}

\newcommand{\opt}{\mathrm{opt}}

\newcommand{\WLD}{\mbox{\tiny$\mathrm{WLD}$}}
\newcommand{\AWLD}{\mbox{\tiny$\mathrm{AWLD}$}}
\newcommand{\LB}{\mbox{\tiny$\mathrm{LB}$}}
\newcommand{\ILBWLD}{I_{\LB}^{\WLD}}
\newcommand{\ILBAWLD}{I_{\LB}^{\AWLD}}
\newcommand{\ILB}{I_{\LB}}
\newcommand{\ILBopt}{\subpsc{I}{\mathrm{LB}}{\,\,\mathrm{opt}}}
\newcommand{\ILBopttilde}{\subpsc{\tilde{I}}{\mathrm{LB}}{\,\,\mathrm{opt}}}

\newcommand{\AWGN}{\mbox{\tiny$\mathrm{AWGN}$}}

\newcommand{\Es}{E_{\mathrm{s}}}
\newcommand{\Nt}{N}
\newcommand{\Nr}{M}
\newcommand{\NbyN}{\Nt\!\times\!\Nt}
\newcommand{\NbyM}{\Nt\!\times\!\Nr}
\newcommand{\MbyN}{\Nr\!\times\!\Nt}

\newcommand{\expec}[1]{\mathsf{E}\!\left[ #1 \right]}
\newcommand{\expecc}[1]{\mathsf{E}[ #1 ]}
\newcommand{\Ex}[2]{\mathsf{E}_{#1}\!\left[ #2 \right]}

\newcommand{\D}{\mbf{D}}
\DeclareRobustCommand\Dp{\subsc{\mbf{D}}{\mathrm{p}}}
\newcommand{\Di}{\supsc{\mbf{D}}{\,-1}}
\DeclareRobustCommand\Dap{\subsc{\mbf{D}}{\mathrm{ap}}}

\DeclareRobustCommand\Dh{\supsc{\mbf{D}}{\,\dag}}

\DeclareRobustCommand\Dparent{\subsc{\mbf{D}}{\parent}}
\DeclareRobustCommand\Dchild{\subsc{\mbf{D}}{\child}}

\newcommand{\E}{\mbf{E}}

\DeclareRobustCommand\Etilde{\tilde{\mbf{E}}}
\DeclareRobustCommand\Etildeh{\supsc{\smash[t]{\tilde{\mbf{E}}}}{\,\dag}}

\DeclareRobustCommand\eparentnoarg{\subsc{e}{\twonegskip\parent}}
\DeclareRobustCommand\eparent[1]{\eparentnoarg\sixnegskip(\twonegskip#1\sixnegskip)}
\DeclareRobustCommand\echildnoarg{\subsc{e}{\twonegskip\child}}
\DeclareRobustCommand\echild[1]{\echildnoarg\fournegskip(\twonegskip#1\sixnegskip)}

\newcommand{\F}{\mbf{F}}
\DeclareRobustCommand\Fh{\supsc{\mbf{F}}{\,\dag}}
\DeclareRobustCommand\Fopt{\supsc{\mbf{F}}{\,\opt}}
\DeclareRobustCommand\Fopth{\supsc{\mbf{F}}{\,\opt\dag}}

\DeclareRobustCommand\Fr{\subsc{\mbf{F}}{\!\mathrm{r}}}

\DeclareRobustCommand\Fropt{\subpsc{\mbf{F}}{\!\mathrm{r}}{\,\opt}}

\DeclareRobustCommand\Fp{\subsc{\mbf{F}}{\!\mathrm{p}}}
\DeclareRobustCommand\Fph{\subpsc{\mbf{F}}{\!\mathrm{p}}{\,\dag}}

\DeclareRobustCommand\Fap{\subsc{\mbf{F}}{\!\mathrm{ap}}}
\DeclareRobustCommand\Faph{\subpsc{\mbf{F}}{\!\mathrm{ap}}{\,\dag}}

\newcommand{\G}{\mbf{G}}

\DeclareRobustCommand\Gi{\supsc{\mbf{G}}{\,-1}}

\DeclareRobustCommand\Gopt{\supsc{\mbf{G}}{\,\opt}}

\DeclareRobustCommand\Gr{\subsc{\mbf{G}}{\mathrm{r}}}

\DeclareRobustCommand\Gp{\subsc{\mbf{G}}{\mathrm{p}}}

\DeclareRobustCommand\Gap{\subsc{\mbf{G}}{\mathrm{ap}}}

\DeclareRobustCommand\Gapi{\subpsc{\mbf{G}}{\mathrm{ap}}{\,-1}}

\newcommand{\I}{\mbf{I}}

\newcommand{\J}{\mbf{J}}
\DeclareRobustCommand\Jh{\supsc{\mbf{J}}{\,\dag}}
\DeclareRobustCommand\Jopt{\supsc{\mbf{J}}{\,\opt}}
\DeclareRobustCommand\Jopth{\supsc{\mbf{J}}{\,\opt\dag}}

\DeclareRobustCommand\Ja{\subsc{\mbf{J}}{1}}

\DeclareRobustCommand\Jb{\subsc{\mbf{J}}{2}}
\DeclareRobustCommand\Jbh{\subpsc{\mbf{J}}{2}{\,\dag}}

\renewcommand{\H}{\mbf{H}}
\newcommand{\Hh}{\supsc{\H}{\,\dag}}

\DeclareRobustCommand\Ha{\subsc{\mbf{H}}{\mathrm{a}}}
\DeclareRobustCommand\Hah{\subpsc{\mbf{H}}{\mathrm{a}}{\,\dag}}

\newcommand{\Hjbar}{\subpsc{\H}{\bar{j}}{}}
\newcommand{\Hjbarherm}{\subpsc{\H}{\bar{j}}{\,\dag}}
\newcommand{\hj}{\subpsc{\mbf{h}}{j}{}}

\renewcommand{\L}{\mbf{L}}
\DeclareRobustCommand\Lh{\supsc{\mbf{L}}{\,\dag}}

\newcommand{\Ltilde}{\tilde{\mbf{L}}}

\DeclareRobustCommand\La[1][]{\subsc{\mbf{L}}{{\mathrm{a}{#1}}}}       
\DeclareRobustCommand\Lah[1][]{\subpsc{\mbf{L}}{{\mathrm{a}{#1}}}{\,\dag}}       
\DeclareRobustCommand\Lai[1][]{\subpsc{\mbf{L}}{{\mathrm{a}{#1}}}{-1}}       

\DeclareRobustCommand\Lp{\subsc{\mbf{L}}{\mathrm{p}}}       
\DeclareRobustCommand\Lph{\subpsc{\L}{\mathrm{p}}{\,\dag}}       
\DeclareRobustCommand\Lptilde{\subsc{\tilde{\mbf{L}}}{\mathrm{p}}}

\DeclareRobustCommand\Lap{\subsc{\mbf{L}}{{\mathrm{ap}}}}
\DeclareRobustCommand\Laph{\subpsc{\mbf{L}}{{\mathrm{ap}}}{\,\dag}}

\DeclareRobustCommand\Laptilde{\subsc{\tilde{\mbf{L}}}{{\mathrm{ap}}}}

\newcommand{\Lu}{\mbf{L}_{u}^{}}
\newcommand{\Luh}{\mbf{L}_{u}^{\dag}}
\newcommand{\Lv}{\mbf{L}_{v}^{}}
\newcommand{\Lvh}{\mbf{L}_{v}^{\dag}}

\newcommand{\n}{\mbf{n}}
\newcommand{\nh}{\mbf{n}^{\dag}}

\newcommand\Omegaa{\subsc{\mbf{\Omega}}{\mathrm{a}}}

\renewcommand{\P}{\mbf{P}}

\renewcommand{\Pi}{\supsc{\mbf{P}}{-1}}

\newcommand\Pa{\subsc{\mbf{P}}{\!\mathrm{a}}}
\DeclareRobustCommand\Pai{\subpsc{\mbf{P}}{\!\mathrm{a}}{\,-1}}
\DeclareRobustCommand\Pah{\subpsc{\mbf{P}}{\!\mathrm{a}}{\,\dag}}
\newcommand{\Ptilde}{\tilde{\mbf{P}}}

\newcommand{\Q}{\mbf{Q}}
\newcommand{\Qh}{\supsc{\mbf{Q}}{\,\dag}}

\newcommand\Qa[1]{\subsc{\mbf{Q}}{\mathrm{a}#1}}
\newcommand\Qah[1]{\subpsc{\mbf{Q}}{\mathrm{a}#1}{\,\dag}}

\newcommand{\Qu}{\mbf{Q}_{u}^{}}

\newcommand{\Quh}{\mbf{Q}_{u}^{\dag}}

\newcommand{\Qtilde}{\tilde{\mbf{Q}}}
\newcommand{\Qtildeh}{\supsc{\tilde{\mbf{Q}}}{\,\dag}}

\DeclareRobustCommand\Qparent{\subsc{\mbf{Q}}{\parent}}
\DeclareRobustCommand\Qchild{\subsc{\mbf{Q}}{\child}}
\DeclareRobustCommand\Qparenth{\subpsc{\mbf{Q}}{\parent}{\,\dag}}
\DeclareRobustCommand\Qchildh{\subpsc{\mbf{Q}}{\child}{\,\dag}}
\DeclareRobustCommand\Qtildeparent{\subsc{\tilde{\mbf{Q}}}{\parent}}

\DeclareRobustCommand\Qtildechild{\subsc{\tilde{\mbf{Q}}}{\child}}
\DeclareRobustCommand\Qtildechildh{\subpsc{\tilde{\mbf{Q}}}{\child}{\,\dag}}

\newcommand{\qj}{\subpsc{\mbf{q}}{j}{}}
\newcommand{\qtildej}{\subpsc{\tilde{\mbf{q}}}{j}{}}

\newcommand\Ra{\subsc{\mbf{R}}{\mathrm{a}}}
\newcommand{\Rtilde}{\tilde{\mbf{R}}}

\renewcommand{\S}{\mbf{S}}

\newcommand{\Si}{\supsc{\mbf{S}}{\,-1}}
\newcommand{\Sih}{\supsc{\mbf{S}}{\,-\dag}}
\newcommand{\Sbi}{\supsc{\breve{\mbf{S}}}{\,-1}}
\newcommand{\Sbih}{\supsc{\breve{\mbf{S}}}{\,-\dag}}

\newcommand\Sa{\subsc{\mbf{S}}{\mathrm{a}}}

\newcommand{\Sai}{\subpsc{\S}{\mathrm{a}}{\,-1}}
\newcommand{\Saih}{\subpsc{\S}{\mathrm{a}}{\,-\dag}}

\newcommand{\Stilde}{\tilde{\mbf{S}}}

\newcommand{\Stildeinv}{\supsc{\tilde{\mbf{S}}}{\,-1}}
\newcommand{\Stildeinvh}{\supsc{\tilde{\mbf{S}}}{\,-\dag}}

\newcommand{\U}{\mbf{U}}

\DeclareRobustCommand\Uh{\supsc{\mbf{U}}{\,\dag}}
\DeclareRobustCommand\Uopt{\supsc{\mbf{U}}{\,\,\opt}}

\newcommand{\V}{\mbf{V}}
\DeclareRobustCommand\Vh{\supsc{\mbf{V}}{\,\dag}}
\DeclareRobustCommand\Vi{\supsc{\mbf{V}}{\,-1}}

\newcommand{\W}{\mbf{W}}
\newcommand{\Wh}{\supsc{\mbf{W}}{\,\dag}}

\newcommand{\Wmmsesym}{M}
\newcommand{\Wmmse}{\mbf{\Wmmsesym}}
\newcommand{\Wmmseh}{\supsc{\Wmmse}{\,\dag}}             

\DeclareRobustCommand\Wp{\subsc{\mbf{W}}{\!\mathrm{p}}}       
\DeclareRobustCommand\Wph{\subpsc{\mbf{W}}{\!\mathrm{p}}{\,\dag}}       
\DeclareRobustCommand\Wpi{\subpsc{\mbf{W}}{\!\mathrm{p}}{\,-1}}
\newcommand{\Wap}{\subsc{\W}{\!\mathrm{ap}}}
\newcommand{\Waph}{\subpsc{\W}{\!\mathrm{ap}}{\,\dag}}

\newcommand\Wtilde{\tilde{\W}}
\newcommand{\Wtildeh}{\supsc{\smash[t]{\tilde{\W}}}{\,\dag}}
\newcommand{\Wptilde}{\subsc{\tilde{\W}}{\!\mathrm{p}}}
\newcommand{\Wptildeh}{\subpsc{\tilde{\W}}{\!\mathrm{p}}{\,\dag}}

\newcommand{\Zp}{\mbf{Z}}
\newcommand{\Zph}{\mbf{Z}_{}^{\dag}}
\newcommand{\Zpi}{\mbf{Z}_{}^{-1}}

\newcommand{\x}{\mbf{x}}
\newcommand{\xh}{\mbf{x}^{\scalebox{.8}{$\scriptstyle \dag$}}}

\newcommand{\xML}{\mbf{x}_{\ML}^{}}
\newcommand{\xWLD}{\mbf{x}_{\WLD}^{}}
\newcommand{\xAWLD}{\mbf{x}_{\AWLD}^{}}

\DeclareRobustCommand\xparent{\subsc{\mbf{x}}{\twonegskip\parent}}

\DeclareRobustCommand\xchild{\subsc{\mbf{x}}{\onenegskip\child}}
\DeclareRobustCommand\xhatchild{\subpsc{\hat{\mbf{x}}}{\onenegskip\child}{}}
\DeclareRobustCommand\xchildh{\subpsc{\mbf{x}}{\onenegskip\child}{\,\dag}}

\newcommand{\y}{\mbf{y}}
\newcommand{\yh}{\mbf{y}^{\scalebox{.8}{$\scriptstyle \dag$}}}
\newcommand{\yt}{\tilde{\mbf{y}}}
\newcommand\ya{\subsc{\y}{\!\mathrm{a}}}
\newcommand\yah{\subpsc{\y}{\!\mathrm{a}}{\,\dag}}

\newcommand{\ytilde}{\tilde{\mbf{y}}}
\newcommand\yp{\subsc{\y}{\!\mathrm{p}}}
\newcommand\yptilde{\subsc{\tilde{\y}}{\!\mathrm{p}}}
\newcommand\yatilde{\subsc{\tilde{\y}}{\!\mathrm{a}}}
\DeclareRobustCommand\yap{\subsc{\y}{\!\mathrm{ap}}}
\DeclareRobustCommand\yaptilde{\subsc{\tilde{\y}}{\!\mathrm{ap}}}

\DeclareRobustCommand\ytildeparent{\subsc{\tilde{\mbf{y}}}{\fournegskip\parent}}

\DeclareRobustCommand\ytildechild{\subsc{\tilde{\mbf{y}}}{\threenegskip\child}}

\newcommand{\z}{\mbf{z}}

\DeclareRobustCommand\Zi{\supsc{\mbf{Z}}{\,-1}}

\newcommand{\beginfirstsupplement}{%
        \clearpage
        \thispagestyle{empty} 
        \renewcommand{\thefigure}{F\arabic{figure}}
        \renewcommand{\thetable}{T\Roman{table}}%
        \renewcommand{\figurename}{\normalsize Supplement Figure}
        \renewcommand{\tablename}{\normalsize\textsc{Supplement Table}}
        \appendices
}

\newcommand{\beginnewsupplement}{%
        \clearpage
        \thispagestyle{empty} 
}

\begin{document}
\bstctlcite[@auxout]{BSTcontrol} 

%
\title{Low-Complexity Soft-Output MIMO Detectors Based on Optimal Channel Puncturing\\
\thanks{Parts of this work have been presented at IEEE ICC 2020~\cite{2020_ICC_AWLD}.}
\thanks{This work is supported by the University Research Board
(URB) at the American University of Beirut.}
\thanks{M. M. Mansour is with the Department of Electrical and
Computer Engineering, American University of Beirut, Beirut 1107 2020,
Lebanon (e-mail: mmansour@aub.edu.lb; mmansour@ieee.org).}
\thanks{This work has been submitted to \textsc{IEEE Transactions on Wireless Communications} for possible publication. Copyright may be transferred without notice, after which this version may no longer be accessible.}
}

\author{Mohammad M. Mansour,~\IEEEmembership{Senior Member,~IEEE}}

\maketitle
\vspace{-0.45in}

%
\begin{abstract}\vspace{-0.125in}
Channel puncturing transforms a multiple-input multiple-output (MIMO) channel into a sparse lower-triangular form using the so-called WL decomposition scheme in order to reduce tree-based detection complexity. We propose computationally efficient soft-output detectors based on two forms of channel puncturing: \textit{augmented} and \textit{two-sided}. The augmented WL detector (AWLD) employs a punctured channel derived by triangularizing the true channel in augmented form, followed by left-sided Gaussian elimination. The two-sided WL detector (dubbed WLZ) employs right-sided reduction and left-sided elimination to puncture the channel. We prove that augmented channel puncturing is optimal in maximizing the lower-bound on the achievable information rate (AIR) based on a new mismatched detection model. We show that the AWLD decomposes into an MMSE prefilter and channel gain compensation stages, followed by a regular WL detector (WLD) that computes least-squares soft-decision estimates. Similarly, WLZ decomposes into a pre-processing reduction step followed by WLD. AWLD attains the same performance as the existing AIR-based partial marginalization (PM) detector, but with less computational complexity. We empirically show that WLZ attains the best complexity-performance tradeoff among tree-based detectors.
\end{abstract}\vspace{-0.15in}

\begin{IEEEkeywords}\vspace{-0.15in}
MIMO detectors, MMSE, achievable information rate, partial marginalization, channel puncturing
\end{IEEEkeywords}

%
\section{Introduction}\label{s:intro}
Modern communication systems rely on multiple-input multiple-output (MIMO) antenna configurations with large dimensions to support the aggressive targets set on spectral efficiencies. However, achieving the ideal performance promised by MIMO technology requires detectors whose complexity grows exponentially in MIMO dimensions. To support low-latency communications while providing high throughput rates, computationally efficient designs of MIMO detectors that do not incur substantial performance loss are essential.

MIMO detection is a classical problem in communications, and the literature is rich with schemes that provide various performance-complexity tradeoffs in the design space (e.g.,~\cite{2018_Xu_two_decades, 2009_larsson_MIMO_detection_methods}). The benchmark for performance in the sense of generating `good' soft decisions on the transmitted bits is maximum likelihood (ML) detection, which provides optimal performance but with exponential complexity. Alternatively, the benchmarks for low-complexity detection are the zero-forcing (ZF) and minimum mean-square error (MMSE) schemes, which decouple the transmit layers through linear filtering to generate log-likelihood ratios (LLRs) for each symbol bit in parallel, or sequentially with decision feedback.

Tree-search based detectors such as sphere decoding~\cite{2003_damen_on_ML_detection}, list decoding~\cite{2003_hochwald_achieving_nearcapacity}, and other variants map the detection problem into a search problem for the closest signal vector. They find the closest vector in $\Nt$-dimensional signal space to the received vector by forming a search-tree and recursively enumerating symbols across all layers from the parent down to the leaves. Such schemes suffer from non-deterministic search-time complexity (see~\cite{2014_sphereP2_mansour,2018_Kato_MCS,2020_He_parellel_tree_search}). To simplify the search process, fixed-complexity schemes such as~\cite{barbero2008fixing,2006_Wenk_ISCAS,2010_Studer} limit the search steps to a set of survivor paths. While these schemes are efficient in finding the ML path, they do not necessarily find all the best competing paths that are needed to generate soft decisions for each symbol bit.

An alternative concept is partial marginalization (PM)~\cite{2008_larsson_fixed_complexity,2011_persson_partial_marg}, which exhaustively enumerates only over a small subset of $\nu$ carefully chosen parent layers out of $\Nt$, and approximately marginalizes over the other $\Nt\!-\!\nu$ child layers using ZF with decision-feedback (ZF-DF) estimates. While the bit LLRs for parent symbols are easy to compute, computing bit LLRs for child symbols is complicated by three facts: 1) for each bit hypothesis of the child symbols, a separate ZF-DF process is needed, which is compute-intensive for large $\Nt$; 2) the LLRs are prone to error propagation for large $\Nt$ due to decision feedback; 3) the quality of the LLRs is very sensitive to the choice of the $\nu$ parent layers. In~\cite{2006_siti_novel_LORD}, the closely related layered orthogonal lattice detector (LORD) scheme mitigates the first drawback by operating with $\nu\!=\!1$ and computing bit LLRs for the parent symbol only; $\Nt$ independent searches using $\Nt$ trees are performed to compute the bit LLRs for all symbols by choosing a new symbol as a parent in each tree.

To overcome the second drawback, the so-called WL detection (WLD) \footnote{The WL decomposition is defined to be a decomposition of the matrix $\H$ as $\W\H\!=\!\L$, where $\W$ is a (non-unitary) filtering matrix and $\L$ is a sparse lower-triangular matrix. A detector that applies WL decomposition to the channel matrix $\H$ and detects symbols based on $\L$ is called a WL detector.} scheme~\cite{2014_mansour_SPL_WLD} first applies a (non-unitary) filtering matrix $\W$ to transform the channel into sparse lower-triangular form $\L$. It then enumerates across one parent layer and detects symbols in all other child layers in parallel via least-squares (LS) estimates without decision feedback. The channel matrix is ``punctured'' to have a special structure that breaks the connections among child nodes, while retaining connections only to the parents. Essentially, all child nodes become leaves, and hence, marginalization is exact in the LS sense. An immediate consequence is that the LS estimates of the counter hypotheses of each leaf symbol bit can be easily derived from the LS estimate itself~\cite{2014_sphereP1_mansour}. A closely related concept is the achievable information rate (AIR)-PM detector~\cite{2012_rusek_optimal_channel_short,2017_hu_softoutput_AIR}, which derives a ``shortened'' channel similar to the WLD's punctured structure using information-theoretic optimizations. Other optimal linear detectors are presented in~\cite{2019_Yang_optimal_linear_detection}.

In this paper, we show that the concepts of channel puncturing of~\cite{2014_mansour_SPL_WLD} and AIR-PM-based channel shortening of~\cite{2017_hu_softoutput_AIR} are related. After introducing the system model and reviewing tree-based detection in Sec.~\ref{s:system_model}, we present a matrix characterization of one-sided and two-sided channel puncturing based on Gaussian elimination and lattice reduction in Sec.~\ref{sec:channel_puncturing}. In Sec.~\ref{sec:WLD}, we present the WLD detection model and derive a lower bound on the achievable rate of the WLD detector, as well as a bound on the quality of its hard decision estimate, and show that these bounds approach capacity and the hard ML decision as the puncturing order increases. In Sec.~\ref{sec:augmented_WLD}, we propose a new \emph{augmented} WLD (AWLD) detection scheme, in which an augmented channel, rather than the true channel, is punctured. We derive a lower bound on the AIR of the AWLD detector and characterize its gap to capacity. In Sec.~\ref{s:optimality_AWLD_modified_detection_model}, we propose an alternate mismatched detection model compared to~\cite{2012_rusek_optimal_channel_short},
and use it to derive optimal punctured channels that maximize the AIR. We prove that the AWLD detector is optimal under this model, and is in fact equivalent to the AIR-PM detector of~\cite{2017_hu_softoutput_AIR}. The AWLD detector decomposes into an MMSE prefilter and channel gain compensation stages, followed by a WLD detector. Hence, AIR-optimal channel puncturing can be achieved using simple QL decomposition followed by Gaussian elimination. In Secs.~\ref{sec:eff_decomp_algorithms}-\ref{s:AWLD_MIMO_detection_algorithms}, we present computationally efficient matrix decomposition, puncturing, and MIMO detection algorithms based on the proposed schemes. Empirical simulation results are presented in Sec.~\ref{s:sim}. Finally, Sec.~\ref{sec:conclusion} concludes the paper. The supplementary material includes proofs, pseudo-codes of all proposed algorithms, and enlarged figures.

\emph{Notation}: $i\!=\!\sqrt{-1}$; $\mathcal{Z},\mathcal{R},\mathcal{C},\mathcal{G}\!=\!\mathcal{Z}\!+\!i\mathcal{Z}$ are the sets of integers, reals, complex numbers, and Gaussian integers; $\mbf{a}\!=\![a_k]$ column vector with elements $a_k$; $\mbf{A}\!=\![a_{kj}]$ matrix with elements $a_{kj}$; $[\mbf{A}]_k\!=\![a_{k1},\cdots,a_{kk}]$; $[\mbf{A}]_{\bar{k}}\!=\![a_{k1},\cdots,a_{k,k-1}]$; $[\mbf{A}]_{\bar{1}}\!=\!\emptyset$; $\mbf{0}_{\MbyN}\!=\!\MbyN$ zero matrix; $\I_{\Nt}\!=\!\NbyN$ identity matrix; $\mbf{e}_k^{}\!=\!\nth{k}$ column of $\I$; $\expec{\cdot}\!=\!\text{expectation}$; $\mathcal{CN}(\mbf{m},\mbf{C})$ denotes circularly-symmetric complex Gaussian distribution with mean $\mbf{m}$ and covariance matrix $\mbf{C}$; $(\cdot)^{\tran}\!=\!\text{transpose}$; $(\cdot)^{\dag}$ $=$ Hermitian transpose; $\Re{\cdot},\Im{\cdot}$ $=$ real, imaginary part;
$\diagg{\cdot}$ $=$ matrix diagonal; $\det{(\cdot)}$ $=$ determinant; $\fnormm{\cdot}\!=\!L_2$ norm; $\fnormm{\cdot}_{\mathrm{F}}$ $=$ Frobenius norm; $\mbf{A}^{1/2}$ $=$ matrix square-root; $\mbf{A}\!\succeq\!\mbf{B}$ denotes $(\mbf{A}\!-\!\mbf{B})$ positive semidefinite; $\cong$ denotes equality up to an additive constant.

%
\vspace{-0.1in}
\section{System Model and Layered Detection}\label{s:system_model}
Let $\H\!\in\!\mathcal{C}^{\MbyN}$ model a MIMO communication channel with $\Nt$ transmit antennas and $\Nr\!\geq\!\Nt$ receive antennas. The transmit signal $\x\!=\![x_n]\!\in\!\X^{\Nt\times 1}$ is composed of $\Nt$ symbols $x_n$ drawn from constellation $\X$ with average energy $\expecc{x_nx_n^{\dag}}\!=\!\Es$ and size $\abs{\mathcal{X}}\!=\!Q$. Each symbol $x_n$ is mapped from $B\!=\!\subsc{\log}{2}Q$ bits $x_{n,b}\!\in\!\{\pm 1\}$ as $x_n\!=\!\subpsc{(x_{n,b})}{b=1}{\,B}$. Assuming $\H$ is perfectly known only at the receiver, the receive signal $\y\!\in\!\mathcal{C}^{\Nr\times 1}$ is modeled using the input-output relation\vspace{-0.175in}
\begin{align}\label{eq:system_model}
    \y = \H\x + \mbf{n},
\end{align}\\[-2.5em]
where the noise term $\mbf{n}\!\sim\! \mathcal{CN}(\mbf{0}_{\Nr\times 1}^{},N_0^{}\I_{\Nr}^{})$ and $N_0$ is the noise variance. The conditional probability $ p(\y|\x)$ and metric $\mu(\y|\x)$ according to~\eqref{eq:system_model} are\vspace{-0.175in}
\begin{align}
    p(\y|\x)
    &=
    \tfrac{1}{(\pi N_0)^{\Nr}}\! \exp{\left( \mu(\y|\x) \right)}, \label{eq:true_detection_prob}
    \\[-0.5em]
    \mu(\y|\x)
    &=
    -\tfrac{1}{N_0}\fnormmsq{\y\!-\!\H\x} \label{eq:true_metric}
    \\[-0.5em]
    &=
    -\tfrac{1}{N_0}(\yh\y \!-\! 2\Re{\yh\H\x} \!+\! \xh\Hh\H\x) \label{eq:true_metric_expanded}
    \\[-0.5em]
    &\cong
    2\Re{\yh\H\x} \!-\! \xh\Hh\H\x. \label{eq:true_metric_expanded_sufficient}
\end{align}\\[-2.5em]
Using the observation $\y$ and assuming no prior information on $\x$ (i.e., $P(x_{n,b}\!=\!+1)\!=\!P(x_{n,b}\!=\!-1)\!=\!\tfrac{1}{2}$), the ML detector generates the LLR of the $\nth{b}$ bit $x_{n,b}$ of the $\nth{n}$ symbol $x_n$ in $\x$ as\vspace{-0.15in}
\begin{align}
    L(x_{n,b}|\y)
    &=
    \ln
    \frac{\sum_{\x:x_{n,b}=+1}\exp\left(\mu(\y|\x)\right)}
         {\sum_{\x:x_{n,b}=-1}\exp\left(\mu(\y|\x)\right)}.
\label{eq:LLR_exact_ML_def}
\end{align}\\[-2.25em]
To avoid computing sums of exponentials in~\eqref{eq:LLR_exact_ML_def}, the ML detector with Max-Log approximation (MLM) recursively applies the Jacobian approximation $\ln(e^c\!+\!e^d)\!\approx\!\max(c,d)$~\cite{2005_moon_error_correcting_codes}
to the exponentials in~\eqref{eq:LLR_exact_ML_def}, and approximates $L(x_{n,b}|\y)$ by $\Lambda(x_{n,b}|\y)$ as\vspace{-0.15in}
\begin{align}
    \Lambda(x_{n,b}|\y)
    &=
    \max_{\x:x_{n,b}=+1}\mu(\y|\x) -
    \max_{\x:x_{n,b}=-1}\mu(\y|\x).
\label{eq:LLR_exact_MLM_def}
\end{align}\\[-2.25em]
In the absence of any structure on $\H$ or any further simplifying assumptions, computing the sums in~\eqref{eq:LLR_exact_ML_def} or the max terms in~\eqref{eq:LLR_exact_MLM_def} have exponential complexities
in $\Nt$.\\[-1.75em]

%
\vspace{-0.1in}
\subsection{Tree-based Layered Detection}\label{sec:tree_detection}
Detecting symbols and generating bit LLRs can be done efficiently on a tree. By triangularizing $\H$ and associating symbols with edges and partial Euclidean distances with nodes, symbols can be detected by searching the tree for a path from the root to a leaf with minimal weight.

Let $\H\!=\!\Q\L$ denote the thin QL decomposition (QLD)~\cite{2013_golub_matrix} of $\H$, where $\Q\!\in\!\mathcal{C}^{\MbyN}$ has orthonormal columns ($\Qh\Q\!=\!\I$) and $\L\!\in\!\mathcal{C}^{\NbyN}$ is a square lower-triangular matrix with real and positive diagonal elements. We write $\y\!-\!\H\x$ in terms of $\Q,\L$ as $\y\!-\!\H\x\!=\!\Q(\Qh\y\!-\!\L\x) \!+\! (\I\!-\!\Q\Qh)\y$. Since $\Q\!\perp\!(\I\!-\!\Q\Qh)$, i.e., $\Qh(\I\!-\!\Q\Qh)\!=\!\mbf{0}$, the squared-distance in~\eqref{eq:true_metric} can be expanded as $\fnormmsq{\y\!-\!\H\x} \!=\!\fnormmsq{\Q(\Qh\y\!-\!\L\x)} \!+\! \fnormmsq{(\I\!-\!\Q\Qh)\y}$. Since $\Qh\Q\!=\!\I$, then $\Q$ does not scale Euclidean distances. Also the term $\fnormmsq{(\I\!-\!\Q\Qh)\y}$ is independent of $\x$ and hence is irrelevant for detection. Thus, it suffices to work with the quantity $\fnormmsq{\yt\!-\!\L\x}$ rather than $\fnormmsq{\y\!-\!\H\x}$ in~\eqref{eq:true_metric}, with $\yt\!=\!\Qh\y$.

With proper layer ordering and partial marginalization, the tree can be searched by enumerating only over a subset of $\nu$ parent layers, rather than all the layers. Let $\xparent\!=\!\supsc{[x_1,\cdots,x_{\nu}]}{\,\tran}$ and $\xchild\!=\!\supsc{[x_{\nu+1},\cdots,x_{\Nt}]}{\,\tran}$ denote the parent and child symbol vectors, respectively. Let $\ytildeparent,\ytildechild$ be similarly defined from $\ytilde$. Define the variables $w_k$ and $z_k$ as\vspace{-0.15in}
\begin{IEEEeqnarray}{rClrCl}
    w_k &=&
    \tilde{y}_k
    -\!\!
    \sum_{j=1}^{\min\{k,\nu\}}\!\! l_{kj}x_j,
    &
    \qquad z_k &=&
    w_k
    -\!\!
    \sum_{j=\nu+1}^{k-1}\!\!l_{kj}x_j,  \label{eq:wk_zk_def}
\end{IEEEeqnarray}\\[-2.25em]
for $k\!=\!1,\cdots,\Nt$.  Note that $w_k$ depends only on $\xparent$, while $z_k$ depends on both $\xparent$ and $\xchild$. The weight of a parent node ($1\!\leq \!k\! \leq \! \nu$) and a child node ($\nu\!+\!1 \!\leq \!k\! \leq \! \Nt$) are given by\vspace{-0.15in}
\begin{IEEEeqnarray}{rClCrCCCl}
    \eparent{w_k} &=& -\tfrac{1}{N_0}\abs{w_k}^2,&\quad& 1&\leq &k& \leq& \nu,
    \label{eq:mu_w_def}
    \\[-0.5em]
    \echild{z_k,x_k}&=& -\tfrac{1}{N_0}\abs{z_k - l_{kk}x_k}^2,&\quad& \nu+1&\leq &k& \leq& \Nt.
    \label{eq:mu_z_def}
\end{IEEEeqnarray}\\[-2.5em]
The weight of a path associated with symbols $\x\!=\![\xparent;~\xchild]$ is\vspace{-0.15in}
\begin{align}
    \mu(\ytilde|\x)
    &=
    \sum_{k=1}^{\nu}\!\eparent{w_k}
    +
    \sum_{k=\nu+1}^{\Nt}\!\!\! \echild{z_k,x_k}
    \triangleq
    \muparent{\ytildeparent|\xparent} + \muchild{\ytildechild|\xparent,\xchild}. \label{eq:path_weight_def}
\end{align}\\[-2.25em]
Maximizing $\mu(\ytilde|\x)$ over all $\x$ such that $x_{n,b}$ is $s\!=\!\pm 1$, for $b\!=\!1,\cdots,B$, $n\!=\!1,\cdots,\Nt$, can be expressed using~\eqref{eq:path_weight_def} as\vspace{-0.2in}
\begin{align}
    \max_{{\x:  x_{n,b}=s} \atop 1\leq n \leq \nu}
    \mu(\ytilde|\x)
    &=
    \max_{{\mbf{x}_{\parent}:} \atop {x_{n,b}=s}}
    \bigl\{
        \muparent{\ytildeparent|\xparent}
        +
        \max_{\mbf{x}_{\child}}
        \muchild{\ytildechild|\xparent,\xchild}
    \bigr\},\label{eq:max_mu_parent_exact}
    \\[-0.25em]
    \max_{{\x:  x_{n,b}=s} \atop \nu+1\leq n \leq \Nt}
    \mu(\ytilde|\x)
    &=
    \max_{\mbf{x}_{\parent}}
    \bigl\{
        \muparent{\ytildeparent|\xparent}
        +
        \max_{{\mbf{x}_{\child}:} \atop {x_{n,b}=s}}
        \muchild{\ytildechild|\xparent,\xchild}
    \bigr\}.\label{eq:max_mu_child_exact}
\end{align}\\[-2.5em]
The inner $\max$ operations in~\eqref{eq:max_mu_parent_exact}-\eqref{eq:max_mu_child_exact} can be approximated by successively solving using ZF-DF for symbols in $\xchild$ having $\xparent$ as parents. Let $\slice{z}_{\!\mathcal{X}}$ and $\slice{z}_{\!\subpsc{\mathcal{X}}{b}{\,(s)}}$ denote slicing to the closest symbol to $z$ in $\X$ and $\subpsc{\mathcal{X}}{b}{\,(s)}\!\triangleq\!\{x_n\!\in\!\mathcal{X}\!:x_{n,b}\!=\!s\}$, respectively. When the hypothesis is for a parent symbol bit ($1\!\leq\!n\!\leq\!\nu$), ZF-DF on child symbols proceeds as follows:\vspace{-0.15in}
\begin{IEEEeqnarray*}{rClrClrCl}
    k & \!=\! & \nu\!+\!1,\cdots,\Nt\!:
    & ~~\hat{z}_k & \!=\! & w_k \!-\!\! \sum_{j=\nu+1}^{k-1}\!\!l_{kj}\hat{x}_j,
    & ~~\hat{x}_k & \!=\! & \slice{\hat{z}_k/l_{kk}}_{\!\mathcal{X}}.
\end{IEEEeqnarray*}\\[-2.25em]
Set $\xhatchild\!=\!\supsc{[\hat{x}_{\nu+1},\cdots,\hat{x}_{\Nt}]}{\,\tran}$ to be the child symbol vector estimate. On the other hand, for a child symbol bit hypothesis ($\nu\!+\!1\!\leq\!n\!\leq\!\Nt$), ZF-DF on child symbols $k\!=\!\nu\!+\!1,\cdots,\Nt$ proceeds as:\vspace{-0.15in}
\begin{IEEEeqnarray*}{rClrClrCl}
    k & \!<\! & n\!:
    & ~~\hat{z}_k & \!=\! & w_k \!-\!\! \sum_{j=\nu+1}^{k-1}\!\!l_{kj}\hat{x}_j,
    & ~\hat{x}_k & \!=\! & \slice{\hat{z}_k/l_{kk}}_{\!\mathcal{X}};
    \\[-0.5em]
    k & \!=\! & n\!:
    & ~~\hat{z}_k &\!=\! & w_k \!-\!\! \sum_{j=\nu+1}^{k-1}\!\!l_{kj}\hat{x}_j,
    & ~\subpsc{\hat{x}}{k;b}{\,(s)} & \!\triangleq\! &
    \slice{\hat{z}_k/l_{kk}}_{\!\subpsc{\mathcal{X}}{b}{\,(s)}};
    \\[-0.5em]
    k & \!>\! & n\!:
    & ~~\hat{z}_k & \!=\! & w_k \!-\!\!
    \sum_{{j=\nu+1}:\atop {j\neq n}}^{k-1}\!\!l_{kj}\hat{x}_j
    \!-\!l_{kn}\subpsc{\hat{x}}{n;b}{\,(s)},
    & ~\hat{x}_k & \!=\! & \slice{\hat{z}_k/l_{kk}}_{\!\mathcal{X}}.
\end{IEEEeqnarray*}\\[-2.00em]
Let $\subpsc{\hat{\mbf{x}}}{2\,n;b}{\,(s)}\!=\!\supsc{[x_{\nu+1},\cdots,\subpsc{\hat{x}}{n;b}{\,(s)},\cdots,x_{\Nt}]}{\,\tran}$ be the resulting child symbol vector estimate. Therefore, the inner $\max$ operations in~\eqref{eq:max_mu_parent_exact}-\eqref{eq:max_mu_child_exact} are approximated as\vspace{-0.15in}
\begin{flalign}
    \max_{\mbf{x}_{\child}}\muchild{\ytildechild|\xparent,\xchild}  &\geq \!\!
    \sum_{k=\nu+1}^{\Nt}\!\! \max_{x_k} \echild{\hat{z}_{k},x_k}
    =\!\!\!
    \sum_{k=\nu+1}^{\Nt}\!\!\! \echild{\hat{z}_{k},\hat{x}_k}
    \triangleq
    \muhatchild{\ytildechild|\xparent,\xhatchild},
    \label{eq:max_mu_xc_parent_approx}
    \\[-0.5em]
    \max_{{\mbf{x}_{\child}:} \atop {x_{n,b}=s}}
    \!\! \muchild{\ytildechild|\xparent,\xchild}  &\geq \!\!
    \sum_{{k=\nu+1}: \atop k\neq n}^{\Nt}\!\!\!\! \max_{x_k} \echild{\hat{z}_{k},x_k}
        \!+\!\!
        \max_{x_n: \atop x_{n,b}=s}\! \echild{\hat{z}_{n},x_n}\notag
    \\[-0.5em]
    &=\!\!\!
    \sum_{{k=\nu+1}:\atop k\neq n}^{\Nt}\!\!\!\! \echild{\hat{z}_{k},\hat{x}_k}
        \!+\!
        \echild{\hat{z}_{n},\subpsc{\hat{x}}{n;b}{\,(s)}}
        =
    \muhatchild{\ytildechild|\xparent,\subpsc{\hat{\mbf{x}}}{2\,n;b}{\,(s)}},
    \label{eq:max_mu_xc_child_approx}
\end{flalign}\\[-2.00em]
and~\eqref{eq:max_mu_parent_exact}-\eqref{eq:max_mu_child_exact} are approximated as \vspace{-0.15in}
\begin{align}
    \max_{{\x:  x_{n,b}=s} \atop 1\leq n \leq \nu}
    \!\mu(\ytilde|\x)
    &\geq
    \max_{{\mbf{x}_{\parent}:} \atop {x_{n,b}=s}}
    \!
    \bigl\{
        \muparent{\ytildeparent|\xparent}
        +
        \muhatchild{\ytildechild|\xparent,\xhatchild}
    \bigr\},\label{eq:max_mu_parent_approx}
    \\[-0.25em]
    \max_{{\x:  x_{n,b}=s} \atop \nu+1\leq n \leq \Nt}
    \!\!\!
    \mu(\ytilde|\x)
    &\geq
    \max_{\mbf{x}_{\parent}}
    \bigl\{
        \muparent{\ytildeparent|\xparent}
        +
        \muhatchild{\ytildechild|\xparent,\subpsc{\hat{\mbf{x}}}{2\,n;b}{\,(s)}}
    \bigr\}.\label{eq:max_mu_child_approx}
\end{align}\\[-3.25em]

The above maxima are not optimal because of the ZF-DF operations on the child layers in~\eqref{eq:max_mu_xc_parent_approx}-\eqref{eq:max_mu_xc_child_approx}. However, if the $l_{kj}$ terms are 0 for $k\!=\!\nu\!+\!2,\cdots,\Nt$ and $j\!=\!\nu\!+\!1,\cdots,k\!-\!1$ in the $z_k$ summation in~\eqref{eq:wk_zk_def}, then $z_k\!=\!w_k$, $e(z_k,x_k)
\!=\! \tfrac{-1}{N_0}\!\supsc{\abs{w_k \!-\! l_{kk}x_k}}{\,2}$, and the maximizations in~\eqref{eq:max_mu_xc_parent_approx}-\eqref{eq:max_mu_xc_child_approx} become exact in this case:\vspace{-0.15in}
\begin{align*}
    \max_{\mbf{x}_{\child}}
    \sum_{k=\nu+1}^{\Nt}\!\!\!
    \echild{w_k,x_k}
    \!&=\!\!
    \sum_{k=\nu+1}^{\Nt}\!\!\!
    \max_{x_k}
    \echild{w_k,x_k}
    \!=\!\!\!
    \sum_{k=\nu+1}^{\Nt}\!\!\!
    \echild{w_k,\hat{x}_k},
    \\[-0.25em]
    \max_{{\mbf{x}_{\child}:} \atop  x_{n,b}=s}
    \!
    \sum_{k=\nu+1}^{\Nt}\!\!\!
    \echild{w_k,x_k}
    \!&=\!\!\!
    \sum_{{k=\nu+1}: \atop k\neq n}^{\Nt}\!\!\! \max_{x_k} \echild{w_{k},x_k}
        \!+\!\!\!
        \max_{x_n: \atop x_{n,b}=s}\! \echild{w_{n},x_n}
    =
    \sum_{{k=\nu+1}:\atop k\neq n}^{\Nt}\!\!\!\! \echild{w_{k},\hat{x}_k}
        \!+\!
        \echild{w_{n},\subpsc{\hat{x}}{n;b}{\,(s)}}.
\end{align*}\\[-2.00em]
In addition, all intermediate complex products involving the zeroed entries $l_{kj}$ are not needed.\vspace{-0.15in}

%
\subsection{Single-Tree and Multi-Tree Approaches}
Soft-output detection is essentially a multi-point search problem for the ML point and all its counter-ML hypotheses. Tree-search algorithms used to generate bit LLRs for channels partitioned into parent and child symbol layers follow either a single-tree or a multi-tree approach to find these points. For single-tree, a pre-processing step chooses $\nu$ ordered layers as parents and $\Nt\!-\!\nu$ ordered layers as children; one tree is used to solve for both parent and child bit LLRs. For multi-tree, $\Nt/\nu$ trees are used to solve only for parent bit LLRs, such that a different combination of layers is chosen as parents for each tree.

Both approaches use enumeration over the parent layers, and marginalization over the child layers. Marginalization complicates LLR generation of child bits for single-tree because it has to be repeated for every child bit hypothesis and for every candidate parent symbol vector. Also, the quality of the LLRs under the single-tree approach is very sensitive to the choice of parent layers and overall ordering of layers. On the other hand, in the multi-tree approach the distinct layer orderings of each tree constitute an added diversity that can be leveraged to globally optimize the closest points locally searched by each tree and their metrics across all the trees.

%
\vspace{-0.1in}
\section{Channel Puncturing}\label{sec:channel_puncturing}
Motivated by the observation from the last section to improve the efficiency and reduce the computational complexity of the detection process by nulling entries below the main diagonal of $\L$, we next investigate possible puncturing schemes that are applicable to integer LS problems.

Consider the lower-triangular matrix shown in Fig.~\ref{fig:puncturing_pattern}. To null all entries below the diagonal and to the right of the $\nth{\nu}$ column of $\L\!=\![l_{kj}^{}]$ ($l_{kj}^{}\!\leftarrow\!0$ for $\nu\!+\!1\!<\!k\!\leq\!\Nt$ and $\nu\!<\!j\!<\!k$) for some $\nu$, $1\!\leq \nu\!\leq\Nt\!-\!1$, we partition $\L$ conformally as\vspace{-0.15in}
\begin{align}
    \mbf{L}_{\Nt\times\Nt}
    &=
    \begin{bNiceMatrix}
        \mbf{P}_{\nu\times\nu} & \mbf{0}_{\nu\times(\Nt-\nu)}^{}  \\
        \mbf{R}_{(\Nt-\nu)\times \nu}^{} & \mbf{S}_{(\Nt-\nu)\times(\Nt-\nu)}^{}
    \end{bNiceMatrix},\label{eq:L_nu_partitioned} 
\end{align}\\[-2.00em]
where $\P\!\in\!\mathcal{C}^{\nu\times \nu}$ and $\S\!\in\!\mathcal{C}^{(\Nt-\nu)\times (\Nt-\nu)}$ are complex square lower-triangular matrices of sizes $\nu$ and $\Nt\!-\!\nu$, respectively, having real diagonal elements, and $\mbf{R}\!\in\!\mathcal{C}^{(\Nt-\nu)\times \nu}$ is a complex rectangular matrix. The target of puncturing is to diagonalize $\mbf{S}$. Hence, without loss of generality, we focus on techniques to diagonalize $\mbf{S}$ that do not alter Euclidean distances of the form $\fnormmsq{\y\!-\!\mbf{L}\x}$. Henceforth, $\L$ is assumed to be non-singular.

Using two-sided unitary transformations $\Wp$ and $\mbf{Z}$ of size $\Nt$, it is well-known that $\mbf{S}$ or all of $\L$ can be reduced to diagonal form $\D\!=\Wp\mbf{L}\mbf{Z}$ via an SVD-like decomposition~\cite{2013_golub_matrix}. The left transformation $\Wp$ must be unitary in order to preserve $L_2$-norms and not alter noise statistics:\vspace{-0.15in}
\begin{IEEEeqnarray}{rClClCrCl}
\fnormmsq{\Wp\x} &=& \xh\Wph\Wp\x &=& \fnormmsq{\x} &\Rightarrow& \Wph\Wp &=& \I,\qquad \label{eq:Wp_col_orthonormality_cond}\\[-0.5em]
\expecc{\Wp\n\nh\Wph} &=& N_0\Wp\Wph &=& N_0\I &\Rightarrow& \Wp\Wph &=& \I.\qquad
\label{eq:Wp_row_orthonormality_cond}
\end{IEEEeqnarray}\\[-2.25em]
The right transformation $\mbf{Z}$ must preserve the (Gaussian) integer nature of the unknown $\x$; that is, if for some $\y\!\in\!\mathcal{C}^{\Nt}$\vspace{-0.15in}
\begin{align}
\mbf{z}^{\star}
\!&=\!
\argmin_{\z\in\mbf{Z}\X^{\Nt}}\,\fnormmsq{\y - \mbf{L}\z},\quad\text{then}\quad
\mbf{Z}^{-1}\mbf{z}^{\star} \!=\! \argmin_{\x\in\X^{\Nt}}\,\fnormmsq{\y - \mbf{L}\mbf{Z}\x} \in \X^{\Nt}.\label{eq:Zp_integer_sol_cond}
\end{align}\\[-2.25em]
Hence $\Zp$ has to be \emph{unimodular}, i.e., an integer matrix in $\mathcal{G}^{\Nt\times\Nt}$ with integer inverse having $\abs{\det{\Zp}}\!=\!1$. Also, $\Zp$ must be lower-triangular in order to induce a parent-child tree structure using forward substitution, and hence cannot be unitary ($\Zph$ is upper triangular, while $\Zpi$ is lower triangular; hence they cannot be equal). However, if $\Zp$ is not unitary, $\Wp$ being unitary and applied from the left cannot alone null an element below the main diagonal of $\mbf{S}$ without creating a non-zero entry in its upper-triangular counterpart, hence altering the lower-triangular structure of $\mbf{S}$. Therefore, both $\Wp$ and $\Zp$ cannot be unitary, and~\eqref{eq:Wp_col_orthonormality_cond}-\eqref{eq:Wp_row_orthonormality_cond} cannot be satisfied.\\[-2.0em]

\begin{figure}[t]
  \centering
  \pgfmathsetmacro{\scl}{0.25}%

\tikzstyle{background}=[rectangle,
                        thin,
                        dashed,
                        draw=black!90,
                        fill=yellow!10,
                        inner sep=0.5mm,
                        rounded corners=0.5mm]

\tikzstyle{dummy_yellow}=[circle,
                   thick,
                   minimum size=0.8cm,
                   draw=yellow!10,
                   fill=yellow!10,
                   scale=\scl]

\tikzstyle{dummy_white}=[circle,
                   thick,
                   minimum size=0.8cm,
                   draw=white,
                   fill=white,
                   scale=\scl]

\tikzstyle{black_dot}=[circle,
                       minimum size=0.8cm,
                       thick,
                       draw=black!100,
                       fill=black!60,
                       scale=\scl]

\tikzstyle{blue_dot}=[circle,
                   thick,
                   minimum size=0.8cm,
                   draw=blue!80,
                   fill=blue!20,
                   scale=\scl]

\begin{tikzpicture}[scale=1]
    \matrix[row sep=.125cm,column sep=.125cm,left delimiter={[},right delimiter={]}](L) at (0,0) {
    \node(n11)[black_dot] {};
    &
    \node(n12)[dummy_yellow] {};
    & & & & & &
    \\
    \node(n21)[black_dot] {};
    &
    \node(n22)[black_dot] {};
    & & & & & &
    \\
    \node(n31)[black_dot] {};
    &
    \node(n32)[black_dot] {};
    &
    \node(n33)[black_dot] {};
    & & & & &
    \\
    \node(n41)[black_dot] {};
    &
    \node(n42)[black_dot] {};
    &
    \node(n43)[blue_dot] {};
    &
    \node(n44)[black_dot] {};
    & & & &
    \\
    \node(n51)[black_dot] {};
    &
    \node(n52)[black_dot] {};
    &
    \node(n53)[blue_dot] {};
    &
    \node(n54)[blue_dot] {};
    &
    \node(n55)[black_dot] {};
    & & &
    \\
    \node(n61)[black_dot] {};
    &
    \node(n62)[black_dot] {};
    &
    \node(n63)[blue_dot] {};
    &
    \node(n64)[blue_dot] {};
    &
    \node(n65)[blue_dot] {};
    &
    \node(n66)[black_dot] {};
    & &
    \\
    \node(n71)[black_dot] {};
    &
    \node(n72)[black_dot] {};
    &
    \node(n73)[blue_dot] {};
    &
    \node(n74)[blue_dot] {};
    &
    \node(n75)[blue_dot] {};
    &
    \node(n76)[blue_dot] {};
    &
    \node(n77)[black_dot] {};
    &
    \\
    \node(n81)[black_dot] {};
    &
    \node(n82)[black_dot] {};
    &
    \node(n83)[blue_dot] {};
    &
    \node(n84)[blue_dot] {};
    &
    \node(n85)[blue_dot] {};
    &
    \node(n86)[blue_dot] {};
    &
    \node(n87)[blue_dot] {};
    &
    \node(n88)[black_dot] {};
    \\
    };
    \begin{pgfonlayer}{background}
        \node [background,
                    fit=(n11)(n21)(n22),
                    label=above:$\mbf{P}$] {};
        \node [background,
                    fit=(n31)(n81)(n82),
                    label=below:$\mbf{R}$] {};
        \node [background,
                    fit=(n33)(n83)(n88),
                    label=below:$\mbf{S}$] {};
    \end{pgfonlayer}
    \matrix[row sep=.125cm,column sep=.125cm,left delimiter={[},right delimiter={]}](Lp) at (5,0) {
    \node(m11)[black_dot] {};
    &
    \node(m12)[dummy_white] {};
    & & & & & &
    \\
    \node(m21)[black_dot] {};
    &
    \node(m22)[black_dot] {};
    & & & & & &
    \\
    \node(m31)[black_dot] {};
    &
    \node(m32)[black_dot] {};
    &
    \node(m33)[black_dot] {};
    & & & & &
    \\
    \node(m41)[black_dot] {};
    &
    \node(m42)[black_dot] {};
    & &
    \node(m44)[black_dot] {};
    & & & &
    \\
    \node(m51)[black_dot] {};
    &
    \node(m52)[black_dot] {};
    & & &
    \node(m55)[black_dot] {};
    & & &
    \\
    \node(m61)[black_dot] {};
    &
    \node(m62)[black_dot] {};
    & & & &
    \node(m66)[black_dot] {};
    & &
    \\
    \node(m71)[black_dot] {};
    &
    \node(m72)[black_dot] {};
    & & & & &
    \node(m77)[black_dot] {};
    &
    \\
    \node(m81)[black_dot] {};
    &
    \node(m82)[black_dot] {};
    & & & & & &
    \node(m88)[black_dot] {};
    \\
    };
    \draw[->>,very thick] (L.east)+(0.35cm,0) -- node [yshift=1.5ex] {$\Wp$} ($(Lp.west) + (-0.35cm,0)$);

    \draw[decorate,decoration={brace,raise=5pt},yshift=20pt]  (m11.west) -- (m12.east) node [black,midway,yshift=0.4cm]
{\footnotesize $\nu=2$};

\end{tikzpicture}
  \vspace{-0.2in}
  \caption{Puncturing an $8\!\times\! 8$ matrix $\L$ into $\Lp$ using $\Wp$ for $\nu=2$.}\label{fig:puncturing_pattern}
\end{figure}

%
\vspace{-0.1in}
\subsection{One-Sided Puncturing Transformations}\label{sec:one_sided_puncturing}
The matrix $\L$ in~\eqref{eq:L_nu_partitioned} can be punctured into $\Lp\!\in\!\mathcal{C}^{\NbyN}$ using a left puncturing matrix $\Wp\!\in\!\mathcal{C}^{\NbyN}$ only ($\Zp\!=\!\I$) as follows:\vspace{-0.15in}
\begin{align}
    \Wp &= \Dp \diagg{\L}
        \begin{bNiceMatrix}
            \I & \mbf{0}  \\
            \mbf{0} & \Si
        \end{bNiceMatrix}, \label{eq:WP_def}
        \\[-0.25em]
    \Lp &= \Wp\L
        =
        \Dp \diagg{\L}
        \begin{bNiceMatrix}
            \I & \mbf{0}  \\
            \mbf{0} & \Si
        \end{bNiceMatrix}
        \begin{bNiceMatrix}
            \mbf{P} & \mbf{0} \\
            \mbf{R} & \mbf{S}
        \end{bNiceMatrix}
        =
        \Dp \diagg{\L}
        \begin{bNiceMatrix}
            \mbf{P} & \mbf{0}  \\
            \Si\mbf{R} & \I
        \end{bNiceMatrix},  \label{eq:LP_def}
\end{align}\\[-1.75em]
where $\Dp\!\in\!\mathcal{R}^{\NbyN}$ is a (normalizing) diagonal matrix. Since $\Wp$ is not unitary, both conditions~\eqref{eq:Wp_col_orthonormality_cond}-\eqref{eq:Wp_row_orthonormality_cond} are not met. We can relax~\eqref{eq:Wp_row_orthonormality_cond} by choosing $\Dp$ so that $\Wp$ satisfies $\diagg{\Wp\Wph}\!=\!\I_{\Nt}$ instead. Hence\vspace{-0.15in}
\begin{align}
        \Dp
        &=
        \supsc{\diagg{\L}}{-1}\!
        \begin{bNiceMatrix}
            \I & \mbf{0}   \\
            \mbf{0} & \mbf{\Omega}
        \end{bNiceMatrix},\qquad
        \Wp
        =
        \begin{bNiceMatrix}
            \I & \mbf{0}  \\
            \mbf{0} & \mbf{\Omega}\,\Si
        \end{bNiceMatrix},\qquad
        \Lp
        =
        \begin{bNiceMatrix}
            \mbf{P} & \mbf{0}  \\
            \mbf{\Omega}\Si\mbf{R} & \mbf{\Omega}
        \end{bNiceMatrix},
        \label{eq:Dp_Wp_Lp_formula}
\end{align}\\[-2.25em]
where\vspace{-0.195in}
\begin{align}\label{eq:Sigma}
    \mbf{\Omega}&=
    \supsc{\diagg{\Si\Sih}}{-1/2}.
\end{align}\\[-4.1em]

Note that $\Wp$ is a non-singular lower-triangular matrix with $\nu$ ones and $\Nt\!-\!\nu$ positive real numbers on the diagonal. Also, since $\mbf{\Omega}$ normalizes $\Si$ so that $\diagg{\Wp\Wph}\!=\!\I_{\Nt}$, then $\subsc{\fnormm{\Wp}}{\mathrm{F}}\!=\!\sqrt{\Nt}$ and the remaining $\Nt\!-\!\nu$ eigenvalues $\lambda$ of $\Wp$ (i.e., diagonal elements of $\mbf{\Omega}\,\Si$) satisfy $0\!<\!\lambda\!\leq\!1$. It follows that $\sqrt{\Nt}\!\geq\!\sigma_{\max} \!\geq\! \lambda_{\max} \!=\! 1$ and $0 \! < \! \sigma_{\min} \!\leq\! \lambda_{\min} \!\leq\! 1$, where $\sigma_{\max}$ ($\sigma_{\min}$) and $\lambda_{\max}$ ($\lambda_{\min}$) are the maximum (minimum) singular values and eigenvalues of $\Wp$, respectively.
\begin{figure}
  \centering
  \includegraphics[scale=1]{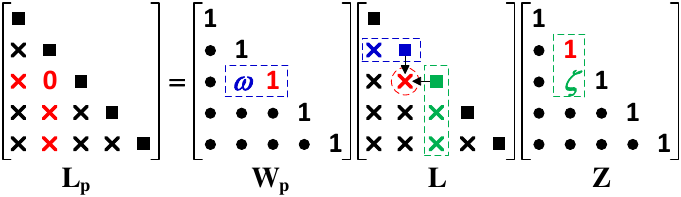}
  \vspace{-0.175in}
  \caption{Puncturing entry $l_{32}$ of $\L$ using two-sided transformations.}\label{fig:left_right_puncturing}
\end{figure}

%
\vspace{-0.1in}
\subsection{Two-Sided Puncturing Transformations}\label{sec:two_sided_puncturing}
Note that the lower-triangular matrix $\Si$ in the left non-unitary transformation $\Wp$ in~\eqref{eq:WP_def} is equivalent to a Gaussian elimination matrix. Using an integer Gauss \emph{reduction} matrix $\mbf{Z}$ as a right transformation can help approximate~\eqref{eq:Wp_col_orthonormality_cond} better by first reducing the lower-triangular entries of $\L$ in $\mbf{S}$ using integer multiples of the diagonal elements and then completely eliminating the remainder using $\Wp$ from the left. The reduction step by $\mbf{Z}$ from the right to reduce $l_{kj}$ by an integer multiple of $l_{kk}$ into $\tilde{l}_{kj}$, followed by an elimination step by $\Wp$ from the left to null $\tilde{l}_{kj}$ using $l_{kk}$ are expressed as (see Fig.~\ref{fig:left_right_puncturing})\vspace{-0.15in}
\begin{IEEEeqnarray}{rCrClrCl}
    \text{reduction }\mbf{Z}_{kj}^{}
    &:&
    \zeta_{kj} &\!=\!&  \bigl\lfloor{\tfrac{l_{kj}}{l_{kk}}}\bigr\rceil,~
    &
    \tilde{l}_{kj}  &\!=\!& l_{kj} \!-\! \zeta_{kj} l_{kk}, \quad \label{eq:Zkj_operation}\\
    \text{elimination }\subsc{\mbf{W}}{\!\!\mathrm{p}\,kj}
    &:&
    \omega_{kj} &\!=\!&  \tfrac{\tilde{l}_{kj}}{l_{kk}},~
    &
    l_{kj}  &\!=\!& \tilde{l}_{kj} \!-\!  \omega_{kj} l_{kk},\quad\label{eq:Wpkj_operation}
\end{IEEEeqnarray}\\[-2.25em]
for $k\!=\! \nu\!+\!2,\cdots,\Nt$ and $j\!=\!\nu\!+\!1,\cdots,k\!-\!1$, where $\slice{z}\!=\!\slice{\Re{z}}\!+\!i\slice{\Im{z}}$ and $\slice{a}\!=\!\flr{a+1/2}$ for $a\!\in\!\mathcal{R}$. In particular, since $|{a \!-\! \lfloor{\tfrac{a}{b} \!+\!\tfrac{1}{2}}\rfloor b }| \!\leq\! \tfrac{\abs{b}}{2}$ for $a,b\!\in\!\mathcal{R}$, then~\eqref{eq:Zkj_operation} results in\vspace{-0.15in}
\begin{align}
    {\smash[tb]{\bigl|\tfrac{\tilde{l}_{kj}}{l_{kk}} \bigr|} }
    =
    \bigl|\bigl(l_{kj} - \bigl\lfloor{\tfrac{l_{kj}}{l_{kk}}}\bigr\rceil l_{kk}\bigr)/l_{kk}\bigr|
    \leq \abs{\tfrac{1}{2} + i\tfrac{1}{2}} = \tfrac{1}{\sqrt{2}}.\label{eq:lkj_lkk_ratio}
\end{align}\\[-2.5em]
In matrix form, operations~\eqref{eq:Zkj_operation}-\eqref{eq:Wpkj_operation} become\vspace{-0.15in}
\begin{IEEEeqnarray}{rClrClr}
    \mbf{Z}_{kj}^{}
    &\!=\!& \I_{\Nt} - \zeta_{kj}^{} \mbf{e}_{k}^{}\mbf{e}_{j}^{\tran}
    \in \mathcal{G}^{\Nt\times\Nt},~~
    & k & \neq & j,~~&
    \zeta_{kj}^{}\in\mathcal{G},~~~~
    \label{eq:Zkj}\\
    \subsc{\mbf{W}}{\!\!\mathrm{p}\,kj}
    &\!=\!& \I_{\Nt} - \omega_{kj}^{} \mbf{e}_{k}^{}\mbf{e}_{j}^{\tran} \in\mathcal{C}^{\Nt\times\Nt},~~
    & k & \neq & j,~~&
    \omega_{kj}^{}\in\mathcal{C}.~~~~
    \label{eq:Wpkj}
\end{IEEEeqnarray}\\[-2.25em]
Note that $\mbf{Z}_{kj}^{-1}\!=\! \I_{\Nt} \!+\! \zeta_{kj}^{} \mbf{e}_{k}^{}\mbf{e}_{j}^{\tran} \!\in\! \mathcal{G}^{\Nt\times\Nt}$. The matrices $\Zp$ and $\Wp$ are formed from the products of the $(\Nt\!-\!\nu\!-\!1)(\Nt\!-\!\nu)/2$ matrices in~\eqref{eq:Zkj} and~\eqref{eq:Wpkj}, respectively. $\Wp$ is then normalized using a diagonal matrix $\Dp$ to satisfy $\diagg{\Wp\Wph}\!=\!\I_{\Nt}$: \vspace{-0.15in}
\begin{align}
    \mbf{Z} \!=\!\!\prod_{k=\nu+2}^{\Nt} \prod_{j=\nu+1}^{k-1} \!\!\mbf{Z}_{kj}^{},\qquad
    \Wp \!= \Dp\!\!\prod_{k=\nu+2}^{\Nt} \prod_{j=\nu+1}^{k-1} \!\!\subsc{\mbf{W}}{\!\!\mathrm{p}\,kj}. \label{eq:Z_Wp}
\end{align} \\[-3.25em]

%
\begin{lemma}[\!\!\!\cite{Lemeire_1975}]\label{lem:lemeire_bound} If $\mbf{T}\!=\![t_{kj}^{}]\!\in\!\mathcal{C}^{\Nt\times\Nt}$ is a nonsingular lower-triangular matrix, then\vspace{-0.15in}
\begin{align}
    \fnormm{\mbf{T}^{-1}}_{2,\mathrm{F}}
    \leq
    \tfrac{1}{(\rho+2)\delta}
    \sqrt{\smash[b]{(\rho+1)^{2\Nt}+2\Nt(\rho+2)-1}},\label{eq:lemeire_bound}
\end{align}\\[-2.25em]
where $\rho\!=\!\max_{k<j}^{}\abss{\smash[tb]{t_{kj}^{}}}/\abss{t_{kk}^{}}$ and $\delta\!=\!\min_{k}^{}\abss{t_{kk}^{}}$.
\end{lemma}
\begin{IEEEproof} See~\cite{Lemeire_1975}.
\end{IEEEproof}
We use Lemma~\ref{lem:lemeire_bound} to show that the norm of $\Si$ (and hence $\Wp$ in~\eqref{eq:Dp_Wp_Lp_formula}) tends to decrease by applying $\mbf{Z}$. Let $\breve{\mbf{Z}}$ be the principal submatrix obtained by deleting the first $\nu$ rows and columns of $\mbf{Z}$, and let $\breve{\mbf{S}}\!=\!\mbf{S}\breve{\mbf{Z}}$. The reduction step in~\eqref{eq:Zkj_operation} ensures that the magnitudes of the lower-diagonal elements of $\breve{\mbf{S}}$ are $\leq\! \tfrac{l_{kk}}{\sqrt{2}}$, while not altering the diagonal elements. The quantity $\subsc{\fnormm{\Sbi \!-\! \diagg{\Sbi}}}{\mathrm{F}}$ measures the `weight' of the lower-triangular portion of $\Sbi$. Applying~\eqref{eq:lemeire_bound} for $\mbf{T}\!=\!\breve{\mbf{S}}$, we obtain\vspace{-0.125in}
\begin{align}
    \subpsc{\fnormm{\Sbi \!-\! \diagg{\Sbi}}}{\mathrm{F}}{\,2}
    \leq
    \frac
    {
        (\rho\!+\!1)^{2(N-\nu)}
        \!-\! \rho(\rho\!+\!2)(\Nt\!-\!\nu)
        \!-\! 1
    }
    {(\rho\!+\!2)^2 \delta^2},\label{eq:Sinv_gap_to_identity}
\end{align}\\[-2.25em]
with $\rho\!=\!1/\sqrt{2}$ and $\delta\!=\!\min_{k>\nu}^{}l_{kk}^{}$. As $\rho$ decreases, this upper bound decreases, and hence $\breve{\mbf{S}}^{-1}$ becomes more diagonal. Therefore, with $\breve{\mbf{\Omega}}\!=\!\supsc{\diagg{\Sbi\Sbih}}{-1/2}$, $\breve{\mbf{\Omega}}\Sbi$ becomes closer to the identity, and when used in lieu of $\mbf{\Omega}\Si$ in~\eqref{eq:Dp_Wp_Lp_formula} makes $\Wp$ closer to $\I_{\Nt}$.

To reduce $\rho$ below $1/\sqrt{2}$, the reduction step in~\eqref{eq:Zkj_operation} can be changed by scaling the ratio $l_{kj}/l_{kk}$ by a power-of-2 so that\vspace{-0.25in}
\begin{align}
    \zeta_{kj}
    &\!=\!
    2^{-c}
    \bigl\lfloor{
        2^{c}\tfrac{l_{kj}}{l_{kk}}
    }\bigr\rceil,\label{eq:zeta_kj}
\end{align}\\[-2.75em]
for some integer $c\!\geq\!0$. In this case,~\eqref{eq:lkj_lkk_ratio} becomes\vspace{-0.15in}
\begin{align}
    {\smash[tb]{\bigl|\tfrac{\tilde{l}_{kj}}{l_{kk}} \bigr|} }
    \!=\!
    \bigl|\!
        \bigl(l_{kj} \!-\!
        \bigl\lfloor
            {\tfrac{l_{kj}}{l_{kk}/2^c}}
        \bigr\rceil
        \tfrac{l_{kk}}{2^c}\bigr)/l_{kk}
    \!\bigr|
    \!\leq\!
    \abs{\tfrac{1}{2^{c+1}} \!+\! \tfrac{i}{2^{c+1}}}
    \!=\!
    \tfrac{1}{2^c\sqrt{2}},\label{eq:lkj_lkk_ratio_2^b}
\end{align}\\[-2.25em]
which gives $\rho\!=\!1/2^{c+1/2}$. Since $2^c\zeta_{kj}$ is an integer, it follows that $2^c\mbf{Z}_{kj}\!\in\!\mathcal{G}^{\Nt\times\Nt}$ and the integer condition~\eqref{eq:Zp_integer_sol_cond} still holds.

Note that~\eqref{eq:Zkj_operation} is similar to the first lattice reduction condition of~\cite{LLL_1982,Agrell_2002,2020_Lyu_low_complexity_LLL}. However, the Lov\'{a}sz condition~\cite{LLL_1982}\vspace{-0.25in}
\begin{align}
    l_{kk}^2 + \abss{l_{k+1,k}^{}}^2 \geq \gamma \cdot l_{k+1,k+1}^2,\quad  \tfrac{1}{4} < \gamma < 1,
    \label{eq:Lovasz_cond}
\end{align}\\[-2.75em]
cannot be enforced for $k\!=\!\nu\!+\!1,\cdots,\Nt\!-\!1$ because it requires permuting the columns of $\mbf{L}$, which destroys the lower-triangular structure of $\mbf{Z}$.

%
\vspace{-0.15in}
\section{WLD MIMO Detection}\label{sec:WLD}\vspace{-0.05in}
In this section, we develop the detection model of the WLD detector and characterize its AIR using single-sided puncturing. The analysis for two-sided puncturing is similar.

Starting with $\fnormmsq{\Qh\y\!-\!\L\x}$ and applying $\Wp$ in~\eqref{eq:Dp_Wp_Lp_formula}, the equivalent metric to~\eqref{eq:true_metric} computed by the WLD detector is\vspace{-0.15in}
\begin{align*}
    -\tfrac{1}{N_0}\fnormmsq{\Qh\y\!-\!\L\x}~
    \smash[tb]{\xrightarrow{~\Wp~}}
    ~\mup{\y|\x}
    \!=\!-\tfrac{1}{N_0}\fnormmsq{\Wp(\Qh\y\!-\!\L\x)}.
\end{align*}\\[-2.25em]
By expanding $\mup{\y|\x}$ and dropping the irrelevant term $-\tfrac{1}{N_0}\fnormmsq{\Wp\Qh\y}$, we obtain\vspace{-0.15in}
\begin{align}
    \mup{\y|\x}
    &\!=\!
    -\tfrac{1}{N_0}\fnormmsq{\yp\!-\!\Lp\x} \cong
    2\Re{\yh\Fp\x} \! - \! \xh\Gp\x, \label{eq:wld_metric}
\end{align}\\[-2.5em]
where $\yp\!=\!\Wp\Qh\y$, $\Lp\!=\!\Wp\L$,\vspace{-0.15in}
\begin{align}
    \Fp
    =
    \tfrac{1}{N_0}\Q\Wph\Lp, ~\quad \text{and} ~\quad
    \Gp
    =
    \tfrac{1}{N_0}\Lph\Lp = \Hh\Fp.  \label{eq:Fp_Gp_def}
\end{align}\\[-2.75em]
The corresponding detection model becomes\vspace{-0.15in}
\begin{align}
    \subsc{p}{\mathrm{p}}(\y|\x)
    &=
    \exp{(2\Re{\yh\Fp\x} \! - \! \xh\Gp\x)}, \label{eq:WLD_detection_prob}
\end{align}\\[-2.75em]
instead of the true conditional probability in~\eqref{eq:true_detection_prob}. Based on~\eqref{eq:WLD_detection_prob}, the achievable information rate of the WLD detector is lower-bounded by~\cite{2006_arnold_simulation_based}\vspace{-0.15in}
\begin{align}
    \ILBWLD
        &= \Ex{\mbf{Y},\mbf{X}}{\ln( \subsc{p}{\mathrm{p}}(\y|\x) )}
            -\Ex{\mbf{Y}}{\ln( \subsc{p}{\mathrm{p}}(\y) )},\label{eq:ILB_WLD}
\end{align}\\[-2.75em]
where the expectations are taken over the true channel statistics, and $\subsc{p}{\mathrm{p}}(\y)
    =
    \int \subsc{p}{\mathrm{p}}(\y|\x) p(\x)\,\mathrm{d}\x$ with $p(\x)$ being the prior distribution of $\x$.

%
\begin{theorem}\label{thm:ILB_wld} Assuming Gaussian inputs $\x\!\sim\! \mathcal{CN}\!(\mbf{0},\Es\I_{\Nt})$, and let $\beta\!=\! \tfrac{E_s}{N_0}$ be the SNR, the lower-bound on the AIR in~\eqref{eq:ILB_WLD} attained by the WLD detector is given by\vspace{-0.15in}
\begin{align}
    \ILBWLD
        \!=\!  \ln \det ( \I \!+\! \beta\Lph\Lp ) \!-\!
            \Trr{(\I \!-\! \Wp\!\Wph)\supsc{(\I \!+\! \beta\Lp\Lph)}{\,-1} }.\label{eq:ILB_WLD_expression}
\end{align}\\[-4.5em]
\end{theorem}
\begin{IEEEproof} Following the approach in~\cite{2012_rusek_optimal_channel_short}, we first compute the probability $\subsc{p}{\mathrm{p}}(\y) \!=\! \int \subsc{p}{\mathrm{p}}(\y|\x) p(\x)\,\mathrm{d}\x$ for $\subsc{p}{\mathrm{p}}(\y|\x)$ in~\eqref{eq:WLD_detection_prob} and $p(\x)\!=\! \tfrac{1}{\pi^\Nt \Es^\Nt} \exp{( -\tfrac{\normm{\x}^2}{\Es})}$. We then compute the expectations in~\eqref{eq:ILB_WLD} over the true channel statistics as\vspace{-0.175in}
\begin{align*}
    \Ex{\mbf{Y},\mbf{X}}{\ln( \subsc{p}{\mathrm{p}}(\y|\x) )}
    &=
    2\Es\Ree{\Trr{\Fph\H}} \!-\! \Es\Trr{\Gp}
    =
    \Es\Trr{\Gp},
    \\[-0.5em]
    -\Ex{\mbf{Y}}{\ln( \subsc{p}{\mathrm{p}}(\y) )}
    &=
    \Nt\ln\Es
    +
    \ln\det( \Gp \!+\! \tfrac{1}{\Es}\I)
    -
    \Trr{ \Fph[ \Es\H \Hh\!+\! N_0\I ] \Fp
            \supsc{[ \Gp \!+\! \tfrac{1}{\Es}\I]}{\,-1}  }.
\end{align*}\\[-2.75em]
Substituting for $\Fp\!=\!\tfrac{1}{N_0}\Q\Wph\Lp$ and $\Gp\!=\!\tfrac{1}{N_0}\Lph\Lp$, and applying the matrix inversion lemma~\cite{2011_zhang_matrix_theory}, followed by some standard simplification steps, the result follows.
\end{IEEEproof}

For $\nu\!=\!1$, the sorted singular values $\{\sigma_k\}_{k=1}^{\Nt}$ of $\Lp$ satisfy an interlacing property with respect to the diagonal elements of $\mbf{\Omega}$ in~\eqref{eq:Dp_Wp_Lp_formula}. Let  $\omega_1 \!<\!\omega_2 \!<\! \cdots \!<\! \omega_{\Nt-\nu}$ be the sorted diagonal elements of $\mbf{\Omega}$, and let $\mbf{v}$ be the first column of $\Lp$, then~\cite{Jessup_Sorensen_1994}\vspace{-0.2in}
\begin{align}
    0 \!<\! \sigma_1 \!<\! \omega_1 \!<\!\cdots \!<\!\sigma_{\Nt-1} \!<\! \omega_{\Nt-1} \!<\! \sigma_{\Nt} \!<\! \omega_{\Nt-1}^{} \!+\! \fnormm{\mbf{v}}.
    \label{eq:interlacing}
\end{align}\\[-3em]
Property~\eqref{eq:interlacing} can be used to bound $\ILBWLD$ in Theorem~\ref{thm:ILB_wld} for $\nu\!=\!1$ since $\ln \det (\I \!+\! \beta\Lph\Lp) \!=\! \sum_{k=1}^{\Nt}\ln(1\!+\!\beta\sigma_k^2)$, $\Trr{(\I \!+\! \beta\Lp\Lph)^{-1}} \!=\! \sum_{k=1}^{\Nt}\!1/(1\!+\!\beta\sigma_k^2)$, and $\Trr{\Wp\!\Wph}\!=\!\Nt$. The details are omitted due to lack of space.

Note that for $\nu\!=\!\Nt\!-\!1$, we have $\Wp\!=\!\I$ and $\Lp\!=\!\L$, and hence $ \ILBWLD
\!=\!  \ln \det \!\left( \I \!+\! \beta\Lh\L \right)$, which is the capacity of the channel. In fact, as $\nu$ increases from 1, the metrics computed by the WLD detector approach the hard-decision ML metrics as shown by the following lemma.

%
\begin{lemma}\label{lem:WLD_distance_bound} If $\x_{\ML}\!=\!\argmin_{\x\in\X^{\Nt}}\!\fnormm{\yt\!-\!\L\x}$ and $\xWLD\!=\!\argmin_{\x\in\X^{\Nt}}\!\fnormm{\Wp(\yt\!-\!\L\x)}$ where $\H\!=\!\Q\L$ and $\yt\!=\!\Qh\y$, then\vspace{-0.25in}
\begin{align}
   \fnormm{\yt\!-\!\L\x_{\ML}}
   \leq
   \fnormm{\yt\!-\!\L\x_{\WLD}}
   &\leq
   \kappa(\Wp)\fnormm{\yt\!-\!\L\x_{\ML}},\label{eq:WLD_distance_bound}
   \\[-0.5em]
   \fnormm{\Wp(\yt\!-\!\L\x_{\WLD})}
   &\leq
   \sigma_{\max}{(\Wp)}\fnormm{\yt\!-\!\L\x_{\ML}}, \label{eq:WLD_distance_bound2}
\end{align}\\[-2.75em]
where $\kappa(\Wp)\!=\!\sigma_{\max}(\Wp)/\sigma_{\min}(\Wp)$ is the condition number of $\Wp$, and $\sigma_{\max}{(\Wp)},\sigma_{\min}{(\Wp)}$ are the largest and smallest singular values of $\Wp$, respectively.
\end{lemma}
\begin{IEEEproof} The first inequality in~\eqref{eq:WLD_distance_bound} follows from the definition of the ML solution. For the second, we have\vspace{-0.225in}
\begin{align}
   \fnormm{\yt\!-\!\L\x_{\WLD}}
   =
   \fnormm{\Wpi\Wp(\yt\!-\!\L\x_{\WLD})}\notag
   &\leq
   \sigma_{\max}(\Wpi)\fnormm{\Wp(\yt\!-\!\L\x_{\WLD})}\notag
   \\[-0.25em]
   &\leq
   \sigma_{\max}(\Wpi)\fnormm{\Wp(\yt\!-\!\L\x_{\ML})}\label{eq:WLD_distance_bound_step}
   \\[-0.25em]
   &\leq
   \sigma_{\max}(\Wpi)\sigma_{\max}(\Wp)\fnormm{\yt\!-\!\L\x_{\ML}},\notag
\end{align}\\[-2.75em]
from which~\eqref{eq:WLD_distance_bound} follows. Note that both~\eqref{eq:WLD_distance_bound2} and~\eqref{eq:WLD_distance_bound_step} follow because $\fnormm{\Wp(\yt\!-\!\L\x_{\WLD})}\!\leq\!\fnormm{\Wp(\yt\!-\!\L\x)}$ for any $\x$.
\end{IEEEproof}

Note that the layer orders within the $\nu$ parent layers and within the $\Nt\!-\!\nu$ child layers are irrelevant. What matters is which layers are selected to form the parent set. This is formalized using the following lemma.
\begin{lemma}\label{lem:wld_perm_distances} Let $\Ja$ and $\Jb$ be permutation matrices of sizes $\nu$ and $\Nt\!-\!\nu$, respectively. If the columns of $\H$ are permuted by $\J\!=\!\begin{bsmallmatrix} \Ja & \mbf{0} \\ \mbf{0} & \Jb\end{bsmallmatrix}$, then the distance metric computed by the WLD detector in~\eqref{eq:wld_metric} does not change, i.e.,\vspace{-0.2in}
\begin{align}\label{eq:permuted_wld_distance_equality}
    \fnormmsq{\Wp(\Qh\y\!-\!\L\x)}
    &=
    \fnormmsq{\Wptilde(\Qtildeh\y\!-\!\Ltilde\supsc{\J}{\,-1}\x)},
\end{align}\\[-2.75em]
where $\H\!=\!\Q\L$, $\Wp$ ($\Wptilde$) is the puncturing matrix of $\L$ ($\Ltilde$), and $\H\J\!=\!\Qtilde\Ltilde$.
\end{lemma}
\begin{IEEEproof}
Let $\x\!=\![\xparent;~\xchild]$, $\Q\!=\![ \Qparent ~ \Qchild]$, $\Qtilde\!=\![ \Qtildeparent ~ \Qtildechild ]$, $\L\!=\!\begin{bsmallmatrix} \P &  \\ \mbf{R} & \mbf{S}\end{bsmallmatrix}$, and $\Ltilde\!=\!\begin{bsmallmatrix} \smash[t]{\Ptilde}\vphantom{\P} &  \\ \Rtilde & \Stilde\end{bsmallmatrix}$ be partitioned corresponding to $\nu$ parent layers and $\Nt\!-\!\nu$ child layers. Then $\Wp\!=\!\begin{bsmallmatrix} \I &  \mbf{0} \\ \mbf{0} & \mbf{\Omega} \mbf{S}^{-1}\end{bsmallmatrix}$ and $\Wptilde\!=\!\begin{bsmallmatrix} \I &  \mbf{0} \\ \mbf{0} & \tilde{\mbf{\Omega}} \Stildeinv\end{bsmallmatrix}$, where $\mbf{\Omega}\!=\!\supsc{\diagg{\Si \Sih}}{-1/2}$ and $\tilde{\mbf{\Omega}}\!=\!\supsc{\diagg{\Stildeinv\Stildeinvh}}{-1/2}$. The partitions of $\Ltilde$ are related to those of $\L$ since $\Q\L\J\!=\!\Qtilde\Ltilde$. Furthermore, $\tilde{\mbf{\Omega}}\!=\! \Jbh \mbf{\Omega} \Jb$ since $\Qchildh\Qtildechild\Qtildechildh\Qchild
\!=\! \I$. Substituting back in both squared-norms in~\eqref{eq:permuted_wld_distance_equality}, and performing simplifications, it follows that both sides are equal to $\fnormmsq{\Qparenth\y \!-\! \P \xparent} \!+\! \fnormmsq{\mbf{\Omega}(\Si\Qchildh{}\y \!- \! \Si\mbf{R}\xparent \!-\! \xchild)}$.
\end{IEEEproof}
%
\begin{corollary} Let $\J$ be any permutation matrix, $\H\J\!=\!\Qtilde\Ltilde$, and $\Wptilde$ the puncturing matrix of $\Ltilde$. Then, the number of distinct solutions of  $\xWLD\!=\!\argmin_{\x}\fnormm{\Wptilde(\Qtildeh\y\!-\!\Ltilde\supsc{\J}{\,-1}\x)}$ for all possible values of $\J$ depends only on the number of parent layer combinations, and is at most $\Nt\choose \nu$.\hspace{1em plus 1fill}\IEEEQEDhere
\end{corollary}

Finally, the bound $\ILBWLD$ for Gaussian inputs can be used as a criterion for parent layer selection, but the complexity of possible combinations grows as $\Nt\choose \nu$. Alternatively, a less sensitive approach to parent layer selection is to do multiple detection rounds, each time choosing $\nu$ new layers as parents and generating bit LLRs for these parent symbols only.
\vspace{-0.125in}

%
\section{Augmented WLD (AWLD) MIMO Detection}\label{sec:augmented_WLD}\vspace{-0.05in}
The lower bound on the AIR in~\eqref{eq:ILB_WLD_expression} attained by the WLD is not optimal. Motivated by the result for the optimal receiver filter derived in~\cite{2012_rusek_optimal_channel_short} in the context of channel shortening for ISI channels, which involves an MMSE filter compensated by receiver tree processing, we introduce in this section an alternate form of puncturing using augmented channels. Instead of basing the detection metric in~\eqref{eq:true_metric} on $\H$, we form the augmented vector $\ya\!=\!\tfrac{1}{\sqrt{N_0}}[\y;~\mbf{0}_{\Nt\times1}]$ and matrix\vspace{-0.2in}
\begin{align}
    \Ha = \begin{bNiceMatrix}
        \tfrac{1}{\sqrt{N_0}}\H_{\MbyN}  \\
        \tfrac{1}{\sqrt{\Es}}\I_{\Nt}^{}
      \end{bNiceMatrix}
      \qquad\text{(size $(\Nr\!+\!\Nt)\!\times\!\Nt$)},
      \label{eq:Ha_def}
\end{align}\\[-2.0em]
in a manner analogous to the square-root MMSE of~\cite{2000_hassibi_square-root_MMSE}, and reformulate $\mu(\y|\x)$ in~\eqref{eq:true_metric} using $\Ha,\ya$ rather than $\H,\y$ as\vspace{-0.175in}
\begin{align}
    \mu(\y|\x)
    &=
    \tfrac{1}{\Es}\fnormmsq{\x}
    \!+
    2
    \Re{\!\tfrac{1}{\sqrt{N_0}}[\yh~\mbf{0}]
        \begin{bNiceMatrix}
            \tfrac{1}{\sqrt{N_0}}\H  \\
            \tfrac{1}{\sqrt{\Es}}\I_{\Nt}^{}
        \end{bNiceMatrix}
        \!\x\!
       }
    - \xh(\tfrac{1}{N_0}\Hh\H + \tfrac{1}{\Es})\x
    -\tfrac{1}{N_0}\fnormmsq{\y}
    \nonumber
    \\[-0.25em]
    &=
    \tfrac{1}{\Es}\fnormmsq{\x}
    +2
    \Ree{\yah
        \Ha
        \x
        }
    -\xh\Hah\Ha\x
    -\fnormmsq{\ya}
    \nonumber
    \\[-0.5em]
    &=
    \tfrac{1}{\Es}\fnormmsq{\x}
    -
    \fnormmsq{\ya - \Ha\x}.
    \label{eq:true_distance}
\end{align}\\[-2.75em]
We next expand the squared-distance in~\eqref{eq:true_distance} in terms of the projection matrix $\subsc{\mbf{P}}{\!\Ha}\!=\!\Ha(\Hah\Ha)^{-1}\Hah$ onto the column space of $\Ha$ and its orthogonal complement $\subpsc{\mbf{P}}{\!\Ha}{\,\perp}\!=\!\I_{\Nr+\Nt}\!-\!\Ha(\Hah\Ha)^{-1}\Hah$ as\vspace{-0.175in}
\begin{align}
    \fnormmsq{\ya \!-\! \Ha\x}
    &=
    \fnormmsq{\subsc{\mbf{P}}{\!\Ha}\!(\ya \!-\! \Ha\x)}
    +
    \fnormmsq{\subpsc{\mbf{P}}{\!\Ha}{\,\perp}\ya}. \label{eq:true_distance_orth_expansion}
\end{align}\\[-2.75em]
Let $\Qa{}\La$ be the thin QL decomposition of $\Ha$ partitioned as\vspace{-0.15in}
\begin{align}
    \Ha &=   \begin{bNiceMatrix}
                \tfrac{1}{\sqrt{N_0}}\H  \\
                \tfrac{1}{\sqrt{\Es}}\I_{\Nt}^{}
            \end{bNiceMatrix}
        =   \Qa{}\La
        =   \begin{bNiceMatrix}
                \Qa{1}  \\
                \Qa{2}
            \end{bNiceMatrix}
            \La
        =   \begin{bNiceMatrix}
                \Qa{1}\La  \\
                \Qa{2}\La
            \end{bNiceMatrix},\label{eq:Ha_QL_decomp_blocks}
\end{align}\\[-2.0em]
where $\Qa{}$ is an $(\Nr\!+\!\Nt)\!\times\!\Nt$ matrix with orthonormal columns (i.e., $\Qah{}\Qa{}\!=\!\I_{\Nt}$ but not unitary \emph{since} $\Qa{}\Qah{}\!\neq\!\I_{\Nr+\Nt}$), $\La$ is $\NbyN$ lower-triangular, and $\Qa{1},\Qa{2}$ are respectively the upper $\MbyN$ and lower $\NbyN$ block matrices of $\Qa{}$. Note that neither the rows nor the columns of $\Qa{1}$ and $\Qa{2}$ are orthonormal. From the partitions in~\eqref{eq:Ha_QL_decomp_blocks}, it follows that \vspace{-0.175in}
\begin{align}
    \H
    &=
    \sqrt{N_0}\Qa{1}\La,
    \label{eq:Ha_QL_decomp_Ha_Qa1La}
    \\[-0.65em]
    \tfrac{1}{\sqrt{\Es}}\I_{\Nt}^{}
    &=
    \Qa{2}\La = \La\Qa{2}.
    \label{eq:Ha_QL_decomp_Ha_Qa2La}
\end{align}\\[-2.75em]
However,~\eqref{eq:Ha_QL_decomp_Ha_Qa1La} is \emph{not} the QL decomposition of $\H$.~\eqref{eq:Ha_QL_decomp_Ha_Qa2La} implies that $\Qa{2}$ is a lower-triangular matrix proportional to the inverse of $\La$, i.e, $\Lai \!=\! \sqrt{\Es}\Qa{2}$. Then, from~\eqref{eq:Ha_QL_decomp_blocks} we have\vspace{-0.15in}
\begin{align*}
    \tfrac{1}{N_0}\Hh\H \!+\! \tfrac{1}{\Es}\I_{\Nt}
    &= \Hah\Ha
    = \Lah\La,
\end{align*}\\[-3.25em]
from which it follows that\vspace{-0.2in}
\begin{align}
    \fnormmsq{\ya \!-\! \Ha\x}
    &=
    \fnormmsq{\La(\Wmmse\y \!-\! \x)}
    +
    \fnormmsq{(\I\!-\!\Qa{}\Qah{})\ya},    \label{eq:true_distance2}
\end{align}\\[-3em]
where $\Wmmse$ is the standard $\NbyM$ MMSE filter matrix,\vspace{-0.175in}
\begin{align}
    \Wmmse
    &= \Hh  \supsc{[ \H \Hh \!+\! \alpha\I_{\Nr} ]}{\,-1}
     =      \supsc{[ \Hh \H \!+\! \alpha\I_{\Nt} ]}{\,-1} \Hh
    \label{eq:Wmmse_formula_left_right}
    \\[-0.5em]
    &=
    \tfrac{1}{N_0}(\Hah\Ha)^{-1}\Hh
    =
    \tfrac{1}{N_0}(\Lah\La)^{-1}\Hh
    \label{eq:Wmmse_Ha_La}
    \\[-0.5em]
    &=
    \sqrt{\beta}\Qa{2} \Qah{1},
    \label{eq:Wmmse_Qa2_Qa1}
\end{align}\\[-2.5em]
with $\alpha \!=\! \tfrac{1}{\beta} \!=\! \tfrac{N_0}{E_s}$. Substituting~\eqref{eq:true_distance2} back in~\eqref{eq:true_distance}, we obtain\vspace{-0.175in}
\begin{align}
   \mu(\y|\x)
   \!=\!
   \tfrac{1}{\Es}\fnormmsq{\x}
   \!-\!
   \fnormmsq{\La(\Wmmse\y \!-\! \x)}
   -
   \fnormmsq{(\I\!-\!\Qa{}\!\Qah{})\ya}.
   \label{eq:true_distance3}
\end{align}\\[-4.5em]

Note that in~\eqref{eq:true_distance3}, the term $\fnormmsq{\x}$ appears explicitly, while tree processing is solely based on $\La$ in $\fnormmsq{\La(\Wmmse\y \!-\! \x)}$. We therefore puncture $\La$ using an appropriate puncturing matrix $\Wap$ similar to puncturing $\L$ in~\eqref{eq:LP_def} using $\Wp$. For a given puncturing order $\nu$, we conformally partition $\La$ similar to~\eqref{eq:L_nu_partitioned} and obtain the partition blocks $\Pa$ of size $\nu\!\times\!\nu$, $\Ra$ of size $(\Nt\!-\!\nu)\!\times\!\nu$, and $\Sa$ of size $(\Nt\!-\!\nu)\!\times\!(\Nt\!-\!\nu)$. The resulting punctured augmented matrix, denoted as $\Lap$, is given by \vspace{-0.175in}
\begin{align}
    \Lap
    =
    \Wap\La
    &=
    \Wap
    \begin{bNiceMatrix}
        \Pa & \mbf{0}  \\
        \Ra & \Sa
    \end{bNiceMatrix}
    =
    \begin{bNiceMatrix}
        \Pa & \mbf{0}  \\
        \Omegaa\Sai\Ra & \Omegaa
    \end{bNiceMatrix},
    \label{eq:Lap_def}
\end{align}\\[-2.25em]
where\vspace{-0.175in}
\begin{align}
    \Wap
    &=
    \Dap \diagg{\La}
        \begin{bNiceMatrix}
            \I_{\nu} & \mbf{0}  \\
            \mbf{0} & \Sai
        \end{bNiceMatrix}
    =
        \begin{bNiceMatrix}
            \I_{\nu} & \mbf{0}  \\
            \mbf{0} & \Omegaa\Sai
        \end{bNiceMatrix},
    \label{eq:Wap_formula}
    \\[-0.25em]
    \Dap
    &\!=\!
    \supsc{\diagg{\La}}{-1}
    \begin{bNiceMatrix}
        \I_{\nu} & \mbf{0}   \\
        \mbf{0} & \mbf{\Omega}_{\mathrm{a}}
    \end{bNiceMatrix},
    \label{eq:Dap_formula}
    \\[-0.25em]
    \Omegaa
    &=
    \supsc{\diagg{\Sai \Saih}}{-1/2},\label{eq:Sigma_a}
\end{align}\\[-2.5em]
and $\Dap$ in~\eqref{eq:Dap_formula} is chosen so that $\diagg{\Wap\Waph}\!=\!\I_{\Nt}$.

Next, applying $\Wap$ to filter $\La(\Wmmse\y \!-\! \x)$ in~\eqref{eq:true_distance3} as \vspace{-0.175in}
\begin{align}
    \fnormmsq{\La(\Wmmse\y \!-\! \x)}
    \xrightarrow{~\Wap~}
    \fnormmsq{\Wap(\La\Wmmse\y \!-\! \La\x)}, \label{eq:awld_metric_tmp}
\end{align}\\[-2.75em]
and dropping the irrelevant term $\fnormmsq{(\I\!-\!\Qa{}\!\Qah{})\ya}$ in~\eqref{eq:true_distance3}, the metric computed by the \emph{augmented} WLD (AWLD) detector corresponding to~\eqref{eq:true_distance3} takes the form \vspace{-0.175in}
\begin{align}
    \muap{\y|\x}
    &=
    \tfrac{1}{\Es}\fnormmsq{\x}
    \!-\!
    \fnormmsq{\Wap\La(\Wmmse\y \!-\! \x)}
    =
    \tfrac{1}{\Es}\fnormmsq{\x}
    \!-\!
    \fnormmsq{\yap \!-\! \Lap\x} \label{eq:awld_metric_1}
    \\[-0.5em]
    &\cong
    2\Ree{\yh\Fap\x}  -  \xh\Gap\x + \tfrac{1}{\Es}\xh\x, \label{eq:awld_metric}
\end{align}\\[-3em]
where $\yap = \Wap\La\Wmmse\y = \Lap\Wmmse\y$, \vspace{-0.2in}
\begin{align}
    \Fap
    =
    \Wmmseh \Gap, ~\quad \text{and} ~\quad
    \Gap
    =
    \Laph\Lap. \label{eq:Fap_Gap_def}
\end{align}\\[-3em]
The corresponding AWLD detection model (Fig.~\ref{fig:awld_block_diagram}) becomes\vspace{-0.175in}
\begin{align}
    \subsc{p}{\mathrm{ap}}(\y|\x)
    &=
    \exp{(2\Re{\yh\Fap\x}  -  \xh\Gap\x + \tfrac{1}{\Es}\xh\x)}. \label{eq:awld_detection_prob}
\end{align}\\[-4.5em]

\begin{figure}
  \centering
  \tikzstyle{int}=[draw, fill=blue!20, minimum size=2em]
\tikzstyle{init} = [pin edge={to-,thin,black}]

\begin{tikzpicture}[node distance=2.75cm, auto, >=latex', pin distance = 3mm]
    \node [int, pin={[init]above:{\small$\H,\beta$}}, label=below:{\scriptsize MMSE
    filter}] (a) {{\small$\Wmmse$\normalsize}};

    \node (b) [left of=a,node distance=0.9cm, coordinate] {};

    \node [int, pin={[init]above:{\small$\Ha$}}] (c) [right of=a,  label=below:{\scriptsize Gain compensation}, node distance=2.25cm] {{\small$\Lap$\normalsize}};

    \node [int, pin={[init]above:{\small$\Lap$}}, label=below:{\scriptsize WLD detector}] (d) [right of=c, node distance=3.5cm] {{\small$\arg\max\tfrac{1}{\Es}\fnormmsq{\x}\!-\! \fnormmsq{\yap\!-\!\Lap\x}$\normalsize}};

    \node [coordinate] (end) [right of=d, node distance=2.7cm]{};

    \path[->] (b) edge node {{\small$\y$}} (a);
    \path[->] (a) edge node {{\small$\Wmmse\y$\normalsize}} (c);
    \draw[->] (c) edge node {{\small$\yap$\normalsize}} (d);
    \draw[->] (d) edge node {{\small$\hat{\x}$}} (end);

\end{tikzpicture} 
  \vspace{-0.25in}
  \caption{Block diagram of the AWLD detector, where $\yap\!=\!\Lap\Wmmse\y$.}\label{fig:awld_block_diagram}
\end{figure}

%
\begin{theorem}\label{thm:ILB_AWLD} Under the same assumptions as Theorem~\ref{thm:ILB_wld}, the AIR of the augmented WLD detector based on~\eqref{eq:awld_detection_prob}, with $\Gap,\Fap$ as given in~\eqref{eq:Fap_Gap_def}, is lower-bounded by\vspace{-0.175in}
\begin{align}
    \ILBAWLD
        = \Nt\ln\Es + \ln \det ( \Laph\Lap ).
        \label{eq:ILB_AWLD}
\end{align}\\[-4.75em]
\end{theorem}
\begin{IEEEproof} The lower bound on the AIR of the AWLD detector based on~\eqref{eq:awld_detection_prob} is defined as\vspace{-0.175in}
\begin{align}
    \ILBAWLD
    &= \Ex{\mbf{Y},\mbf{X}}{\ln( \subsc{p}{\mathrm{ap}}(\y|\x) )}
    -\Ex{\mbf{Y}}{\ln( \subsc{p}{\mathrm{ap}}(\y) )},
    \label{eq:ILB_AWLD_def}
\end{align}\\[-3em]
where $\subsc{p}{\mathrm{ap}}(\y)$ is given by\vspace{-0.2in}
\begin{align}
    \subsc{p}{\mathrm{ap}}(\y)
    &=
    \int_{\x\in\C^{\Nt}} \subsc{p}{\mathrm{ap}}(\y|\x) p(\x)
    \,\mathrm{d}\x,
    \label{eq:awld_detection_prob_y}
\end{align}\\[-2.5em]
assuming $\x\!\sim\! \mathcal{CN}(\mbf{0},\Es\I_{\Nt})$. The main difference compared to the proof of Theorem~\ref{thm:ILB_wld} is the effect of the term $\tfrac{1}{\Es}\xh\x$ in~\eqref{eq:awld_detection_prob} when evaluating~\eqref{eq:awld_detection_prob_y} under Gaussian densities, which annihilates the effect of the prior density $p(\x)$ to give \vspace{-0.175in}
\begin{align}
    \subsc{p}{\mathrm{ap}}(\y)
    &\!=\!
    \tfrac{1}{\pi^\Nt \Es^\Nt}\!\!
    \int
        \exp{\left(2\Ree{\yh\Fap\x} \!-\! \xh\Gap\x \right)}
        \,\mathrm{d}\x.
        \label{eq:awld_p_ap_tmp}
\end{align}\\[-2.75em]
With standard manipulations, the expectations in~\eqref{eq:ILB_AWLD_def} become\vspace{-0.175in}
\begin{align*}
    \Ex{\mbf{Y},\mbf{X}}{\ln( \subsc{p}{\mathrm{ap}}(\y|\x) )}
    &=
    \Nt \!-\! \Es\Tr{\Gap} \!+\!2 \Es\Ree{\Trr{\Faph\H}},
    \\[-0.5em]
    -\Ex{\mbf{Y}}{\ln( \subsc{p}{\mathrm{ap}}(\y) )}
    &=
    \Nt\ln\Es \!+\! \ln  \det \left( \Gap \right)
    -
    \Trr{ \Faph [ \Es\H \Hh + N_0\I ] \Fap \Gapi  }.
\end{align*}\\[-2.75em]
Substituting~\eqref{eq:Fap_Gap_def} for $\Gap$ and $\Fap$, and applying~\eqref{eq:Wmmse_formula_left_right} for $\Wmmse$, then
$\Faph[\Es\H\Hh \!+\! N_0\I ]\Fap\Gapi \!=\! \Es\Faph\H \!=\! \Es\Gap\Wmmse\H$. Also, it is easy to show that\vspace{-0.175in}
\begin{align}
\Wmmse\H
    &\!=\!
    [\Hh \H \!+\! \a\I_{\Nt}]^{-1} \Hh\H
    =
    \I \!-\! \a[\a\I_{\Nt} \!+\! \Hh\H]^{-1}\label{eq:WmmseH},
\end{align}\\[-2.75em]
which implies that $\Wmmse\H$ is Hermitian. Hence, $\Trr{\Gap\Wmmse\H}\!=\! \Trr{\Gap[\I \!-\! \a(\a\I \!+\! \Hh\H)^{-1}]}$ is real. Adding the two expectations above results in \vspace{-0.175in}
\begin{align*}
\ILBAWLD
    &=
    \Nt\!\ln\Es \!+\! \ln  \det( \Gap )
    \!-\! \Trr{\!\Gap[\!\tfrac{1}{\Es}\I \!+\! \tfrac{1}{N_0}\Hh\H]^{-1}}
    \!+\! \Nt
    \\[-0.5em]
    &=
    \Nt\!\ln\Es \!+\! \ln  \det ( \Gap )
    \!-\! \Trr{ \Gap(\Lah\La)^{-1}}
    \!+\! \Nt
    \\[-0.5em]
    &=
    \Nt\!\ln\Es \!+\! \ln  \det ( \Gap )
    \!-\! \Trr{ \Waph\Wap}
    \!+\! \Nt,
\end{align*}\\[-2.75em]
from which~\eqref{eq:ILB_AWLD} follows since $\Trr{ \Waph\Wap}\!=\!\Nt$.
\end{IEEEproof}\vspace{-0.01in}

With the punctured structure of the channel matrix $\Lap$ as given in~\eqref{eq:Lap_def}, the gap of $\ILBAWLD$ to AWGN capacity can be determined using the following corollary.

%
\begin{corollary} The gap of the AIR of the AWLD detector to AWGN capacity is\vspace{-0.15in}
\begin{align}\label{eq:gap_capacity_AWLD}
    C^{\AWGN} \!-\! \ILBAWLD
    &=\sum_{k=1}^{\Nt-\nu}\!\ln
        \left(s_{\mathrm{a}\:kk}^{2}\fnormmsq{[\mbf{S}_{\mathrm{a}}^{-1}]_{\bar{k}}}+1\right),
\end{align}\\[-2.25em]
where $s_{\mathrm{a}\:kk}^{}$ is the $\nth{k}$ diagonal element of $\Sa$ in~\eqref{eq:Lap_def}, and $[\Sai]_{\bar{k}}$ is the row vector consisting of the first $k\!-\!1$ elements in row $k$ of $\Sai$ in~\eqref{eq:Wap_formula}, excluding the diagonal element. 
\end{corollary}
\begin{IEEEproof} Applying~\eqref{eq:Lap_def}-\eqref{eq:Sigma_a} in~\eqref{eq:ILB_AWLD}, the $\ln\det$ term splits and the $C^{\AWGN}\!=\!\ln\det( \tfrac{\Es}{N_0}\Hh\H\!+\!\I_{\Nt})$ term emerges.
\end{IEEEproof}

Similar to the WLD case, the gap to capacity vanishes for $\nu\!=\!\Nt-1$. Also, the metrics computed by the AWLD detector approach the hard-decision ML metrics as $\nu$ increases from 1.

%
\begin{lemma}\label{lem:AWLD_distance_bound} Let $\mu(\x) \!=\!
\tfrac{1}{\Es}\fnormmsq{\x} \!-\!\fnormmsq{\La(\Wmmse\y \!-\! \x)}$, $\x_{\ML} \!=\!
\arg\max_{\x} \mu(\x)$, $\omega(\x) \!=\!
\tfrac{1}{\Es}\fnormmsq{\x} \!- \fnormmsq{\Wap\La(\Wmmse\y \!-\! \x)}$, and $\x_{\AWLD}
\!=\! \arg\max_{\x} \omega(\x)$. Then,\vspace{-0.175in}
\begin{align}
   \kappa^2\mu(\xML)
    -
    \eta(\kappa^2 -1)
    \leq
    \mu(\xAWLD)
    \leq
    \mu(\xML)\label{eq:AWLD_distance_bound}\\[-0.5em]
    \omega(\xAWLD)
    \geq
    \eta (1 - \sigma_{\max}(\Wap))
    + \sigma_{\max}(\Wap)\mu(\xML),\label{eq:AWLD_distance_bound2}
\end{align}\\[-2.75em]
where $\kappa\!=\!\sigma_{\max}(\Wap)/\sigma_{\min}(\Wap)$, and $\sigma_{\max}{(\Wap)}$, $\sigma_{\min}{(\Wap)}$ are the largest and smallest singular values of $\Wap$, respectively, $\eta\!=\!\tfrac{\Nt E_{\max}}{\Es}$, and $E_{\max}\!=\!\max_{x\in\X}\abs{x}^2$.
\end{lemma}
\begin{IEEEproof} The proof is similar to~Lemma~\ref{lem:WLD_distance_bound}, and uses the fact that $\omega(\x_{\AWLD})\!\geq\!\omega(\x_{\ML})$. As $\kappa\!\rightarrow\!1$, $\mu(\xAWLD)\!\rightarrow\!\mu(\xML)$.
\end{IEEEproof}

As illustrated in Fig.~\ref{fig:awld_block_diagram}, the AWLD detector includes the WLD detector as a sub-block; the processing steps of MMSE filtering and gain are done prior to WLD detection. Also, it is worth noting that computing the augmented channel requires simple processing comparable to QL decomposition. In particular, matrix inversion is not needed to compute $\Wmmse$ in~\eqref{eq:Wmmse_Ha_La} because the inverse of $\La$ is available from~\eqref{eq:Ha_QL_decomp_Ha_Qa2La}. In addition, using the modular approach of~\cite{2015_mansour_JSP_2x2QAM}, an efficient hardware architecture for an AWLD MIMO detector can be constructed from optimized $2\!\times\!2$ MIMO detector cores. Extensions to include soft-input information, imperfect channel estimation effects, and correlated channels are directly applicable based on~\cite{2017_hu_softoutput_AIR}. Finally, a scheme similar to~\cite{2018_Sarieddeen_Mansour_large_MIMO} can be used for analyzing the diversity gain.
\vspace{-0.125in}

%
\section{AIR-Optimality of the AWLD Detector}\label{s:optimality_AWLD_modified_detection_model}\vspace{-0.05in}
Instead of working with Euclidean-distance based metrics as in~\eqref{eq:true_metric}, the authors in~\cite{2012_rusek_optimal_channel_short} propose replacing $N_0$, $\H$, $\Hh\H$ in~\eqref{eq:true_metric_expanded}
with mismatched parameters $\subsc{N}{\mathrm{r}},\Fr,\Gr$ that are subject to AIR optimization. Hence, instead of the true metric in~\eqref{eq:true_metric_expanded_sufficient} and true probability in~\eqref{eq:true_detection_prob}, the mismatched model of~\cite{2012_rusek_optimal_channel_short} is\vspace{-0.2in}
\begin{align}
    \mu_{\mathrm{r}}(\y|\x)
    &=
    2\Re{\yh\Fr\x} -  \xh\Gr\x, \label{eq:mismatched_prob_rusek_mu}
    \\
    p_{\mathrm{r}}(\y|\x)
    &=
    \exp{\left( 2\Re{\yh\Fr\x}  -  \xh\Gr\x \right)}, \label{eq:mismatched_prob_rusek}
\end{align}\\[-2.75em]
where $\subsc{N}{\mathrm{r}}$ is absorbed into $\Fr$ and $\Gr$. It is shown in~\cite{2012_rusek_optimal_channel_short} that detectors limited to the Euclidean-based model in~\eqref{eq:true_metric_expanded_sufficient}, where $\Gr$ admits a Cholesky factorization proportional to $\Hh\H$, are not optimal from a mutual information perspective because the resulting optimal matrix $\Gr$ to use in~\eqref{eq:mismatched_prob_rusek} may not be positive semidefinite, and hence no such factorization exists. The optimal $\Fr$ and $\Gr$ are derived by maximizing the lower bound on the AIR in two steps, assuming $\Gr$ is Hermitian (and hence has real eigenvalues). First, an explicit expression for $\Fr$ is derived, having the form $\Fropt\!=\!(\H\Hh\!+\!\alpha\I)^{-1}\H(\Gr\!+\!\I)$; this is the MMSE filter compensated by the receiver tree processing through $\Gr\!+\!\I$ (rather than $\Gr$). Next, the corresponding AIR bound with $\Fropt$ substituted, depends on $\Gr$ through the factor $(\Gr\!+\!\I)$. An assumption on the matrix $\Gr$ is imposed to have all its eigenvalues strictly larger than $-1$, so that $\Gr\!+\!\I$ becomes positive semidefinite and hence admits a Cholesky factorization of the form $\mathbf{L}_{\mathrm{r}}^{\herm}\mathbf{L}_{\mathrm{r}}$. Accordingly, the AIR bound depends solely on the lower-triangular matrix $\mathbf{L}_{\mathrm{r}}$. By maximizing this bound, the optimal $\mathbf{L}_{\mathrm{r}}$ is derived, having a shortened (punctured) structure analogous to that of the WLD scheme~\cite{2014_mansour_SPL_WLD}.

In this work, we propose the following modified model\vspace{-0.175in}
\begin{align}
    \mum{\y|\x}
    &=
    2\Re{\yh\F\x} - \xh\G\x + \tfrac{1}{\Es}\xh\x, \label{eq:new_metric}
\end{align}\\[-2.75em]
and $\subsc{p}{\mathrm{m}}(\y|\x)\!=\!\exp(\mum{\y|\x})$, where tree processing is split into an explicit term $\tfrac{1}{\Es}\xh\x$ separate from $\xh\G\x$ for which $\G$ is subject to optimization. The reason is that the optimal $\F$ in this case, as will be shown, takes the form $\Fopt=\supsc{[ \H \Hh + \alpha\I ]}{\,-1} \H \G$, and the resulting AIR lower bound depends on $\G$ directly and not through the term $\G+\I$, as is the case with~\cite{2012_rusek_optimal_channel_short}. Hence, the assumption on $\G$ to have all its eigenvalues strictly larger than $-1$ is dropped. We directly require that $\G$ be positive semidefinite having a Cholesky factorization $\Jh\J$, where $\J$ has the desired punctured lower-triangular form. Under such formulation, we show that the optimal $\F$ and $\G$ coincide exactly with those of the AWLD detector in~\eqref{eq:Fap_Gap_def}.
\vspace{-0.05in}

%
\begin{theorem}\label{thm:optimal_ILB_modified_model} Under the same assumptions as Theorem~\ref{thm:ILB_wld}, the
optimal $\F$ and $\G$ that maximize $\ILB\!=\!\Ex{\mbf{Y},\mbf{X}}{\ln( \subsc{p}{\mathrm{m}}(\y|\x) )} -\Ex{\mbf{Y}}{\ln( \subsc{p}{\mathrm{m}}(\y) )}$, such that $\G$ is positive semidefinite with factor matrices having a punctured structure of order $\nu$, are\vspace{-0.175in}
\begin{align}
    \Fopt
    &=
    \Wmmseh\Gopt
    \quad\text{and}\quad
    \Gopt
    =
    \Jopth\Jopt, \label{eq:Fopt_Gopt_modified_model}
\end{align}\\[-2.5em]
where $\Wmmse$ is the standard $\NbyM$ MMSE filter matrix in~\eqref{eq:Wmmse_formula_left_right}, and $\Jopt$ is the punctured augmented WLD matrix $\Lap$ given in~\eqref{eq:Lap_def}. Accordingly, the lower bound attained by the AWLD detector in~\eqref{eq:ILB_AWLD} is optimal.\\[-2.25em]
\end{theorem}
\begin{IEEEproof} Let $\ILBopt =\!\! \max_{\F,\G:\G\succeq 0}^{}\ILB$ and
$(\Fopt,\Gopt) \!=\! \argmax_{\F,\G:\G\succeq 0}^{}\ILB$. The expectations in the $\ILB$ expression with $\subsc{p}{\mathrm{m}}(\y|\x)\!=\!\exp(\mum{\y|\x})$ and $\subsc{p}{\mathrm{m}}(\y)\!=\!\int \subsc{p}{\mathrm{m}}(\y|\x) p(\x)\,\mathrm{d}\x$ are\vspace{-0.15in}
\begin{align*}
    \Ex{\mbf{Y},\mbf{X}}{\ln( \subsc{p}{\mathrm{m}}(\y|\x) )}
    &=
    \Nt - \Es\Trr{\G} +2 \Es\Ree{\Trr{\Fh\H}},
    \\[-0.5em]
    -\Ex{\mbf{Y}}{\ln( \subsc{p}{\mathrm{m}}(\y) )}
    &=
    \Nt\ln\Es + \ln  \det ( \G )
    -
    \Trr{ \Fh[ \Es\H \Hh + N_0\I  ] \F \Gi  }.
\end{align*}\\[-3.0em]
To determine $\F$ that maximizes $\ILB$, we set the derivative of the terms in the sum of the two expectations involving $\F$ to 0,\vspace{-0.15in}
\begin{align*}
    \diffp*{(2 \Es\Ree{\Trr{\Fh\H}}\!-\!\Trr{ \Fh[\Es\H \Hh \!+\! N_0\I] \F \Gi})}\F
    = 0,
\end{align*}\\[-2.75em]
from which it follows, after some tedious steps, that\vspace{-0.175in}
\begin{align*}
    \Fopt
    \!=\!
    \supsc{[ \H \Hh + \alpha\I ]}{\,-1} \H \G
    = \Wmmseh \G, \qquad \alpha = \tfrac{N_0}{E_s}.
\end{align*}\\[-2.75em]
Substituting $\Fopt$ back in $\ILB$, and noting that $\Fopth\H\!=\!\G(\Wmmse\H)$ is the product of two Hermitian matrices and hence has real trace, we obtain, after further simplifications\vspace{-0.175in}
\begin{align}
    \ILBopttilde
    &\!=\!
    \Nt\!\ln\!\Es + \ln\det (\G)
    - \Es \Trr{(\I \!-\! \Wmmse\H)\G\!} + \Nt. \label{eq:ILBopt_tilde}
\end{align}\\[-2.75em]
Using~\eqref{eq:WmmseH}, it follows that $\Es(\I \!-\! \Wmmse\H)= \Es\a\supsc{[\a\I_{\Nt} + \Hh\H]}{\,-1}\!=\!\supsc{(\Hah\Ha)}{\,-1}$, where $\Ha$ is defined in~\eqref{eq:Ha_def}. Then\vspace{-0.25in}
\begin{align}
    \ILBopttilde
    &\!=\!
    \Nt\!\ln\Es + \ln\det (\Jh\!\J)
    -  \Trr{\supsc{(\Lah\La)}{\,-1}\!\Jh\!\J} + \Nt,
    \label{eq:ILBopt_tilde_J}
\end{align}\\[-2.75em]
where $\Ha\!=\Qa{}\La$ is the QL decomposition of $\Ha$, and $\G\!=\!\Jh\J$ such that $\J$ is a punctured lower triangular matrix of order $\nu$. We next determine $\J$ that maximizes $\ILBopttilde$:\vspace{-0.2in}
\begin{align}
    \Jopt
    &=
    \argmax_{\J}\ILBopttilde. \label{eq:Jopt_def}
\end{align}\\[-2.75em]
Assume $\J$ and $\La$ are conformally partitioned as\vspace{-0.175in}
\begin{align}
    \J
    &=
    \begin{bNiceMatrix}
        \subsc{\J}{1} &   \\
        \subsc{\J}{2} & \subsc{\J}{3}
    \end{bNiceMatrix}
    \qquad\text{and}\qquad
    \La
    =
    \begin{bNiceMatrix}
        \Pa &   \\
        \Ra & \Sa
    \end{bNiceMatrix},\label{eq:J_La_partitioned}
\end{align}\\[-2.2em]
where $\subsc{\J}{1},\Pa$ are $\nu\!\times\!\nu$ lower triangular,  $\subsc{\J}{3}$ is $(\Nt\!-\!\nu)\!\times\!(\Nt\!-\!\nu)$ real diagonal, $\Sa$ is $(\Nt\!-\!\nu)\!\times\!(\Nt\!-\!\nu)$ lower triangular, and $\subsc{\J}{2},\Ra$ are $(\Nt\!-\!\nu)\!\times\!\nu$ matrices. Note that $\subsc{\J}{3}$ is constrained to be a diagonal matrix, not just lower-triangular\footnote{Hence Lemma~\eqref{lem:trace_formula} is not directly applicable to derive the optimal $\J$ that minimizes~\eqref{eq:ILBopt_tilde_J} at this point.}. Then the trace $\Trr{\supsc{(\Lah\La)}{\,-1}\Jh\J}\!=\!\Trr{(\J\Lai)\supsc{(\J\Lai)}{\,\dag}}\!=\!\subpsc{\fnormm{\J\Lai}}{\mathrm{F}}{\,2}$
in~\eqref{eq:ILBopt_tilde_J} can be computed using $\J\Lai$ as follows:\vspace{-0.15in}
\begin{align*}
\hspace{-0.1in}
    \J\Lai
    &\!=
    \begin{bNiceMatrix}
        \subsc{\J}{1} &   \\
        \subsc{\J}{2} & \subsc{\J}{3}
    \end{bNiceMatrix}
    \begin{bNiceMatrix}
        \Pai &   \\
        -\Sai\Ra\Pai & \Sai
    \end{bNiceMatrix}
    \\[-0.5em]
    \subpsc{\fnormm{\J\Lai}}{\mathrm{F}}{\,2}
    &\!=\!
    \subpsc{\fnormm{\subsc{\J}{1}\Pai}}{\mathrm{F}}{\,2}
    \!+\!
    \subpsc{\fnormm{(\subsc{\J}{2} \!-\! \subsc{\J}{3}\Sai\Ra)\Pai}}{\mathrm{F}}{\,2}
    \!+\!
    \subpsc{\fnormm{\subsc{\J}{3}\Sai}}{\mathrm{F}}{\,2}.
\end{align*}\\[-2.75em]
Since the $\ln\det (\Jh\J)$ term in~\eqref{eq:ILBopt_tilde_J} involves the diagonal terms of $\subsc{\J}{1}$ and $\subsc{\J}{3}$ only, then $\ILBopttilde$ can be optimized for $\subsc{\J}{2}$ and $(\subsc{\J}{1},\subsc{\J}{3})$ independently.

Starting with $\subsc{\J}{2}$, we set $\diffp*{\ILBopttilde}{\subsc{\J}{2}}\!=\!\diffp*{\Trr{\supsc{(\Lah\La)}{\,-1}\!\Jh\J}}{\subsc{\J}{2}}\!=\!0$, to obtain $\subpsc{\J}{2}{\,\opt} \! \!=\!  \subsc{\J}{3}\Sai \Ra$. Substituting back in~\eqref{eq:ILBopt_tilde_J}, we get\vspace{-0.175in}
\begin{align}
    \ILBopttilde
    &\!=\!
    \Nt\ln\Es
    + \ln\det (\subsc{\J}{1}^{\dag}\subsc{\J}{1})
    + \ln\det (\subsc{\J}{3}^{\dag}\subsc{\J}{3})
    + \Nt
    - \subpsc{\fnormm{\subsc{\J}{1}\Pai}}{\mathrm{F}}{\,2}
    - \subpsc{\fnormm{\subsc{\J}{3}\Sai}}{\mathrm{F}}{\,2}.
\label{eq:ILBopt_tilde_J2opt}
\end{align}\\[-2.75em]
Moving to $\subsc{\J}{3}$, we set $\diffp*{\ILBopttilde}{\subsc{\J}{3}}\!=\!0$. Noting that $\subsc{\J}{3}$ is real and diagonal, we obtain $2\subpsc{\J}{3}{\,-1}\!-\!2\subsc{\J}{3}\diagg{\Sai\Saih}\!=\!0$, from which it follows that $\subpsc{\J}{3}{\,\opt}\!=\!\supsc{\diagg{\Sai\Saih}}{-1/2}\!=\!\Omegaa$. Substituting back in~\eqref{eq:ILBopt_tilde_J2opt}, we get\vspace{-0.2in}
\begin{align}
    \ILBopttilde
    &=
    \Nt\ln\Es
    + \ln\det ( \subpsc{\mbf{\Omega}}{\mathrm{a}}{\,2}  )
    - \subpsc{\fnormm{\Omegaa\Sai}}{\mathrm{F}}{\,2}
    + \Nt + \ln\det (\subsc{\J}{1}^{\dag}\subsc{\J}{1})
    -
    \Trr{(\subsc{\J}{1}\Pai)\supsc{(\subsc{\J}{1}\Pai)}{\,\dag}}.
    \label{eq:ILBopt_tilde_J2opt_J3opt}
\end{align}\\[-2.75em]
Finally, using Lemma~\ref{lem:trace_formula} below, the optimal $\subsc{\J}{1}$ that maximizes $\ILBopttilde$ with $\Pa$ being lower triangular is
$\subpsc{\J}{1}{\,\opt}\!=\!\Pa$. The resulting $\supsc{\J}{\,\opt}$, with $\subpsc{\J}{1}{\,\opt},\subpsc{\J}{2}{\,\opt},\subpsc{\J}{3}{\,\opt}$ in place, is\vspace{-0.125in}
\begin{align}
    \Jopt
    &=
    \begin{bNiceMatrix}
        \Pa &   \\
        \Omegaa \Sai \Ra & \Omegaa
    \end{bNiceMatrix},\label{eq:Jopt}
\end{align}\\[-1.75em]
which coincides with $\Lap$ as given in~\eqref{eq:Lap_def}. The optimal lower bound $\ILBopt$ attained in~\eqref{eq:ILBopt_tilde_J2opt_J3opt} is\vspace{-0.175in}
\begin{align}
    \ILBopt
    \!&=\!
    \Nt\!\ln\Es
    \!+\! \ln\det ( \subpsc{\mbf{\Omega}}{\mathrm{a}}{\,2}  )
    \!-\! \subpsc{\fnormm{\Omegaa\Sai}}{\mathrm{F}}{\,2}
    \!+\! \Nt \! +\! \ln\det (\Pah\Pa)
    \!-\! \nu =
    \Nt\!\ln\Es
    \!+\! \ln\det ( \Jopth\Jopt),\label{eq:ILBopt_tilde_J2opt_J3opt_J1opt}
\end{align}\\[-2.75em]
since $\subpsc{\fnormm{\Omegaa\Sai}}{\mathrm{F}}{\,2}\!=\!\Nt\!-\!\nu$, which equals $\ILBAWLD$ in~\eqref{eq:ILB_AWLD}.
\end{IEEEproof}

%
\begin{lemma} \label{lem:trace_formula}
Let $\U$ and $\V$ be two non-singular square matrices in $\mathcal{C}^{N\times N}$. Let $f(\U,\V) \!=\! \ln\det(\U\Uh) \!-\! \Trr{(\U\V)\supsc{(\U\V)}{\,\dag}}$ be a real-valued function of complex-valued matrices. Then the optimal $\U$ that maximizes $f$ for a given $\V$ is\vspace{-0.175in}
\begin{align*}
    \Uopt
    &=
    \argmax_{\U} f(\U,\V)
    = \Vi,
\end{align*}\\[-2.5em]
and $f(\Uopt,\V) \!=\! -\sum_{k=1}^{N}\ln \tilde{v}_{kk}^2 \!-\! N$,
where $\tilde{v}_{kk}$ is the $\nth{k}$ diagonal element of the Cholesky factor of $\V\Vh$.
\end{lemma}\vspace{-0.025in}
\begin{IEEEproof} See Supplement~\ref{supplement:proof_of_trace_formula}.
\end{IEEEproof}
\vspace{-0.025in}

\textit{Discussion}: We conclude that punctured augmented channel matrices processed by the AWLD detector are optimal in maximizing the lower bound on the achievable information rate. Their structure matches exactly that of the AIR-PM detector, but most importantly, they can be computed using simple QL decomposition followed by Gaussian elimination, resulting in a significant complexity reduction compared to~\cite{2017_hu_softoutput_AIR}.
\vspace{-0.1in}

%
\section{Efficient Matrix Decomposition Algorithms}\label{sec:eff_decomp_algorithms}
\vspace{-0.05in}

%
\subsection{Matrix-Inverse-Free Puncturing via Gaussian Elimination}\label{sec:WLD_gaussian_elimination}
Directly inverting $\S$ in~\eqref{eq:Dp_Wp_Lp_formula} can be avoided if we apply Gaussian elimination to null the elements below the main diagonal of $\S\!=\![s_{kj}]$ in~\eqref{eq:L_nu_partitioned}. Let \vspace{-0.15in}
\begin{align}\label{eq:Gaus_trans_Ek}
    \E_j^{}\!=\! \I_{\Nt-\nu} - \subpsc{\bm{\tau}}{j}{} \subpsc{\mbf{e}}{j}{\tran} \in\C^{{(\Nt-\nu)}\times {(\Nt-\nu)}},
\end{align}\\[-2.5em]
be a Gauss transformation~\cite{2013_golub_matrix}, where $\subpsc{\mbf{e}}{j}{}$ is the $\nth{j}$ column of $\I_{\Nt-\nu}$, and $\subpsc{\bm{\tau}}{j}{}$ is the Gauss vector\vspace{-0.175in}
\begin{align*}
    \subpsc{\bm{\tau}}{j}{\tran}
    \!=\!
    [\underbrace{0,\cdots,0}_{j},\subpsc{\tau}{j+1}{},\cdots,\subpsc{\tau}{\Nt-\nu}{}],
    ~~
    \subpsc{\tau}{i}{}\!=\!\tfrac{s_{kj}^{}}{s_{jj}^{}},
    ~~
    \mbox{\small $k\!=\!j\!+\!1\!:\!\Nt\!-\!\nu$}.
\end{align*}\\[-2.0em]
Then the operation $\E_j^{}\S$ nulls all the entries below the $\nth{j}$ diagonal element in $\S$. Applying this operation repeatedly for $j\!=\!1,\cdots,\Nt\!-\!\nu\!-\!1$ would null all entries in $\S$ below the main diagonal. Grouping these row operations into \vspace{-0.2in}
\begin{align}
    \E
    &=
    \E_{\Nt-\nu-1}^{}\cdots\E_2^{}\E_1^{}
    =
    {{\prod_{j=1}^{\Nt-\nu-1} \E_j^{}}},
    \label{eq:E_gaus_matrix_def}
\end{align} \\[-2.5em]
results in $\E \mskip1mu \S \!=\!\diagg{\S}$, or $\Si\!=\!\supsc{\diagg{\S}}{-1}\E$ (note that $\E$ is non-unitary). Setting \vspace{-0.175in}
\begin{align}\label{eq:Sigma_E}
    \subsc{\mbf{\Omega}}{\mathrm{E}}
    &=
    \supsc{\diagg{ \mbf{E} \, \supsc{\mbf{E}}{\,\dag} } }{-1/2},
\end{align}\\[-3em]
gives the required product $\mbf{\Omega}\mskip1mu\Si$ in~\eqref{eq:Dp_Wp_Lp_formula} inverse-free as\vspace{-0.2in}
\begin{align}\label{eq:Sigma_E_times_E}
    \mbf{\Omega}\mskip1mu\Si
    &=
    \subsc{\mbf{\Omega}}{\mathrm{E}}\E.
\end{align}
\vspace{-0.65in}

%
\subsection{Eliminating Square-Roots via QDL Decomposition}\label{s:elim_sqrt_op} \vspace{-0.075in}
In forming the QL decomposition $\H\!=\!\Q\L$, the $\nth{j}$ column $\qj$ of $\Q\!=\![\subsc{q}{kj}]$ is obtained by subtracting from the $\nth{j}$ column $\hj$ of $\H$ the orthogonal projection of all other columns of $\H$ (denoted as $\Hjbar$) onto $\hj$, i.e., $\qj\!=\!\hj \!-\!(\Hjbarherm\hj)\Hjbar$. The $\nth{j}$ diagonal element $\subpsc{l}{jj}{}$ of $\L$ is set to the norm of $\qj$, $\subpsc{l}{jj}{}\!=\!\fnormm{\qj}$. Finally, $\qj$ is normalized as $\qj\!=\!\qj/\fnormm{\qj}$.

The square-root operation required to compute $\fnormm{\qj}\!=\!\sqrt{\smash[b]{\sum_{k=1}^{\Nt}{\supsc{\abss{\subsc{q}{kj}}}{\,2}}}}$ can be eliminated by working with squared-norms $\subsc{d}{jj}\!=\!\supsc{\fnormm{\qj}}{\,2}$ instead, and storing them in a diagonal normalizer matrix $\D\!=\![\subsc{d}{jj}]\!\in\! \supsc{\C}{\,\Nt\times \Nt}$, apart from the factors $\Q,\L$. The modified `QDL' decomposition becomes\vspace{-0.16in}
\begin{align}\label{eq:QDL_def}
    \H
    &=
    \Q\L
    =
    (\Q\supsc{\D}{-1/2})
    \mskip1mu \D \mskip1mu
    (\supsc{\D}{-1/2}\L)
    =
    \Qtilde\D\Ltilde,
\end{align}\\[-2.75em]
where $\Qtilde \!=\! \Q \supsc{\D}{\,-1/2} \!\in\! \supsc{\C}{\,\Nr\times \Nt}$ is an unnormalized matrix with orthogonal columns $\Qtildeh\Qtilde \!=\! \Di \!\neq\! \subsc{\mbf{I}}{\Nt}$, $\D\!=\!\supsc{\diagg{\L}}{\,2}$, and $\Ltilde\!=\!\supsc{\D}{\,-1/2}\L \!\in\! \C^{\Nt\times \Nt}$ is an unnormalized unit lower-triangular matrix. Observe now that the column vectors $\qtildej$ of $\Qtilde$ and the diagonal entries $\subpsc{\tilde{l}}{jj}{}$ of $\Ltilde$ both do not involve square-roots also because $\qtildej\!=\!\qj/\supsc{\fnormm{\qj}}{\,2}$ and $\subpsc{\tilde{l}}{jj}{}\!=\!\fnormm{\qj}/\fnormm{\qj}\!=\!1$.

The pseudo-codes of the standard (unnormalized) QL algorithm and QDL algorithm are shown in \algsnameabbr~\ref{algo:qldy} and~\ref{algo:qdly}, respectively. The codes are optimized to produce $\yt\!=\!\Qh\y$ and $\tsup{\y}\!=\!\Qtildeh\y$ indirectly as well by augmenting $\y$ to $\H$ and performing modified Gram-Schmidt operations on $[\y~\H]$.
\vspace{-0.125in}

%
\subsection{Combined Inverse-Free and Square-Root-Free WLD}\label{s:inverse_sqrt_free_wld}
The puncturing matrix $\Wp$ and the punctured lower-triangular matrix $\Lp$ can now be expressed in terms of the QDL factors of $\H\!=\!\Qtilde\D\Ltilde$ as follows. Starting with \vspace{-0.175in}
\begin{align}\label{eq:Qt_D_Lt}
    \Qtilde
    &\!=\!
    \Q \supsc{\D}{\,-1/2}\!,
    ~~
    \D
    \!=\!
    \supsc{\diagg{\L}}{\,2}\!,
    ~~
    \Ltilde
    \!=\!
    \supsc{\D}{\,-1/2}\L,
\end{align}\\[-3em]
and forming the conformal partitions as in~\eqref{eq:L_nu_partitioned},\vspace{-0.125in}
\begin{align}\label{eq:L_D_Lt_partitions}
    \L
    &\!=\!
    \begin{bNiceMatrix}
        \P & \mbf{0}  \\
        \mbf{R} & \S
    \end{bNiceMatrix},
    ~~
    \supsc{\D}{\,1/2}
    \!=\!
    \begin{bNiceMatrix}
        \subpsc{\D}{1}{\,1/2} & \mbf{0} \\
        \mbf{0} & \subpsc{\D}{2}{\,1/2}
    \end{bNiceMatrix},
    ~~
    \Ltilde
    \!=\!
    \begin{bNiceMatrix}
        \Ptilde & \mbf{0} \\
        \Rtilde & \tilde{\mbf{S}}
    \end{bNiceMatrix},
\end{align}\\[-1.75em]
we have $\Ptilde \!=\! \subpsc{\D}{1}{\,-1/2} \P$ with $\diagg{\Ptilde}
\!=\!\I$, $\Rtilde \!=\! \subpsc{\D}{2}{\,-1/2} \mbf{R}$, and $\Stilde \!=\! \subpsc{\D}{2}{\,-1/2} \S$ with $\diagg{\Stilde}
\!=\!\I$. However, the true $\Wp$, $\mbf{\Omega}$, and $\Lp$,\vspace{-0.155in}
\begin{align*}
    \Wp
    &\!=\!
    \begingroup 
        \setlength\arraycolsep{3pt}
    \begin{bNiceMatrix}
            \I & \mbf{0}  \\
            \mbf{0} & \mbf{\Omega}\Si
    \end{bNiceMatrix}
    \endgroup\!,
    ~~
    \mbf{\Omega}
    \!=\!
    \supsc{\diagg{\Si\Sih}}{-1/2}\!,
    ~~
    \Lp
    \!=\!
    \begingroup 
        \setlength\arraycolsep{3pt}
    \begin{bNiceMatrix}
        \P & \mbf{0}  \\
        \mbf{\Omega}\Si\mbf{R} & \mbf{\Omega}
    \end{bNiceMatrix}
    \endgroup,
\end{align*}\\[-2.0em]
in~\eqref{eq:Dp_Wp_Lp_formula}-\eqref{eq:Sigma} require the inverse of $\S$, when only the submatrix $\tilde{\mbf{S}}$ is computed in~\eqref{eq:L_D_Lt_partitions} via the QDL scheme. In addition, $\mbf{\Omega}$ involves square-root operations. We first expand $\Lp$ as follows\vspace{-0.15in}
\begin{align*}
    \Lp
    &=
    \Wp\L
    =
    \Wp \supsc{\D}{\,1/2}
    \Ltilde
    =
    \begin{bNiceMatrix}
        \I & \mbf{0}  \\
        \mbf{0} & \mbf{\Omega}\,\Si
    \end{bNiceMatrix}
    \begingroup 
        \setlength\arraycolsep{2pt}
        \begin{bNiceMatrix}[columns-width = 6mm]
            \subpsc{\D}{1}{\,1/2} & \mbf{0} \\
            \mbf{0} & \subpsc{\D}{2}{\,1/2}
        \end{bNiceMatrix}
    \endgroup
    \begin{bNiceMatrix}
        \Ptilde & \mbf{0} \\
        \Rtilde & \Stilde
    \end{bNiceMatrix}.
\end{align*}\\[-2.15em]
Substituting $\Si\!=\!\Stildeinv\subpsc{\D}{2}{\,-1/2}$, we obtain\vspace{-0.175in}
\begin{align}
    \Wptilde
    &=
    \Wp \supsc{\D}{\,-1/2}
    =
    \begin{bNiceMatrix}
        \subpsc{\D}{1}{\,-1/2} & \mbf{0}  \\
        \mbf{0} & \mbf{\Omega}\,\supsc{\tilde{\S}}{\,-1}\subpsc{\D}{2}{\,-1}
    \end{bNiceMatrix}\label{eq:Wptilde}
    \\
    \Lp
    &=
    \Wp
    \L
    =
    \Wptilde
    \D
    \Ltilde
    =
    \begin{bNiceMatrix}
        \subpsc{\D}{1}{\,1/2}\tilde{\mbf{P}} & \mbf{0}  \\
        \mbf{\Omega}\,\supsc{\tilde{\S}}{\,-1}\tilde{\mbf{R}} & \mbf{\Omega}
    \end{bNiceMatrix}.\label{eq:Lp_Wtp*Lt}
\end{align}\\[-2.0em]
Similarly, we can express $\mbf{\Omega}$ in terms of $\tilde{\S}$ as follows \vspace{-0.175in}
\begin{align}\label{eq:Sigma_Stilde}
    \mbf{\Omega}
    &=
    \supsc{
        \diagg{
            \Stildeinv
            \,\subpsc{\D}{2}{\,-1}
            \Stildeinvh
        }
    }
    {-1/2}.
\end{align}\\[-2.75em]
We next eliminate computing the inverse $\Stildeinv$ in the above equations. Using Gaussian elimination, we apply a sequence of Gauss transformations $\Etilde$ to invert $\Stilde$ similar to~\eqref{eq:Gaus_trans_Ek}. Since $\Stilde$ has unit diagonal, we obtain $\Etilde \, \Stilde \!=\! \diagg{\Stilde} \!=\!\I_{\Nt-\nu}$,
from which it follows that the inverse of $\Stilde$ is simply \vspace{-0.175in}
\begin{align}
    \Stildeinv
    &=
    \Etilde.
    \label{eq:Stilde_inv}
\end{align}\\[-2.75em]
Substituting $\Etilde$ for $\Stildeinv$ in the equations of $\Wptilde$~\eqref{eq:Wptilde}, $\mbf{\Omega}$~\eqref{eq:Sigma_Stilde}, and $\Lp$~\eqref{eq:Lp_Wtp*Lt}, we get:\vspace{-0.125in}
\begin{align}
\hspace{-0.05in}
    \Wptilde
    &=
    \begin{bNiceMatrix}
        \subpsc{\D}{1}{\,-1/2} & \mbf{0}  \\
        \mbf{0} & \mbf{\Omega}\Etilde\subpsc{\D}{2}{\,-1}
    \end{bNiceMatrix},
    \quad
    \mbf{\Omega}
    =
    \supsc{
        \diagg{
            \tilde{\E}
            \,\subpsc{\D}{2}{\,-1}
            \,\supsc{\tilde{\E}}{\,\dag}
        }
    }
    {-1/2},
    \quad
    \Lp
    =
    \begin{bNiceMatrix}
        \subpsc{\D}{1}{\,1/2}\tilde{\mbf{P}} & \mbf{0}  \\
        \mbf{\Omega}\,\tilde{\E}\tilde{\mbf{R}} & \mbf{\Omega}
    \end{bNiceMatrix}.\label{eq:Wptilde_Sigma_Lp_in_terms_tildes}
\end{align}\\[-2.0em]
Note now that the above equations do not involve matrix inversion ($\subsc{\D}{1},\subsc{\D}{2}$ are diagonal matrices).

Moving to the square-roots in~\eqref{eq:Wptilde_Sigma_Lp_in_terms_tildes}, we show that these operations also are not needed by the detector when computing squared-distances. Since $\Wp\Qh\!=\!\Wptilde\D\Qtildeh$ and $\Wp\L\!=\!\Wptilde\D\Ltilde$, then\vspace{-0.15in}
\begin{align}
    \fnormmsq{\Wp(\Qh\y\!-\!\L\x)}
    &=
    \fnormmsq{\Wptilde\D(\Qtildeh\y\!-\!\Ltilde\x)}
    =
    \supsc{(\Qtildeh\y\!-\!\Ltilde\x)}{\,\dag}\Dh\Wptildeh\Wptilde\D(\Qtildeh\y\!-\!\Ltilde\x).
    \label{eq:WLD_inv_sqrt_free_1}
\end{align}\\[-2.5em]
The quantities $\Qtildeh,\Ltilde$ are square-root free, and so is the product \vspace{-0.15in}
\begin{align}
    \Dh\Wptildeh\Wptilde\D =
    \begin{bNiceMatrix}
        \I & \mbf{0}  \\
        \mbf{0} & \Etildeh\supsc{\mbf{\Omega}}{\,2}\Etilde
    \end{bNiceMatrix},\label{eq:Wptildeh_Wptilde}
\end{align}\\[-2.25em]
since $\supsc{\mbf{\Omega}}{\,2}$ and $\Etilde$ do not involve square-roots.

The pseudo-code of the optimized WDL decomposition algorithm is shown in \algnameabbr~\ref{algo:generalized_wdl}. It first performs QDL decomposition on $\H$, followed by Gaussian elimination. The code is further optimized to eliminate computing the matrix products $\Wptilde\D\Qtildeh\y$ and $\Wptilde\D\Ltilde\!=\!\Lp$ in~\eqref{eq:WLD_inv_sqrt_free_1} explicitly. The QDL procedure first generates $\tsup{\y}\!=\!\Qtildeh\y$ as a byproduct to computing $\Ltilde$, $\D$, and $\Qtilde$. Next, starting with $\Wtilde\!=\!\Qtilde$, the Gaussian elimination loop then immediately applies the same operations to null the entries in $\Ltilde$ on $\tsup{\y}$, as well as on the corresponding columns of $\Wtilde$, and updates their resulting squared-norms in $\D$. The generated $\Wtildeh$ equals $\supsc{\mbf{D}}{\,-1/2}\Wptilde\D\Qtildeh$, and the required outputs are formed as $\Wtildeh\H\!=\!\Lptilde$ and $\Wtildeh\y\!=\!\yptilde$ with an extra scaling factor $\supsc{\mbf{D}}{\,-1/2}$. Note that $\Wtildeh$ operates on $\H$ directly, rather than on $\D\Ltilde$ like $\Wptilde$ in~\eqref{eq:Lp_Wtp*Lt} to form $\Lp\!=\!\Wptilde\D\Ltilde$. The output quantities $\Lptilde, \yptilde, \D, \Wtilde$ from the algorithm are then used to compute the metrics in~\eqref{eq:WLD_inv_sqrt_free_1} and~\eqref{eq:wld_metric} as follows\vspace{-0.175in}
\begin{align}
    \fnormmsq{\Wp(\Qh\y\!-\!\L\x)}
    &=
    \fnormmsq{\Wp\Qh\!(\y\!-\!\H\x)}
    \!=\!
    \fnormmsq{\supsc{\D}{\,1/2}\Wtildeh\!(\y\!-\!\H\x)}\notag
    \\[-0.25em]
    \!&=\!
    \fnormmsq{\supsc{\D}{\,1/2}\!(\yptilde\!-\!\Lptilde\x)}
    \!=\!\supsc{(\yptilde\!-\!\Lptilde\x)}{\,\dag}\D(\yptilde\!-\!\Lptilde\x),
    \notag
    \\[-0.25em]
    \mup{\y|\x}
    &=-\tfrac{1}{N_0}\supsc{(\yptilde\!-\!\Lptilde\x)}{\,\dag}\D(\yptilde\!-\!\Lptilde\x).
    \label{eq:mup_WLD_algo}
\end{align} \\[-2.25em]
For reference, an optimized version of the standard (unnormalized) WL algorithm of~\cite{2014_mansour_eurasip_WLD} is listed in \algnameabbr~\ref{algo:generalized_wld}. The outputs $\Lp, \yp, \W$ from this algorithm compute~\eqref{eq:WLD_inv_sqrt_free_1} as follows\vspace{-0.175in}
\begin{align*}
    \fnormmsq{\Wp(\Qh\y\!-\!\L\x)}
    \!=\!
    \fnormmsq{\Wp\Qh(\y\!-\!\H\x)}
    \!=\!
    \fnormmsq{\W(\y\!-\!\H\x)}.
\end{align*}\\[-3em]

\vspace{-0.45in}
%
\subsection{Eliminating Explicit Computation of MMSE Filter Matrix}\label{s:elim_MMSE_filter}\vspace{-0.05in}
For the AWLD detector, the MMSE filter matrix $\Wmmse$ in~\eqref{eq:Wmmse_formula_left_right} is needed to compute the metrics in~\eqref{eq:awld_metric_1}-\eqref{eq:awld_metric}. This $\Wmmse$ is to be pre-multiplied with $\Wap\La\!=\!\Lap$ and applied to $\y$ in~\eqref{eq:awld_metric_1}, or pre-multiplied with $\Laph\Lap$ and then applied to $\y$ in~\eqref{eq:awld_metric}. In either case, working with the quantity $\Wap\La\Wmmse$ suffices. However,~\eqref{eq:Wmmse_Qa2_Qa1} shows that $\Wmmse$ can be obtained from the QL decomposition of $\Ha$ in~\eqref{eq:Ha_def} as $\sqrt{\beta}\Qa{2} \Qah{1}$ without explicitly inverting $\Ha$. But $\La\Qa{2}\!=\!\tfrac{1}{\sqrt{\Es}}\I_{\Nt}^{}$ from~\eqref{eq:Ha_QL_decomp_Ha_Qa2La}, so that $\Wap\La\Wmmse\y$ actually reduces to $\tfrac{1}{\sqrt{N_0}}\Wap\Qah{1}\y$. The product $\tfrac{1}{\sqrt{N_0}}\Qah{1}\y$ can be obtained indirectly from the QL decomposition procedure (\algnameabbr~\ref{algo:qldy}) when applied to $\Ha$ and $\ya\!=\!\tfrac{1}{\sqrt{N_0}}[\y;~\mbf{0}_{\Nt\times 1}]$ as
$\Qah{}\ya\!=\!\tfrac{1}{\sqrt{N_0}}[\Qah{1}~\Qah{2}]\begin{bsmallmatrix} \y &  \\ \mbf{0}\end{bsmallmatrix}\!=\!\tfrac{1}{\sqrt{N_0}}\Qah{1}\y$, in addition to generating $\La$. Finally, applying $\Wap$ to puncture $\La$ can be done using Gaussian elimination as before, with the elimination operations simultaneously applied to $\yatilde\!=\!\tfrac{1}{\sqrt{N_0}}\Qah{1}\y$ to generate the product $\yap\!=\!\tfrac{1}{\sqrt{N_0}}\Wap\Qah{1}\y$. Therefore, the $\mathsf{WL}$ algorithm in \algnameabbr~\ref{algo:generalized_wld}, when applied to $\Ha$ and $\ya$, produces the necessary quantities to compute the metrics in~\eqref{eq:awld_metric_tmp}-\eqref{eq:awld_metric_1}, without any matrix inversion, as\vspace{-0.2in}
\begin{align}
  \fnormmsq{\La(\Wmmse\y \!-\! \x)}
  \!&=\!
  \fnormmsq{\tfrac{1}{\sqrt{N_0}}\Qah{1}\y \!-\! \La\x}
  =
  \fnormmsq{\yatilde \!-\! \La\x}
   \label{eq:eliminating_explicit_mmse_dist_1}
   \\[-0.5em]
   \muap{\y|\x}
    \!&=\!
    \tfrac{1}{\Es}\fnormmsq{\x}
    \!-\!
    \fnormmsq{\Wap\La(\Wmmse\y \!-\! \x)}
    \label{eq:eliminating_explicit_mmse_dist_2}
    \\[-0.5em]
    \!&=\!
    \tfrac{1}{\Es}\fnormmsq{\x}
    \!-\!
    \fnormmsq{\Wap(\yatilde \!-\! \La\x)}
    \label{eq:eliminating_explicit_mmse_dist_3}
    \\[-0.5em]
    \!&=\!
    \tfrac{1}{\Es}\fnormmsq{\x}
    \!-\!
    \fnormmsq{\yap \!-\! \Lap{}\x}.
    \label{eq:eliminating_explicit_mmse_dist_4}
\end{align}\\[-2.5em]
Similarly, the $\mathsf{WDL}$ procedure in \algnameabbr~\ref{algo:generalized_wdl} generates these quantities without any square-root operations (assuming $\sqrt{\Es},\sqrt{N_0}$ are available at the input to form $\Ha,\ya$). The output quantities from the algorithm, now labeled as $\Laptilde, \D, \yaptilde, \Wtilde$, are used to compute the above metrics as\vspace{-0.175in}
\begin{align}
    \fnormmsq{\Wap(\tfrac{\Qah{1}\y}{\sqrt{N_0}}\!-\!\La\x)}
    &=
    \fnormmsq{\supsc{\D}{\,1/2}\Wtildeh(\tfrac{\y}{\sqrt{N_0}}\!-\!\H\x)}
    =
    \fnormmsq{\supsc{\D}{\,1/2}(\yaptilde\!-\!\Laptilde\x)}
    \notag
    \\[-0.5em]
    &=
    (\yaptilde\!-\!\Laptilde\x)^{\dag}\D(\yaptilde\!-\!\Laptilde\x),
    \label{eq:dist_metric_awdl}
\end{align}\\[-5.5em]
\begin{align}
    \muap{\y|\x}
    \!&=\!
    \tfrac{1}{\Es}\fnormmsq{\x}
    \!-\!
    \fnormmsq{\Wap(\tfrac{1}{\sqrt{N_0}}\Qah{1}\y\!-\!\La\x)}
    \!=\!
    \tfrac{1}{\Es}\fnormmsq{\x}
    \!-\!
    (\yaptilde\!-\!\Laptilde\x)^{\dag}\D(\yaptilde\!-\!\Laptilde\x).
    \label{eq:muap_metric_awdl}
\end{align}\\[-5em]

%
\subsection{Combined Two-Sided QLZ and WLZ Decompositions}\label{s:combined_2sided_decomp_alg}\vspace{-0.075in}
The reduction and elimination operations for two-sided decompositions of Sec.~\ref{sec:two_sided_puncturing} can be combined efficiently as
shown in \algnameabbr~\ref{algo:wlzdopt}. The code starts with QL decomposition, and then performs right reduction followed immediately by left elimination operations, analogous to~\eqref{eq:Zkj_operation}-\eqref{eq:Wpkj_operation}. The matrix $\mbf{Z}$ and its inverse are updated with every right operation, the matrix $\W$ and output vector $\yp$ are updated with every left operation, while $\Lp$ is updated after each of these operations. For reference, \algnameabbr~\ref{algo:qlzd} shows the code for QLZ decomposition with right reduction but without elimination. Note again that the generated $\W$ is related to $\Wp$ in~\eqref{eq:Z_Wp} as $\Wh\!=\!\Wp\Qh$.

Table~\ref{tab:summary_decomposition_algorithms} in the supplement summarizes all algorithms presented in this section, and highlights their main features.

\vspace{-0.1in}

%
\section{(A)WLD-Based MIMO Detection Algorithms}\vspace{-0.075in}\label{s:AWLD_MIMO_detection_algorithms}
In this section, we present computationally-efficient soft-output MIMO detection algorithms based on the AWLD, WLD, and WLZ puncturing schemes, and compare them with the LORD algorithm~\cite{2006_siti_novel_LORD}. Supplement Table~\ref{tab:summary_detection_algorithms} summarizes all the algorithms discussed and their features in terms of decomposition and puncturing schemes, metric used for LLR computation, marginalization on child layers, single-tree versus multi-tree, as well as local versus global metric update. The pseudo-codes of all algorithms are available in the supplement.

In general, MLM bit LLRs are computed using~\eqref{eq:LLR_exact_MLM_def} as Max-Log approximations of the exact ML LLRs in~\eqref{eq:LLR_exact_ML_def}, with the true metric $\mu(\y|\x)\!=\!\mu(\ytilde|\x)\!=\!-\tfrac{1}{N_0}\fnormmsq{\yt\!-\!\L\x}$ as defined in~\eqref{eq:true_metric}. For an arbitrary $\L$ partitioned into $\nu$ parent layers and $\Nt\!-\!\nu$ child layers, the exact maximizations in~\eqref{eq:LLR_exact_MLM_def} are expressed as in~\eqref{eq:max_mu_parent_exact}-\eqref{eq:max_mu_child_exact}, and approximated using ZF-DF on child symbols using~\eqref{eq:max_mu_parent_approx}-\eqref{eq:max_mu_child_approx}. For a punctured $\L$, alternative metrics $\mup{\ytilde|\x}$ and $\muap{\ytilde|\x}$ to $\mu(\ytilde|\x)$ are derived in~\eqref{eq:mup_WLD_algo} and~\eqref{eq:muap_metric_awdl} under the optimized WLD and AWLD models. When $\mup{\ytilde|\x}$ and $\muap{\ytilde|\x}$ are used in~\eqref{eq:LLR_exact_MLM_def} instead of $\mu(\ytilde|\x)$, and with $\L$ being punctured, then all child symbols become leaves, decision feedback disappears, and the ZF-DF approximations in~\eqref{eq:max_mu_xc_parent_approx}-\eqref{eq:max_mu_xc_child_approx} turn into exact LS estimates as required to satisfy~\eqref{eq:max_mu_parent_exact}-\eqref{eq:max_mu_child_exact}.
\vspace{-0.175in}

%
\subsection{AWDL MIMO Detection Algorithm}\vspace{-0.1in}
An augmented channel is first formed as in~\eqref{eq:Ha_def}, and then punctured using \algnameabbr~\ref{algo:generalized_wdl} for a given $\nu$, to yield $\Laptilde,\yaptilde,\D$, with the following structure: $\Laptilde \!=\! \begin{bsmallmatrix} \smash[t]{\Ptilde}\vphantom{\P} & \mbf{0}  \\ \Rtilde & \I \end{bsmallmatrix}$, $\yaptilde\!=\![\ytildeparent;~\ytildechild]$, and $\D\!=\! \begingroup 
\setlength\arraycolsep{0pt}\begin{bsmallmatrix} \Dparent & \mbf{0}  \\ \mbf{0} & \Dchild \end{bsmallmatrix}\endgroup$. With $\x\!=\![\xparent;~\xchild]$, the bit LLRs are computed as\vspace{-0.15in}
\begin{align*}
    \Lambda_{\mathrm{ap}}(x_{n,b}|\y)
    \!&=\!
    \max_{\x:x_{n,b}=+1}\muap{\y|\x} \!-\!    \max_{\x:x_{n,b}=-1}\muap{\y|\x},
    \\[-0.25em]
    \muap{\y|\x}
    \!&=\!
    \tfrac{1}{\Es}\fnormmsq{\x} \!-\!(\yaptilde\!-\!\Laptilde\x)^{\dag}\D(\yaptilde\!-\!\Laptilde\x)\hfill
    \triangleq
    \muparent{\ytildeparent|\xparent}
    +
    \muchild{\ytildechild|\xparent,\xchild},
\end{align*}\\[-2.5em]
where $\muparent{\ytildeparent|\xparent}\!=\!\tfrac{1}{\Es}\fnormmsq{\xparent}\!-\!
\subpsc{\fnormm{\ytildeparent \!-\! \Ptilde \xparent}}{\Dparent}{\,2}$ and
$\muchild{\ytildechild|\xparent,\xchild}\!=\!\tfrac{1}{\Es}\fnormmsq{\xchild}\!-\!
\subpsc{\fnormm{\ytildechild \!-\! \Rtilde \xparent \!-\! \xchild}}{\Dchild}{\,2}$.
Next, for any $\xparent$, the leaf symbols $\xchild$ that maximize $\subsc{\mu}{2}$ are obtained through LS by setting the derivative of $\subsc{\mu}{2}$ with respect $\xchild$ to 0. We obtain $\tfrac{1}{\Es}\xchildh \!+\!\supsc{(\ytildechild \!-\! \Rtilde \xparent \!-\! \xchild)}{\,\dag}\Dchild\!=\!0$, from which it follows that \vspace{-0.2in}
\begin{align}
    \max_{\x:x_{n,b}=s}\!\muap{\yt|\x}
    &\!=\!\!
    \max_{\xparent:x_{n,b}=s}
        \!\bigl\{
            \muparent{\ytildeparent|\xparent} \!+\!
            \max_{\xchild} \muchild{\ytildechild|\xparent,\xchild}
        \bigr\}\notag
    \\[-0.5em]
    &\!=\!\!
    \max_{\xparent:x_{n,b}=s}
        \!\!\left\{
            \muparent{\ytildeparent|\xparent} \!+\! \muhatchild{\ytildechild|\xparent,\xhatchild}
        \right\},
    \label{eq:AWLD_LLR_equations_parent}
    \\[-0.5em]
    \xhatchild
    &\!=\!
    \bigl\lfloor
        \supsc{(\Dchild \!-\!\tfrac{1}{\Es}\I_{\Nt-\nu})}{\,-1}
        \Dchild
        (\ytildechild \!-\! \Rtilde \xparent)
    \bigr\rceil_{\!\supsc{\mathcal{X}}{\Nt-\nu}},
    \notag\\[-0.5em]
    \max_{\x:x_{n,b}=s}\!\muap{\yt|\x}
    &\!=\!
    \max_{\xparent}
        \bigl\{
            \muparent{\ytildeparent|\xparent} \!+\!
            \max_{\xchild:x_{n,b}=s} \muchild{\ytildechild|\xparent,\xchild}
        \bigr\}
    \notag
    \\[-0.5em]
    &\!=\!
    \max_{\xparent}
        \bigl\{
            \muparent{\ytildeparent|\xparent} \!+\! \muhatchild{\ytildechild|\xparent,\subpsc{\hat{\mbf{x}}}{2\,n;b}{\,(s)}}
        \bigr\},
    \label{eq:AWLD_LLR_equations_child}
    \\[-0.5em]
    ~\subpsc{\hat{\mbf{x}}}{2\,n;b}{\,(s)}
    & \!\triangleq\!
    \bigl\lfloor
        \supsc{(\Dchild \!-\!\tfrac{1}{\Es}\I_{\Nt-\nu})}{\,-1}
        \Dchild
        (\ytildechild \!-\! \Rtilde \xparent)
    \bigr\rceil_{\!\subpsc{\mathcal{A}}{n,b}{\,(s)}},\notag
\end{align}\\[-2.5em]
for $s\!=\!\pm 1$, where $\subpsc{\mathcal{A}}{n,b}{\,(s)} \!=\! \{ \supsc{[x_{\nu+1},\cdots,x_n,\cdots,x_{\Nt}]}{\,\tran}\!\in\!\mathcal{X}^{\Nt-\nu}:x_{n,b}\!=\!s \}$.

The pseudo-code of the multi-tree version of the AWDL algorithm is shown in \algnameabbr~\ref{algo:awdl_detector_algorithm}. It performs multiple runs, each time grouping a new set of $\nu$ layers as parents to generate bit LLRs using~\eqref{eq:AWLD_LLR_equations_parent} only. The multi-tree WLD algorithm that implements~\eqref{eq:AWLD_LLR_equations_parent} but using the metric $\mupnoarg$ in~\eqref{eq:mup_WLD_algo} is shown in \algnameabbr~\ref{algo:wld_Lv_mul_loc_detector_algorithm}. For reference, the pseudo-code of the multi-tree LORD algorithm that implements~\eqref{eq:max_mu_parent_approx} with the true metric~\eqref{eq:true_metric} is shown in \algnameabbr~\ref{algo:lord_Lv_mul_loc_detector_algorithm}. Because its channel $\L$ is full lower-triangular, LORD applies ZF-DF rather than LS to estimate the child symbols, resulting in a significant increase in computational complexity compared to WLD/AWDL. Note that for all three algorithms, the multiple runs are independent and the metrics computed are used to update just the tracked maxima of the parent layer bits only, and are not globally shared to update the maxima for other bits.

The search space of $\abs{\X}^{\nu}$ parent symbol vectors of the AWDL algorithm can be reduced to $\nu\cdot\abs{\X}$ by enumerating only over the root and applying ZF-DF on the other $\nu\!-\!1$ parents. Using Lemma~\ref{lem:wld_perm_distances}, for a given choice of $\nu$ layers as parents and $\Nt\!-\!\nu$ layers as leaves, the metric of a given symbol vector does not change if the parent layers are permuted and the leaf layers are permuted. Hence, doing $\nu$ runs over the parent layers, each time with a different layer as root, would improve the estimates by updating the maxima being tracked for the bits on \emph{all} $\nu$ parents in each run, and not just those bits of the current root symbol. For the case $\nu\!=\!2$, the search on layer 2 can be limited to a small window of $\eta$ symbols around the ZF solution. Empirical simulations demonstrate that $\eta\!=\!4$ is sufficient to achieve the accuracy of enumerating all $\supsc{\abs{\X}}{\,2}$ parent symbol vectors. The pseudo-code of the windowed-AWDL algorithm is shown in \algnameabbr~\ref{algo:awdl_box_detector_algorithm}.

Finally, the LORD algorithm can be similarly optimized to update the tracked maxima for each bit hypothesis on \emph{all} layers in each run
as shown on \algnameabbr~\ref{algo:lord_Lv_mul_glob_detector_algorithm}. This is possible in this case because Euclidean distances do not change under column permutation of $\H$: $\fnormm{\y\!-\!\H\x}\!=\!\fnormm{\y\!-\!\H\J\J^{\tran}\x}\!=\!\fnormm{\yt\!-\!\L\J^{\tran}\x}$ for any permutation $\J$, where $\H\J\!=\!\Q\L$ and $\yt\!=\!\Qh\y$.
\vspace{-0.225in}

%
\subsection{WLZ-Based MIMO Detection Algorithm}\vspace{-0.075in}
While the metrics of the WLD and AWDL algorithms are not preserved under arbitrary layer permutation as they are for LORD, they are \emph{approximately} preserved under 2-sided WLZ decomposition (Section~\ref{sec:two_sided_puncturing},~\algnameabbr~\ref{algo:wlzdopt}). The pseudo-code of the multi-tree version of the WLZ algorithm shown in \algnameabbr~\ref{algo:wlzd_Lv_mul_glob_detector_algorithm} implements~\eqref{eq:AWLD_LLR_equations_parent} similar to the WLD detection algorithm of \algnameabbr~\ref{algo:wld_Lv_mul_loc_detector_algorithm}, but with $\mupnoarg$ in~\eqref{eq:mup_WLD_algo} being based on the 2-sided WLZ rather than the 1-sided WL decomposition.

%
\vspace{-0.225in}
\section{Simulation Results}
\begin{figure*}[!t]
\centering
\subfloat[Gaussian inputs]
{\includegraphics[scale=0.6]{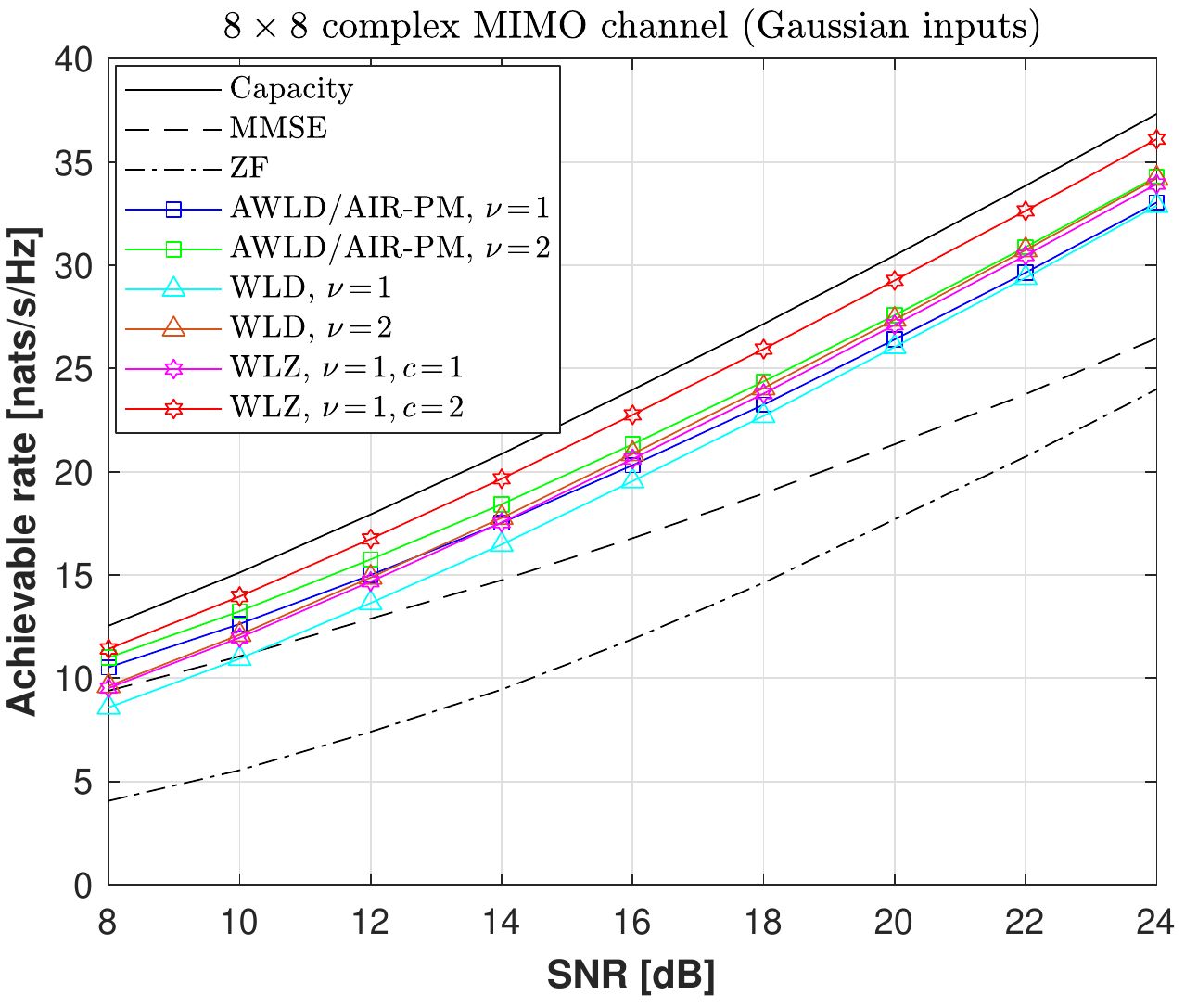}
\label{fig:achievable_rates_8x8}}
\hfil
\subfloat[Finite QAM constellations]{\includegraphics[scale=0.6]{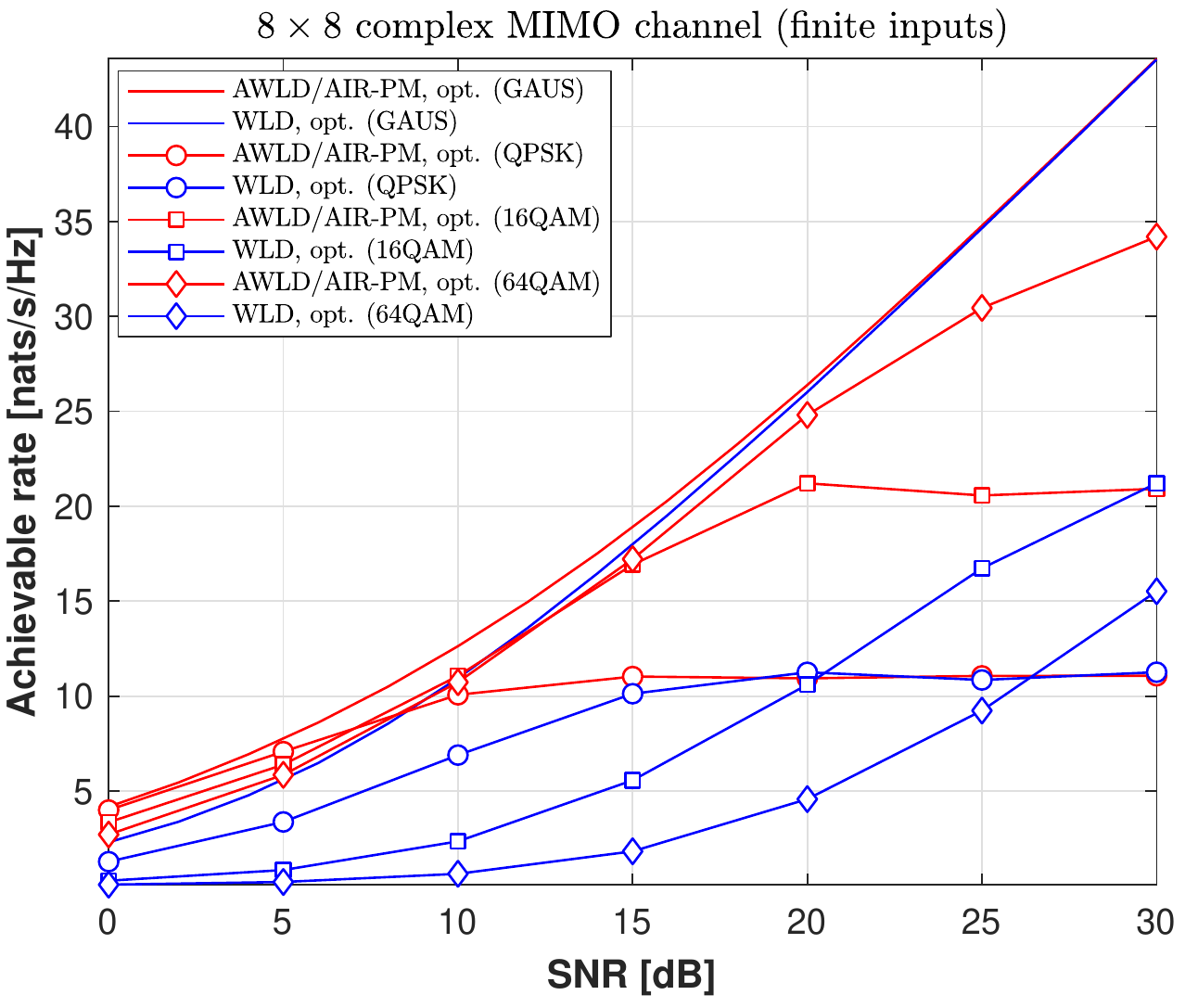}
\label{fig:finite_inputs_rates_8x8}}
\vspace{-0.075in}
\caption{Comparison of AIRs for $8\!\times\! 8$ MIMO channels with (a) Gaussian inputs, and (b) finite QAM constellations.}
\label{fig:achievable_rates}
\end{figure*}

\vspace{-0.075in}\label{s:sim}
A library of MIMO detection algorithms have been implemented and characterized for both algorithmic performance and computational complexity. Fast-fading Rayleigh complex MIMO channels are assumed. In Fig.~\ref{fig:achievable_rates_8x8}, we compare the achievable rates of the proposed WLZ and AWLD detectors against the AIR-PM detector~\cite{2017_hu_softoutput_AIR}, as well as the ZF, MMSE, and WLD~\cite{2014_mansour_SPL_WLD} for $8\!\times\!8$ MIMO channels, assuming Gaussian inputs and with parent layers selected so as to maximize $\ILBWLD$ in~\eqref{eq:ILB_WLD_expression}. The AWLD and WLD are simulated for both $\nu\!=\!1$ and $\nu\!=\!2$ configurations, while WLZ is simulated for $\nu$ and $c\!=\!1,2$. WLZ attains the closest rate to capacity with reduction parameter $c\!=\!2$, followed by AWLD/AIR-PM (which attain the same rate), followed by WLD. This is because as $\rho\!=\!1/2^{c+1/2}$ decreases by increasing $c$, then from Lemma~\ref{lem:lemeire_bound} and~\eqref{eq:Sinv_gap_to_identity}, $\Wp$ gets closer to $\I$ and $\Lp$ approaches the true unpunctured $\L$. Hence from Theorem~\ref{thm:ILB_wld}, the lower bound on the achievable rate $\ILBWLD$ approaches the capacity of the channel.

On the other hand, Fig.~\ref{fig:finite_inputs_rates_8x8} plots the AIR of AWLD and WLD with $\nu\!=\!1$ for finite constellations. The AWLD achieves higher rates than WLD, especially for 64QAM. Parent layers are selected to maximize
$\ILBWLD$ in~\eqref{eq:ILB_WLD_expression} if Gaussian inputs were used. For the AWLD scheme at very low SNR regimes, it attains higher rates for low-order constellations compared to denser constellations. For SNRs beyond $\sim\unit[10]{dB}$, the trend gets reversed, with the AWLD scheme attaining higher rates for denser constellations. Similarly for the WLD scheme. However, the SNRs for which denser constellations start to outperform low-order constellations are much higher than those of the AWLD case; the rate for which 16QAM becomes better than that for QPSK is roughly $\unit[20]{dB}$ for the WLD case, while for the AWLD case, it is roughly $\unit[6]{dB}$. The same applies between 64QAM and 16QAM, but at an impractically very high SNR value (range not shown in the figure). The reason is that the WLD scheme is not optimal, and misses the ML decision for denser constellations at low SNRs more often compared to the AWLD scheme.

\begin{figure*}[!t]
\centering
\subfloat[$4\!\times\!4$, 256QAM, $\nu\!=\!1$]
{
\includegraphics[scale=0.6]{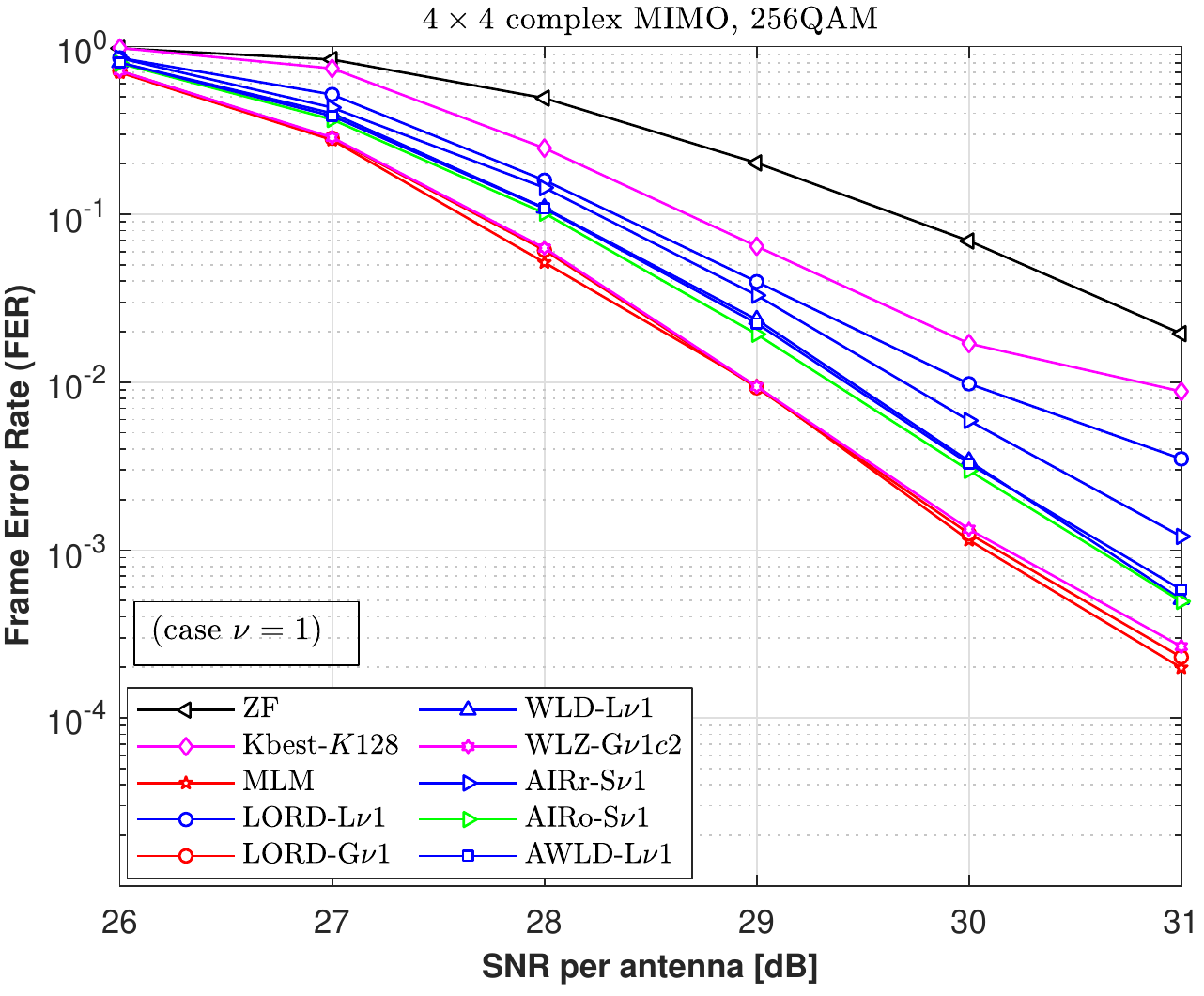}
\label{fig:fer_plot_4x4_256QAM_L1}
}
\hfil
\subfloat[$4\!\times\!4$, 256QAM, $\nu\!=\!2$]
{
\includegraphics[scale=0.6]{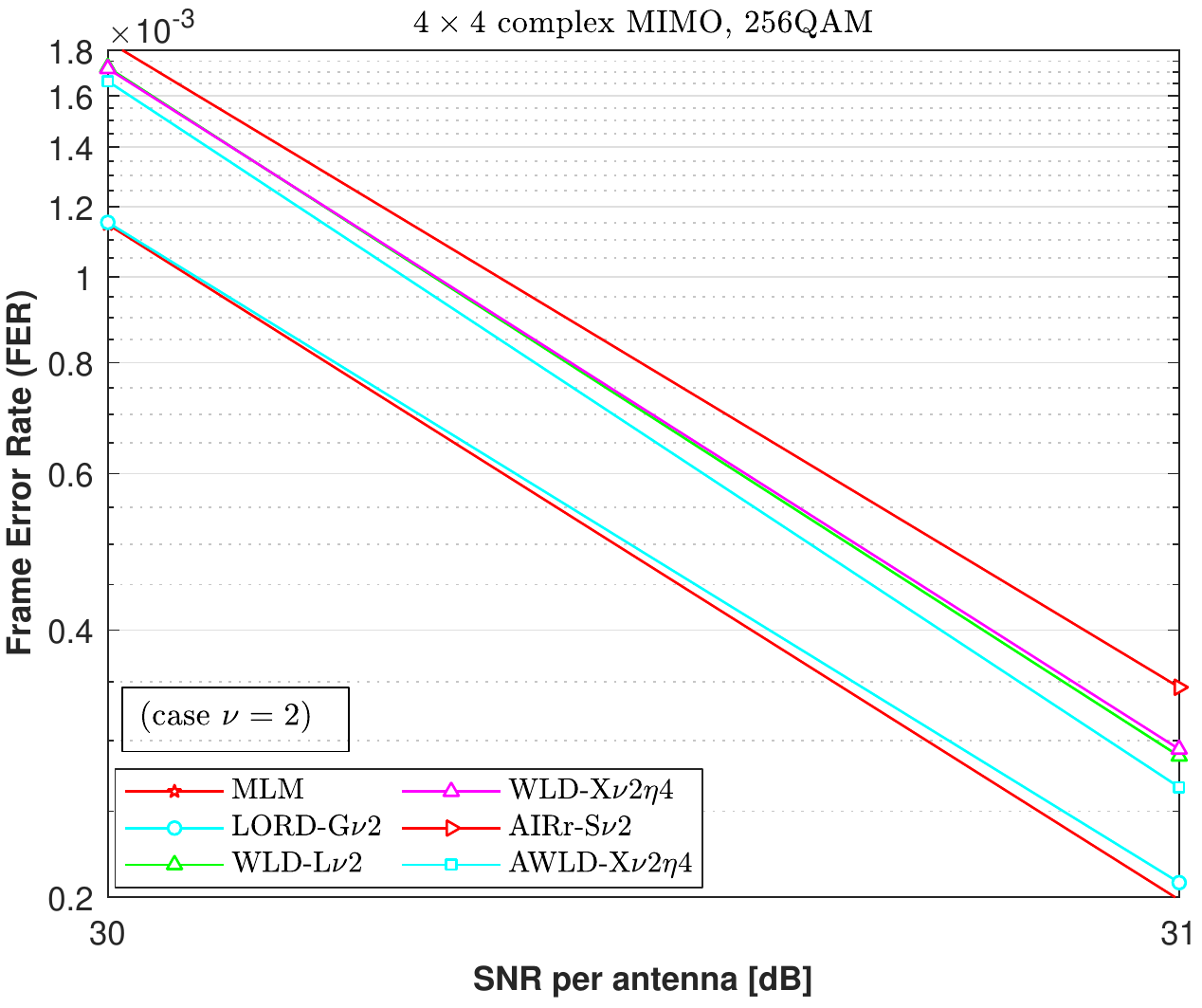}
\label{fig:fer_plot_4x4_256QAM_L2}
}
\\
\subfloat[$8\!\times\!8$, 64QAM, $\nu\!=\!1$]
{
\includegraphics[scale=0.6]{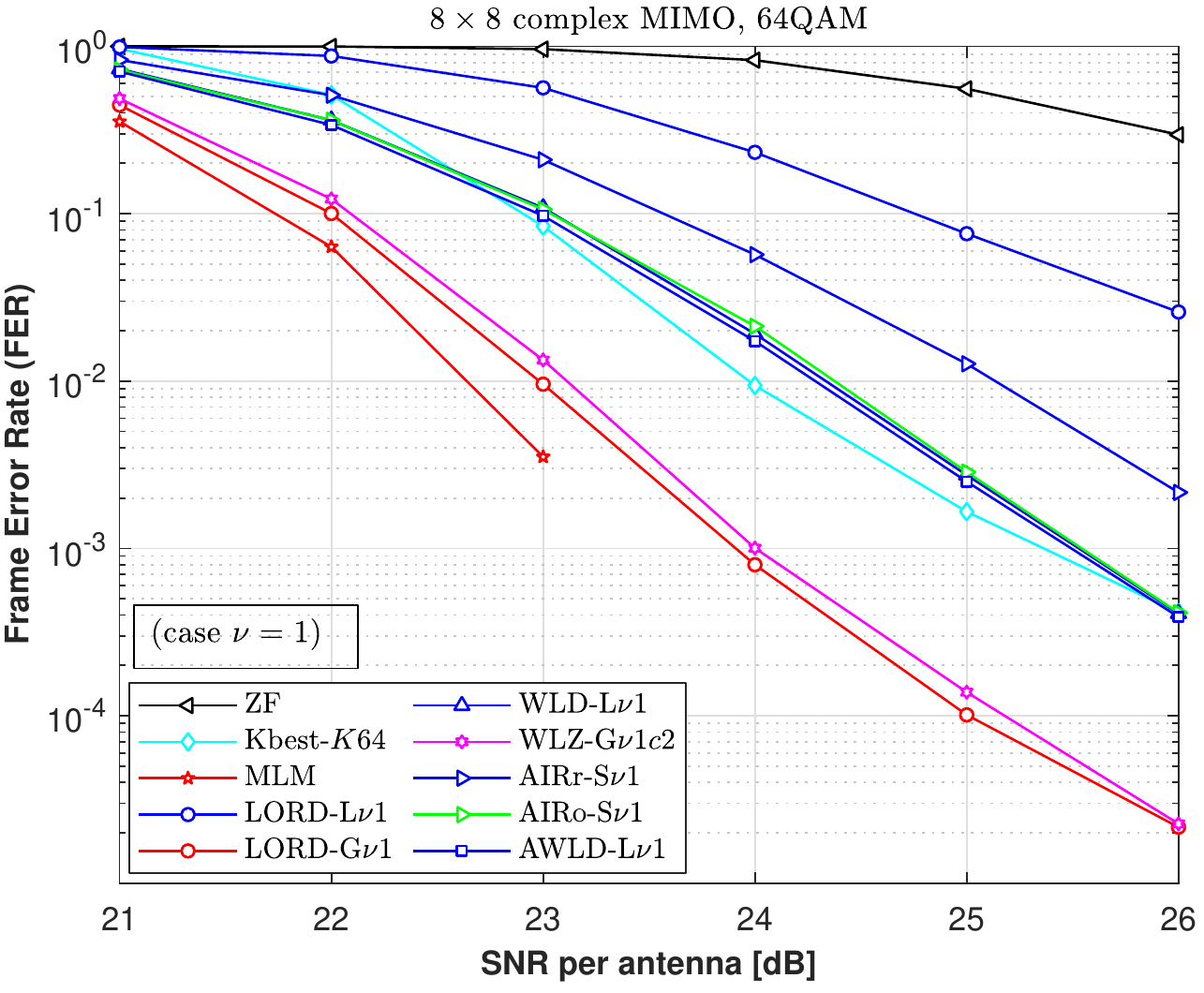}
\label{fig:fer_plot_8x8_64QAM_L1}
}
\hfil
\subfloat[$8\!\times\!8$, 64QAM, $\nu\!=\!2$]
{
\includegraphics[scale=0.6]{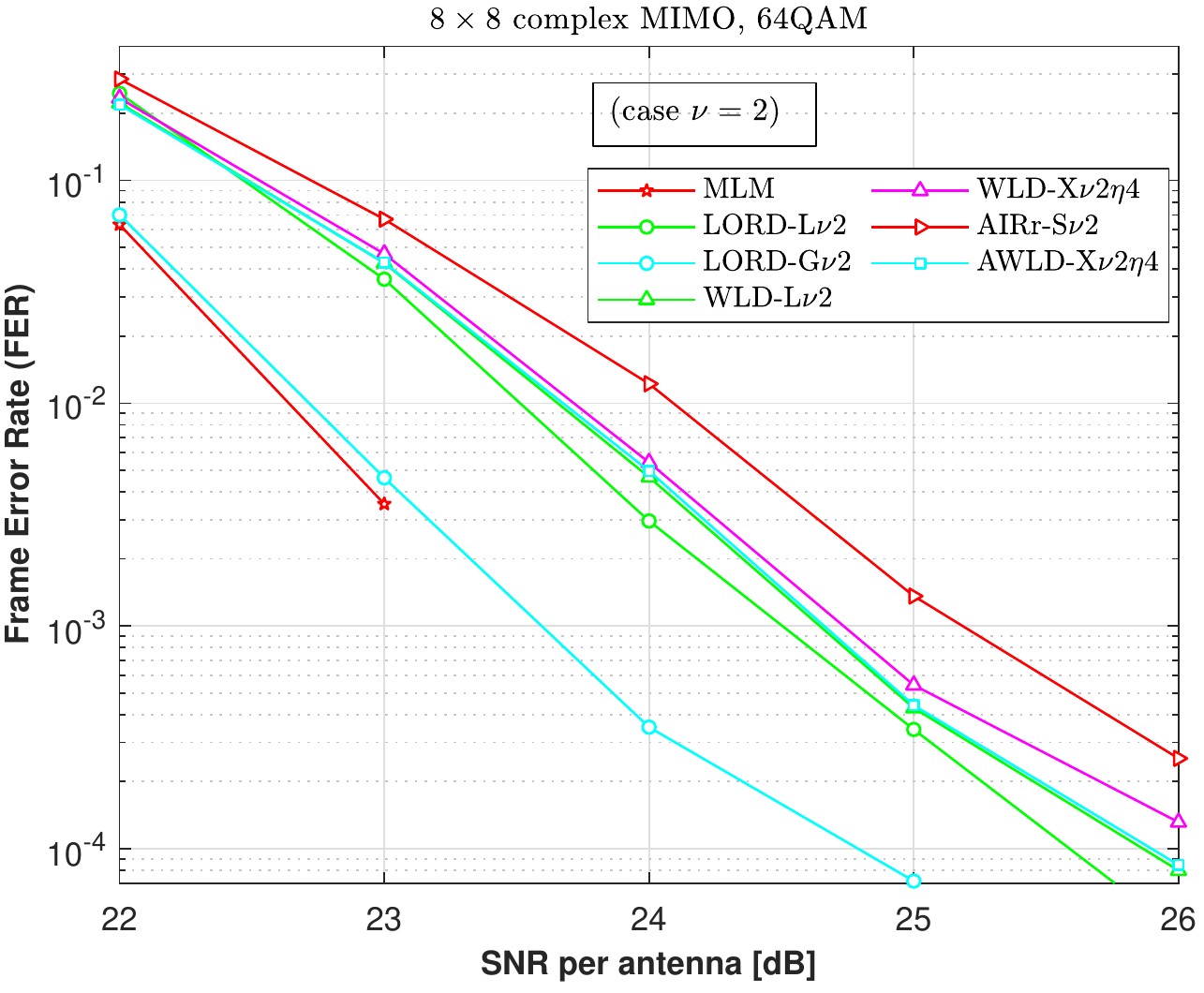}
\label{fig:fer_plot_8x8_64QAM_L2}
}
\\
\subfloat[$12\!\times\!12$, 64QAM, $\nu\!=\!1$]
{
\includegraphics[scale=0.6]{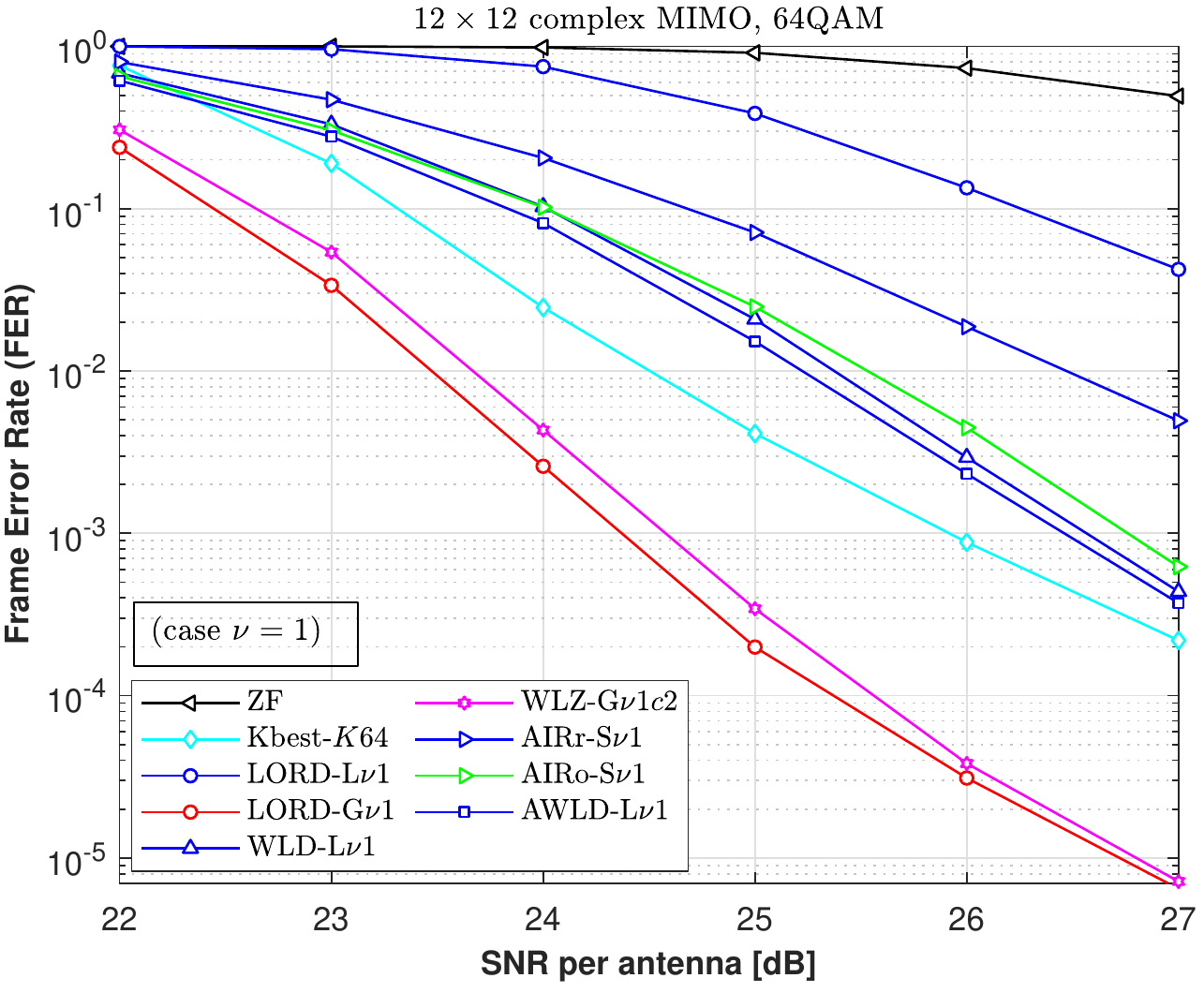}
\label{fig:fer_plot_12x12_64QAM_L1}
}
\hfil
\subfloat[$12\!\times\!12$, 64QAM, $\nu\!=\!2$]
{
\includegraphics[scale=0.6]{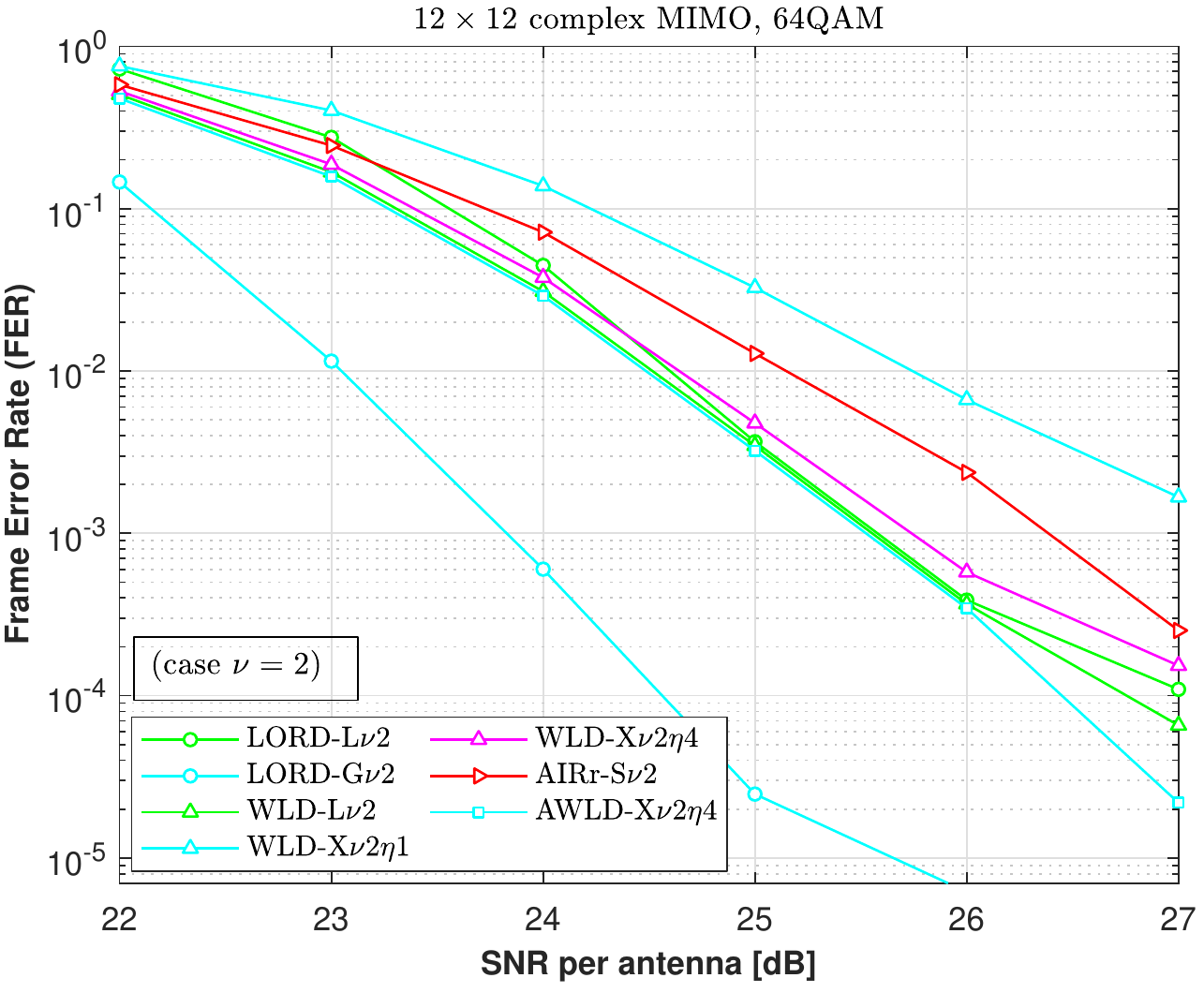}
\label{fig:fer_plot_12x12_64QAM_L2}
}
\caption{Comparisons of FERs vs. SNR for various QAM constellations, puncturing orders, and MIMO dimensions $4,8,\text{and},12$.}
\label{fig:fer_L1}
\end{figure*}

\begin{figure*}[!t]
\centering
\subfloat[$16\!\times\!16$, 16QAM, $\nu\!=\!1$]
{
\includegraphics[scale=0.6]{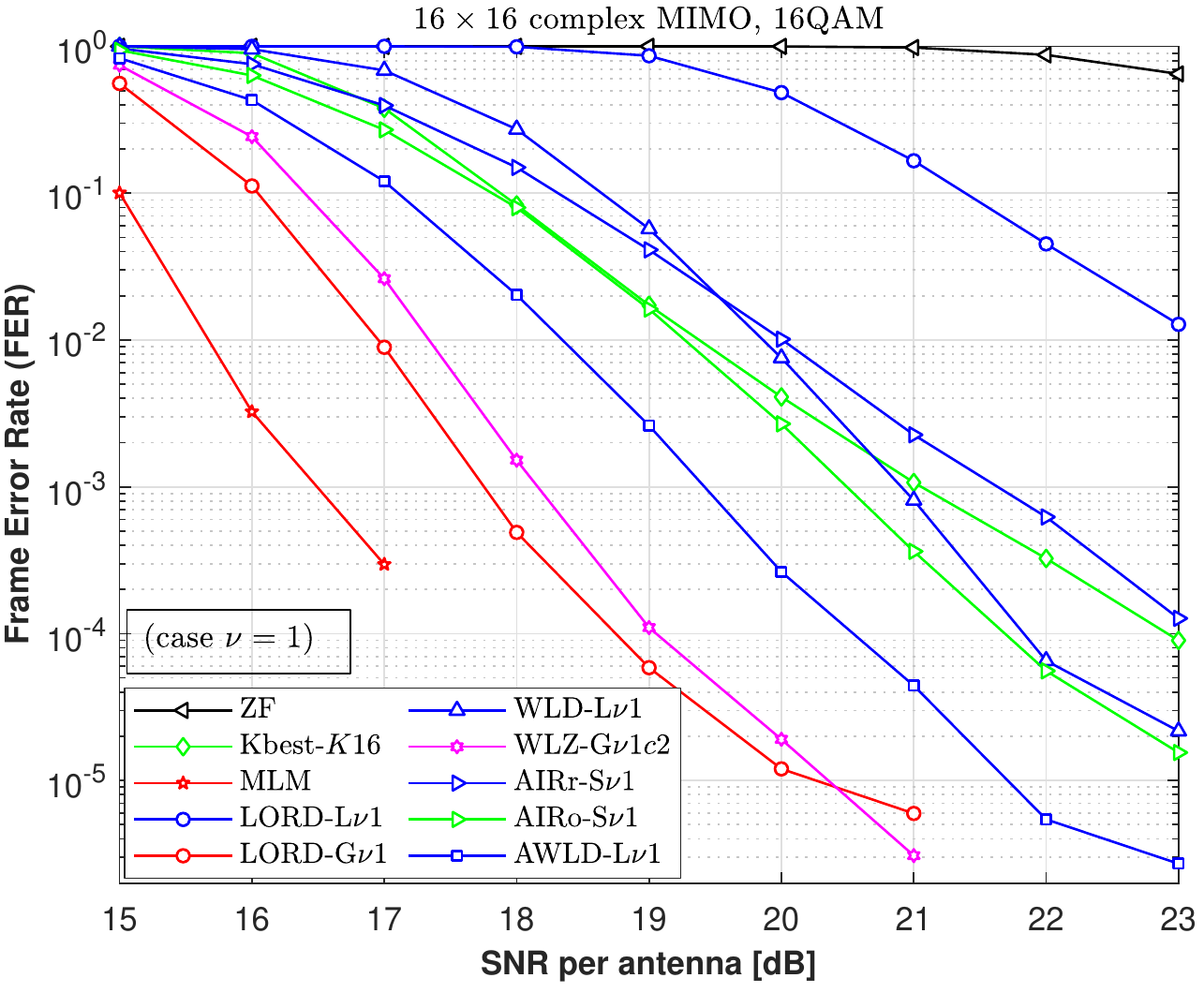}
\label{fig:fer_plot_16x16_16QAM_L1}
}
\hfil
\subfloat[$16\!\times\!16$, 16QAM, $\nu\!=\!2$]
{
\includegraphics[scale=0.6]{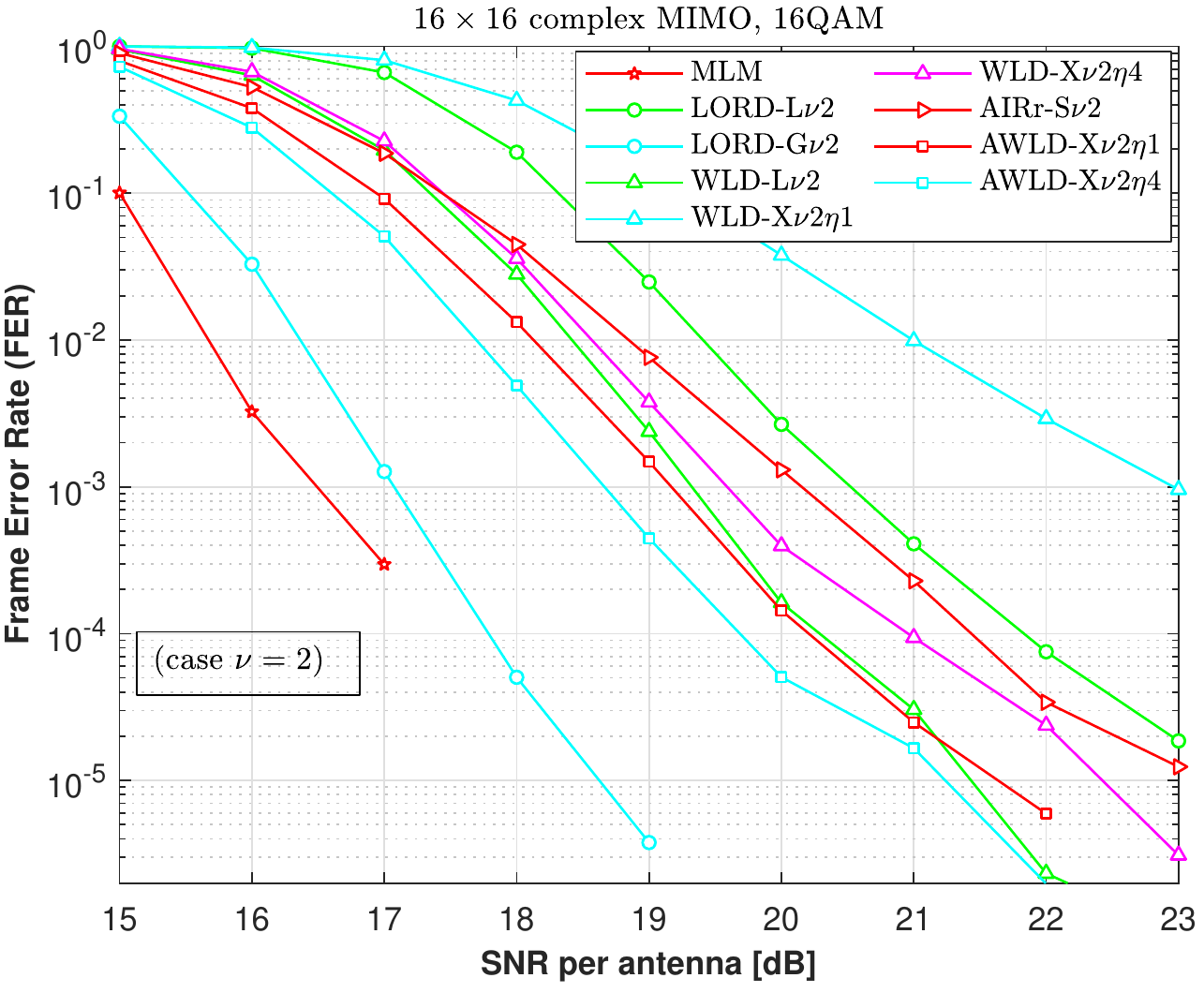}
\label{fig:fer_plot_16x16_16QAM_L2}
}
\\
\subfloat[$16\!\times\!16$, 64QAM, $\nu\!=\!1$]
{
\includegraphics[scale=0.6]{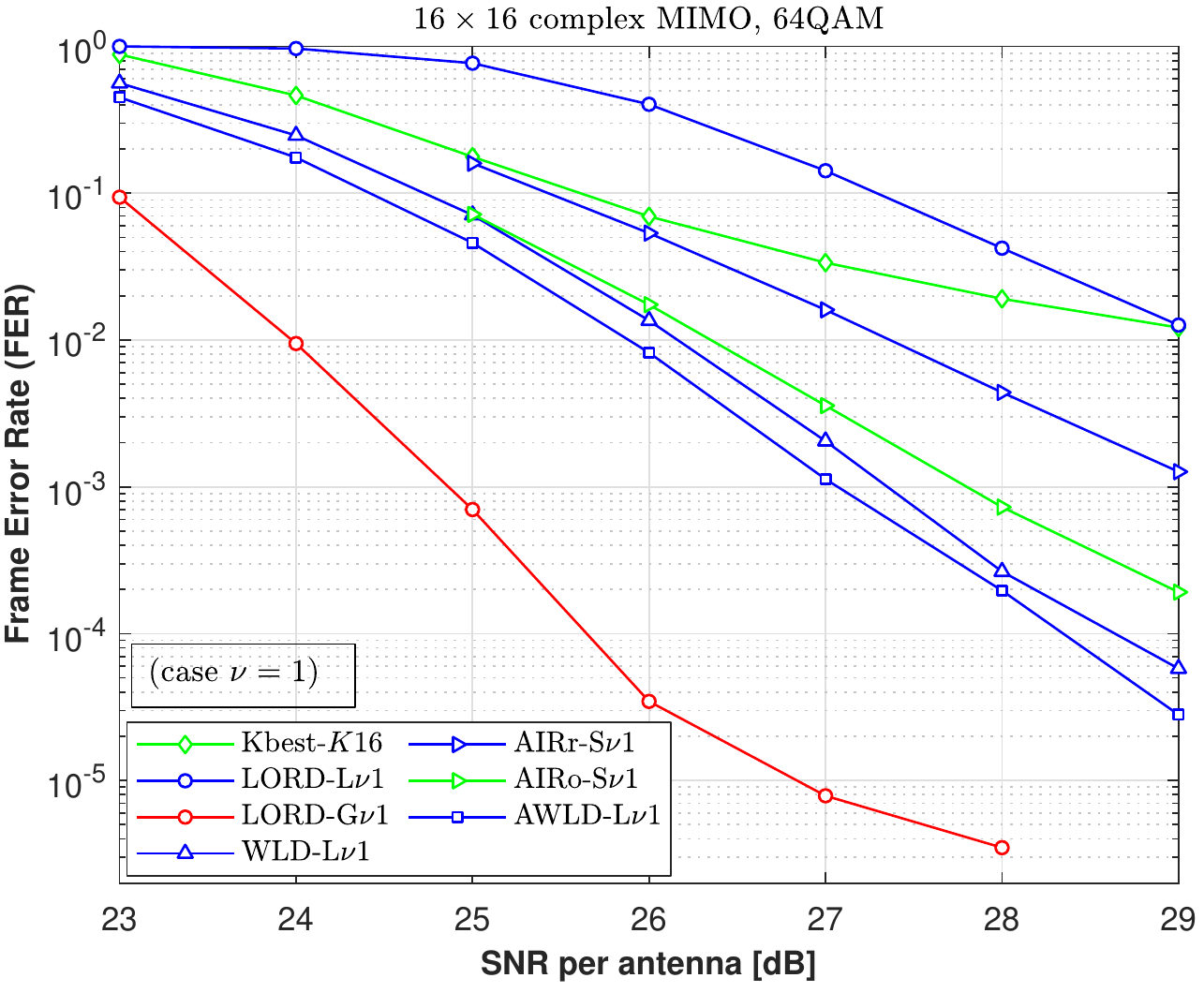}
\label{fig:fer_plot_16x16_64QAM_L1}
}
\hfil
\subfloat[$16\!\times\!16$, 64QAM, $\nu\!=\!2$]
{
\includegraphics[scale=0.6]{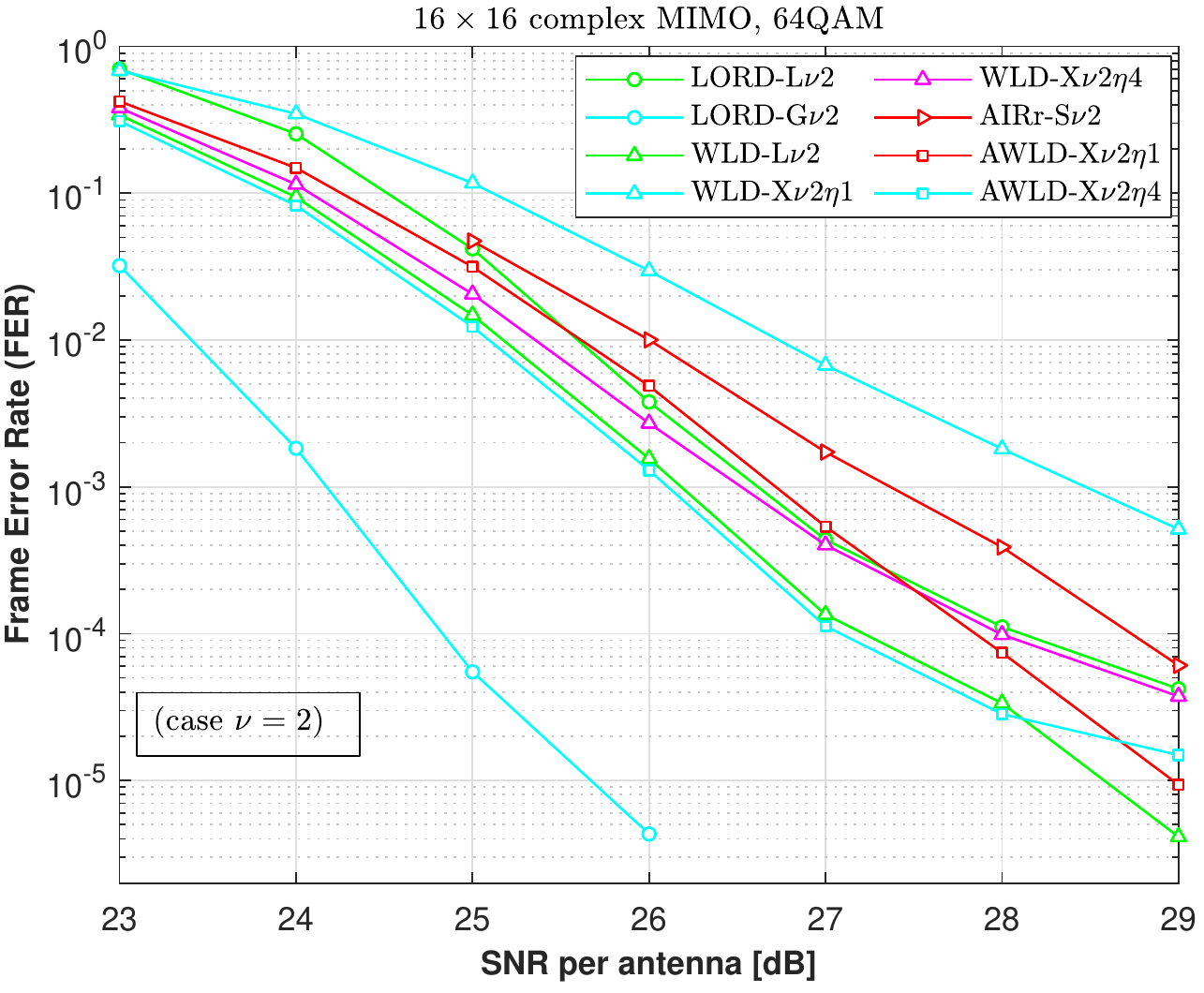}
\label{fig:fer_plot_16x16_64QAM_L2}
}
\\
\subfloat[$32\!\times\!32$, 16QAM, $\nu\!=\!1$]
{
\includegraphics[scale=0.6]{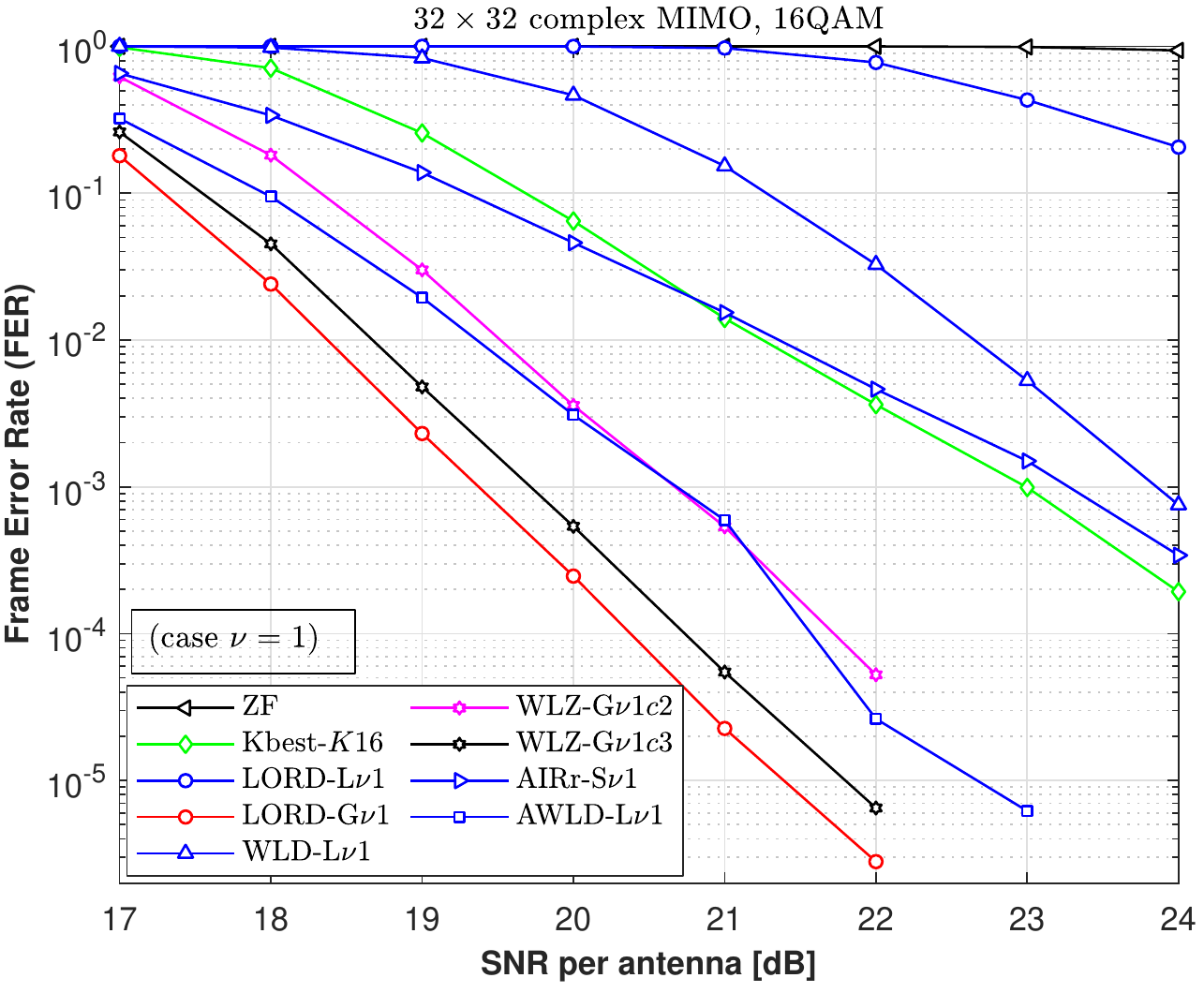}
\label{fig:fer_plot_32x32_16QAM_L1}
}
\hfil
\subfloat[$32\!\times\!32$, 16QAM, $\nu\!=\!2$]
{
\includegraphics[scale=0.6]{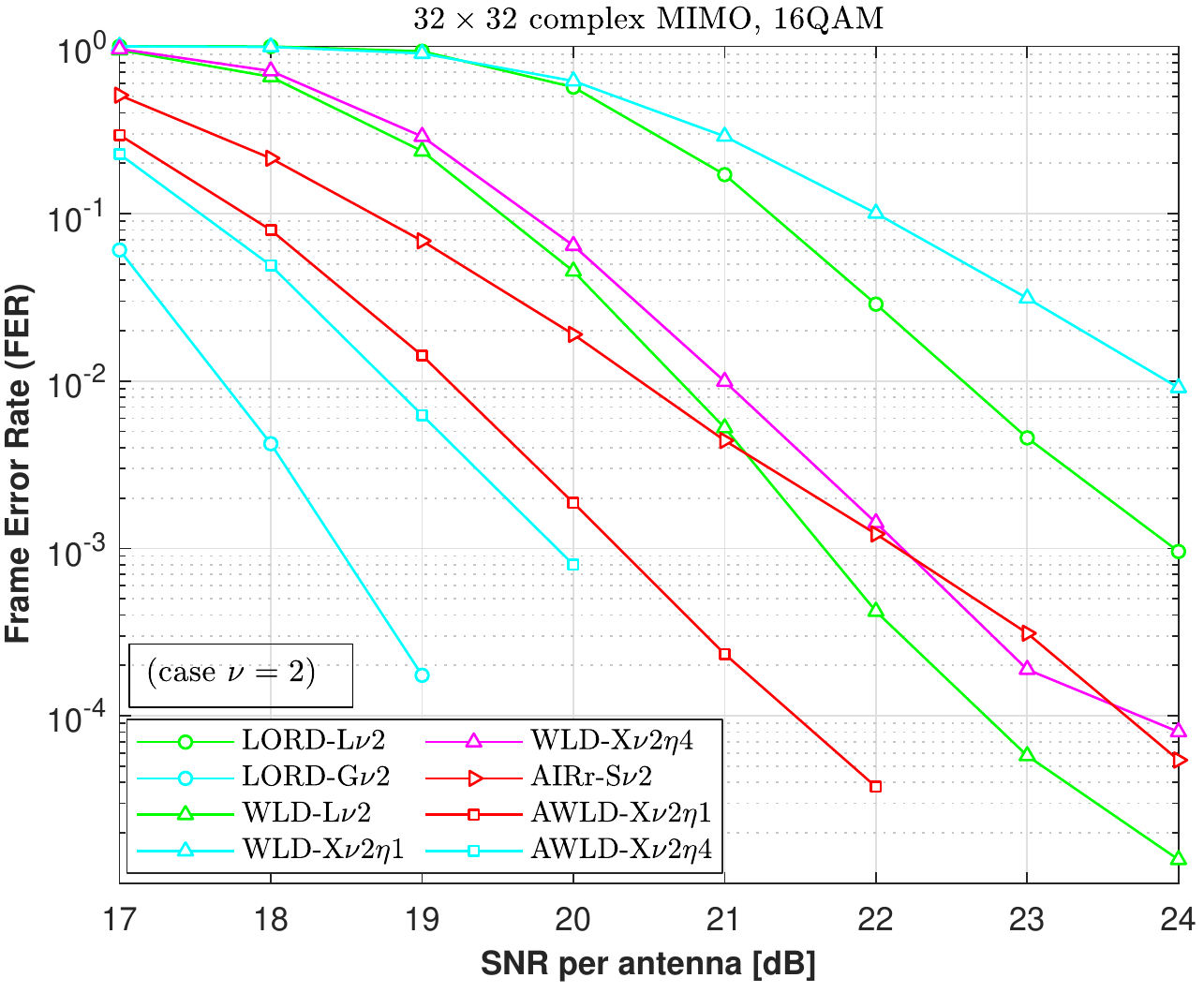}
\label{fig:fer_plot_32x32_16QAM_L2}
}
\caption{Comparisons of FERs vs. SNR for various QAM constellations, puncturing orders, and MIMO dimensions $16$ and $32$}
\label{fig:fer_L2}
\end{figure*}

In Figs.~\ref{fig:fer_L1}-\ref{fig:fer_L2}, we compare the frame error rate (FER) of the proposed WLZ and AWLD detectors against the Max-Log ML (MLM) sphere decoder with optimized pruning~\cite{2014_sphereP2_mansour}, ZF, K-best~\cite{2006_Wenk_ISCAS}, LORD~\cite{2006_siti_novel_LORD}, WLD~\cite{2014_mansour_SPL_WLD}, and AIR-PM~\cite{2017_hu_softoutput_AIR} detectors for various MIMO dimensions, QAM constellations, and puncturing orders. Max-log approximations for exponential sums are used. An LTE rate-1/2 punctured turbo code of length 1024 is used, and 8 turbo decoder iterations are performed. For K-best, sorted-QRD~\cite{Wubben_2001} is used, and the $K$ best competing paths are retained. Counter hypotheses are formed relative to the best survivor path. Counter hypotheses of all leaf bits are updated using the optimization in~\cite{2014_sphereP2_mansour}. Un-updated LLR values are replaced with the minimum LLR in the corresponding symbol.

For LORD, both $\nu\!=\!1,2$ are simulated using the multi-tree approach; $\Nt/\nu$ rounds of $\nu$-layer parent selections, QLDs, and ZF-DF steps on the $\Nt\!-\!\nu$ child layers are performed. LORD-L$\nu1$ and LORD-L$\nu2$ perform local (within-tree) metric updates only (\algnameabbr~\ref{algo:lord_Lv_mul_loc_detector_algorithm}), while LORD-G$\nu1$ and LORD-G$\nu2$ perform global (across all trees) metric updates (\algnameabbr~\ref{algo:lord_Lv_mul_glob_detector_algorithm}). For $\nu\!=\!2$, consecutive layer pairing is done.

For WLD, WLD-L$\nu1$ and WLD-L$\nu2$ perform local metric updates only (\algnameabbr~\ref{algo:wld_Lv_mul_loc_detector_algorithm}). WLD-X$\nu2\eta1$ enumerates on parent 1, does ZF on parent 2, and ZF-DF on child nodes. WLD-X$\nu2\eta4$ enumerates on parent 1, then enumerates over a window of 4 symbols around the ZF solution (ZF-W) on parent 2, and does ZF-DF on child nodes. Both WLD-X algorithms update metrics across tree pairs. Similarly for AWLD; AWLD-L$\nu1$ and AWLD-L$\nu2$ apply \algnameabbr~\ref{algo:awdl_detector_algorithm} using augmented channel puncturing with local metric updates, while AWLD-X$\nu1\eta1$ and AWLD-X$\nu2\eta4$ are similar to their WLD counter parts but apply augmented puncturing (\algnameabbr~\ref{algo:awdl_box_detector_algorithm}).

For AIR-PM, the single-tree approach is used. AIRr-S$\nu1$ randomly selects a parent and orders the other child layers, while AIRo-S$\nu1$ does optimal layer ordering to maximize the AIR assuming Gaussian inputs. AIRr-S$\nu2$ uses two parents with random layer ordering.

For WLZ, 2-sided puncturing and reduction are applied using \algnameabbr~\ref{algo:wlzd_Lv_mul_glob_detector_algorithm}. WLZ-L$\nu1c2$ does local metric updates with one parent and $c\!=\!2$. Similarly, WLZ-G$\nu1c2$ and WLZ-G$\nu1c3$ perform global metric updates with $c\!=\!2,3$, respectively.

Several observations can be made: 1) Multi-tree approaches are superior to single-tree approaches, and are less sensitive to layer ordering. 2) Global metric updates across trees significantly improves performance compared to local within-tree only updates. 3) For trees with more than one parent, there is no need to enumerate across all $\abs{\mathcal{X}}^{\nu}$ parent combinations. Running $\nu$ trees instead, each time enumerating on one parent and doing ZF-W only on parent 2 is as good. 4) Augmented-WLD based algorithms consistently perform better than their WLD counter parts. 5) Two-sided WLZ based algorithms perform better than AWLD and WLD, and almost match the performance of LORD with global metric updates (LORD has dense $\L$, while WLZ has punctured $\L$). 6) Puncturing remains very effective even for large MIMO dimensions.

\begin{figure}[t]
  \centering
  \includegraphics[scale=0.65]{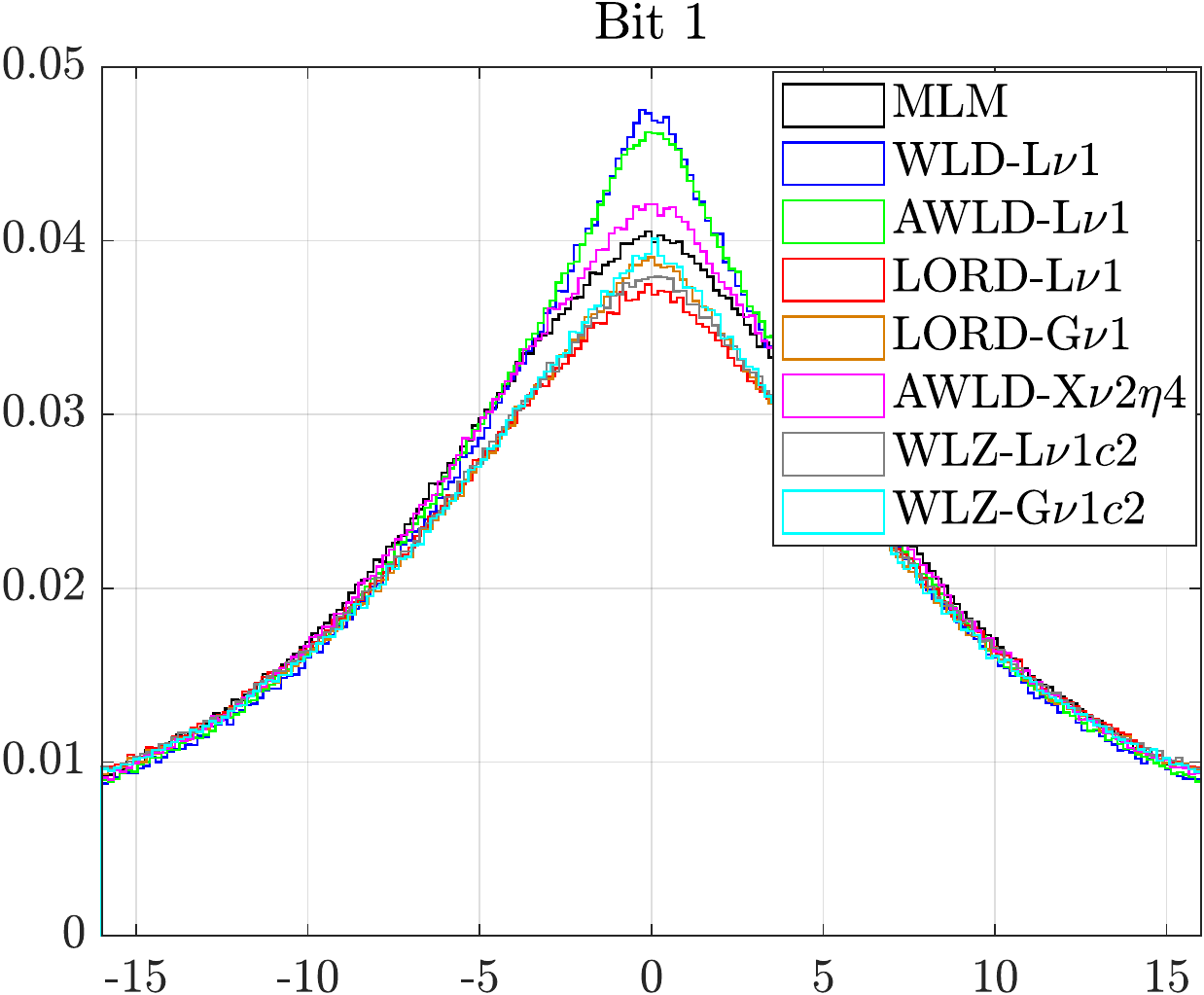}\hspace{-0.00in}
  \includegraphics[scale=0.65]{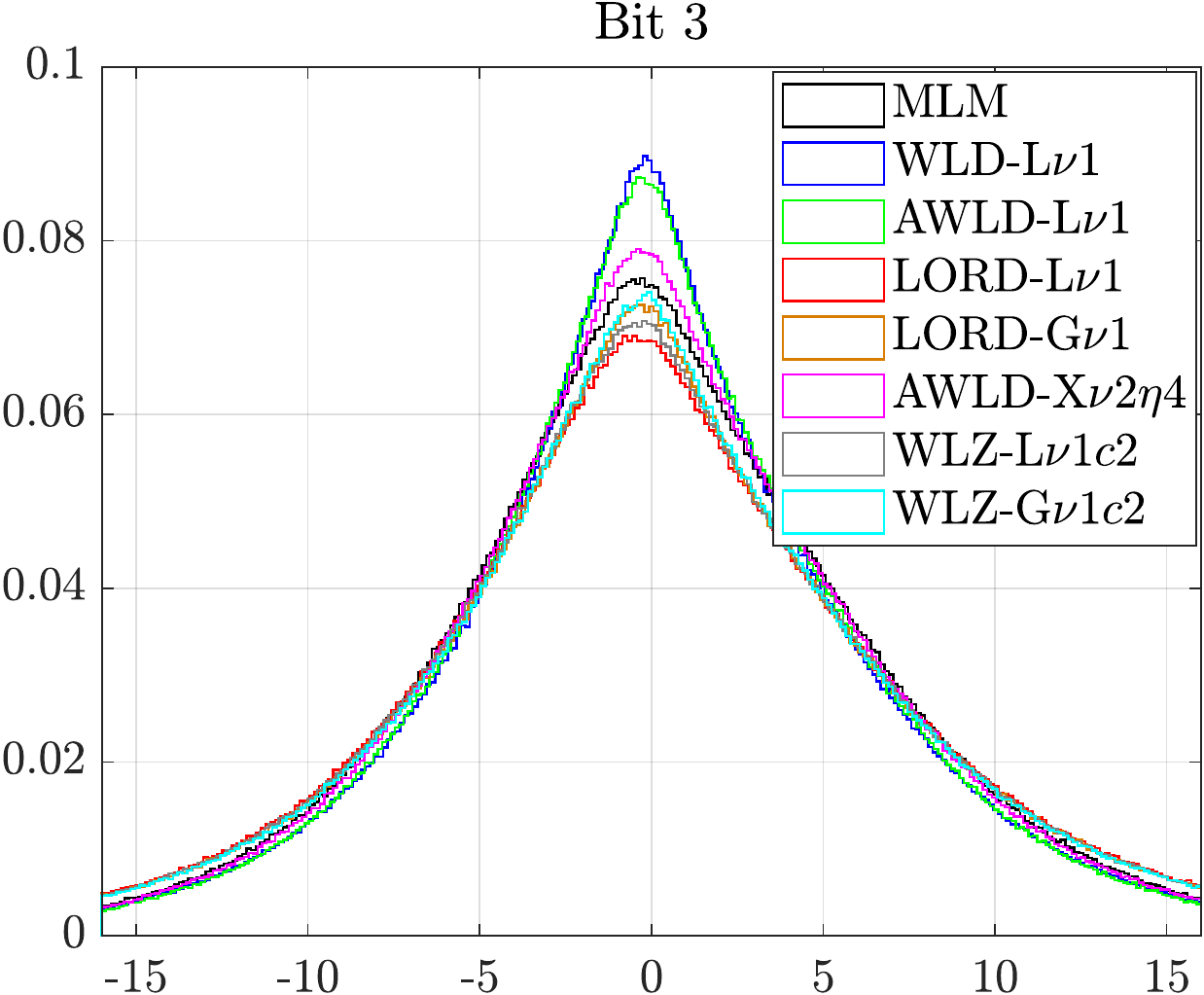}
  \vspace{-0.275in}
  \caption{Distribution of LLRs for bits 1 and 3 of one symbol: $4\!\times\! 4$ complex MIMO channel, 16QAM, $\text{SNR}\!=\!\unit[20]{dB}$.}
  \label{fig:LLR_bit_1_4_dist_plot_4x4_16QAM}
\end{figure}

Figure~\ref{fig:LLR_bit_1_4_dist_plot_4x4_16QAM} plots the LLR distributions of bits 1 and 3 of one symbol in a $4\!\times\! 4$, 16QAM MIMO system at $\text{SNR}\!=\!\unit[20]{dB}$. As shown, AWLD and WLZ track the optimal LLRs very closely.

\begin{figure}
  \centering
  \includegraphics[scale=0.85]{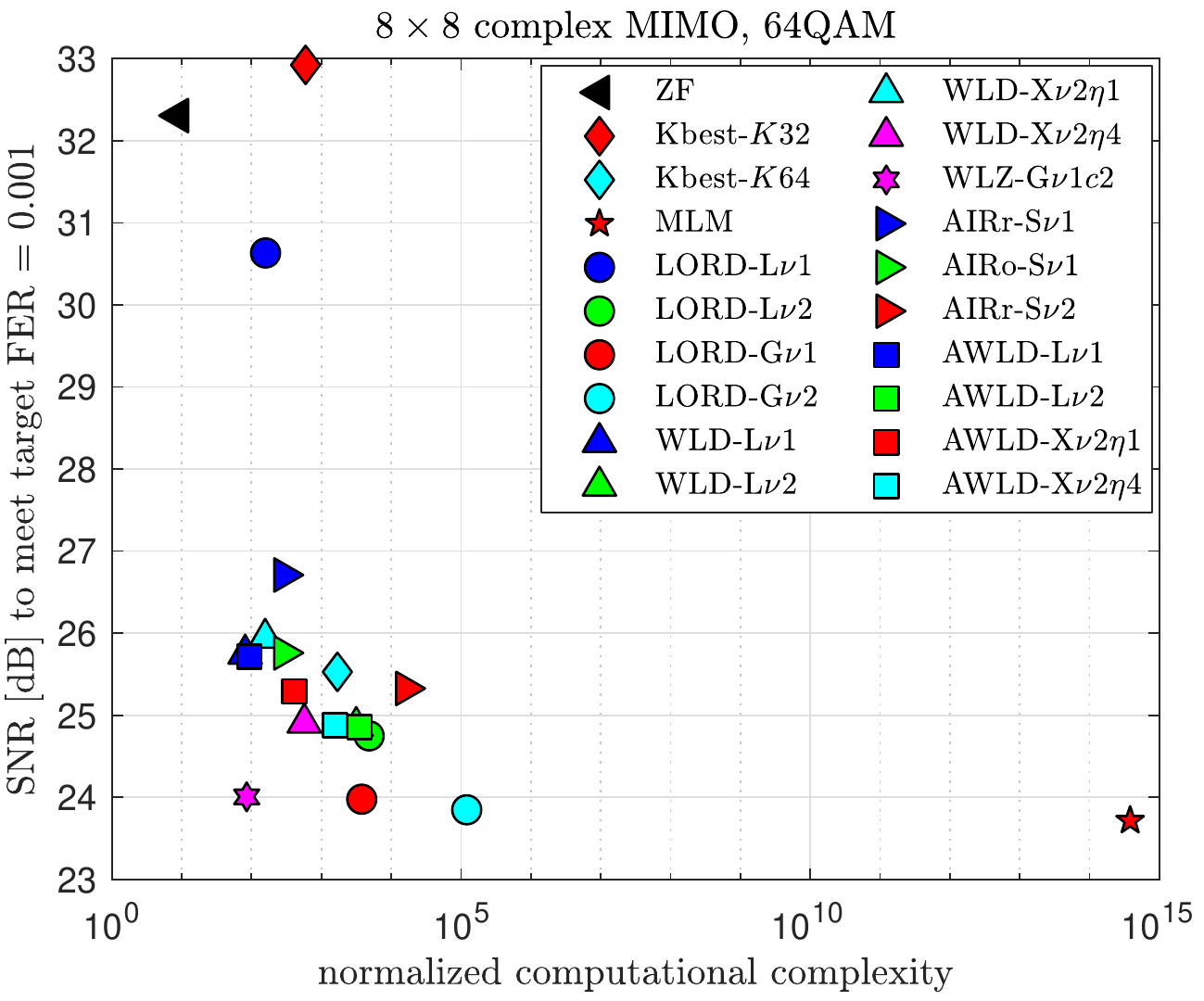}
  \vspace{-0.15in}
  \caption{SNR to meet a target FER of $0.1\%$ versus complexity.}\label{fig:complexity_fer_8x8_64QAM}
\end{figure}

The complexity of various algorithms is benchmarked and compared in Fig.~\ref{fig:complexity_fer_8x8_64QAM} for an $8\!\times\!8$ MIMO system and 64QAM. The figure plots the SNR required to achieve a target FER of $0.1\%$ versus normalized complexity. All algorithms (matrix decomposition, filtering, MMSE, MIMO detection) are first implemented using fixed-point arithmetic, and then profiled in terms of memory storage requirements and kernel mathematical operations. These operations include (both for real and complex quantities, where applicable): multiplication, division, multiply-accumulate, squaring, addition/subtraction, inversion, (inverse) square-root, slicing, look-up table (LUT) operations, comparison operations, vector norm and norm-square, multiplexing, sorting, and permutation. The gate-count complexity of these operations is evaluated by mapping them to a library of pre-characterized logic gates that includes basic adders/subtractors, multipliers, squarers, dividers, multiplexers, memory elements, comparators, slicers, and (inverse) square-rooters. As a result, each operation is characterized with a gate complexity value.

Parallel architectures for all algorithms are developed, and their gate-count complexity is plotted in Fig.~\ref{fig:complexity_fer_8x8_64QAM}. For the MLM algorithm, a serial depth-first tree traversal architecture is developed, and its complexity is reported as gate-count per tree node, multiplied by the number of nodes visited. Since the latter is non-deterministic, the value reported is averaged over 1000 detection trials. For the K-best algorithm, a $K$-wide parallel architecture is developed.

As expected, the ZF and MLM algorithms lie at opposite extremes in the performance-complexity space. The proposed WLZ algorithm offers the best performance-complexity tradeoff among all algorithms. It matches the performance of LORD at roughly $20\mathsf{x}$ less complexity. The savings are primarily due to the eliminated complex multiplications in $\L$ as a result of the puncturing and reduction operations.


%

%
\section{Conclusions}\label{sec:conclusion}
Channel puncturing in augmented and two-sided forms has been investigated in this work as an effective means to reduce computational complexity of tree-based soft-output MIMO detectors. It has been shown that punctured augmented channel matrices processed by the AWLD detector are optimal in maximizing the lower bound on the achievable information rate. Their structure matches exactly that of AIR-PM, but most importantly, they can be derived using simple QL decomposition followed by Gaussian elimination. When used in multi-tree mode with local metric updates, AWLD beats LORD both performance-wise and complexity-wise. However, LORD, when optimized to operate with global across-tree metric updates, attains a significant performance gain that AWLD cannot match because its puncturing matrix is non-unitary, and hence Euclidean-distance based metrics are not preserved under column permutations in multiple trees. This shortcoming is mitigated by employing two-sided puncturing based on right-sided integer reduction and left-sided elimination. The resulting puncturing matrices processed by WLZ are almost unitary, and hence the global across-tree metric update property of LORD is retained. The result is that the proposed WLZ scheme offers the best performance-complexity tradeoff among tree-based detectors. Finally, extensions to include soft-input information, imperfect channel estimation effects, and correlated channels are directly applicable based on~\cite{2017_hu_softoutput_AIR}.

{
\setstretch{1.1}
\bibliographystyle{IEEEtran}
\bibliography{IEEEabrv,bibliography}
}

\begin{IEEEbiography}[{\includegraphics[width=1in,height=1.25in,clip,keepaspectratio]{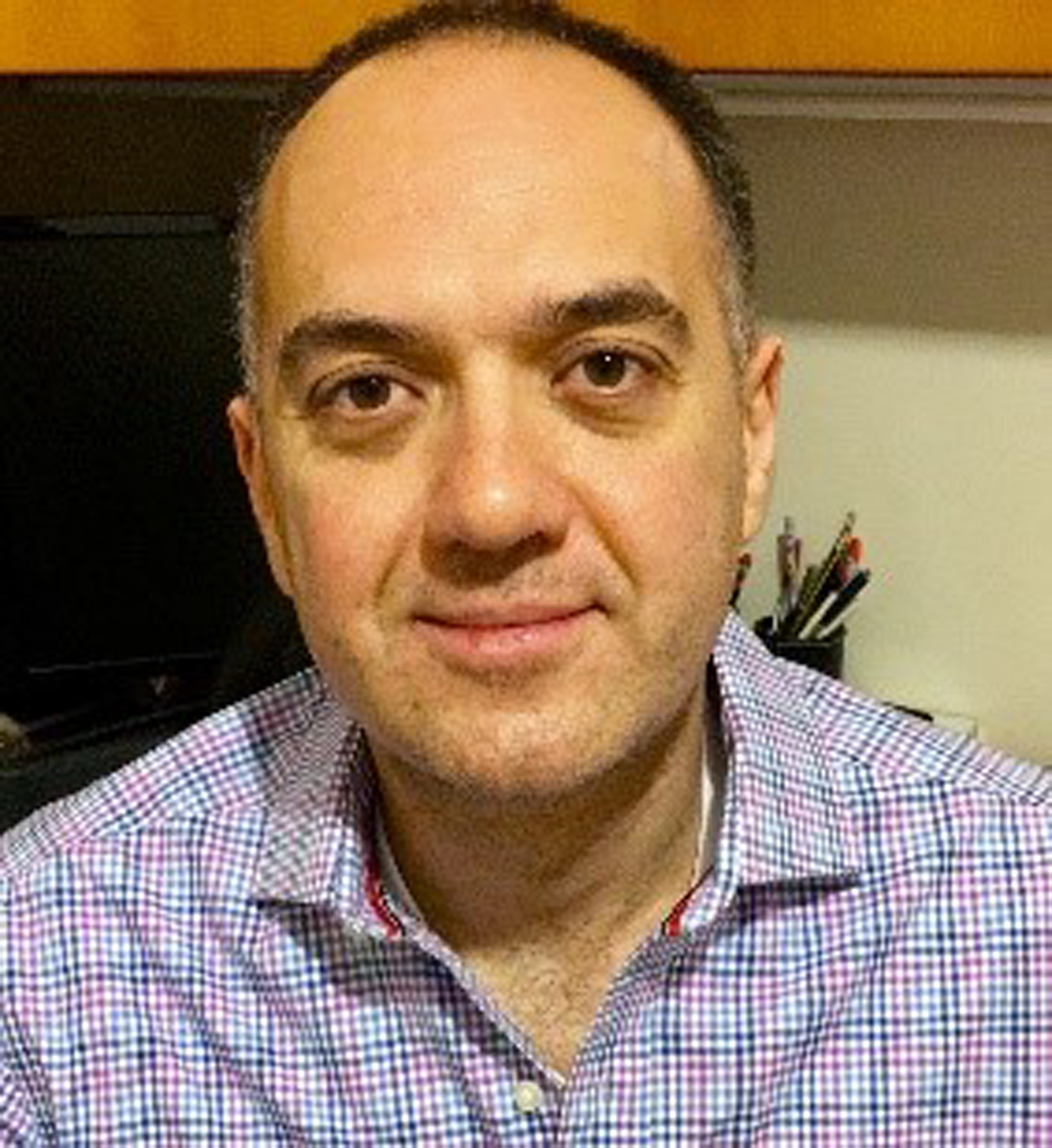}}]{Mohammad M. Mansour}(S'97-M'03-SM'08) received the B.E. (Hons.) and the M.E. degrees in computer and communications engineering from the American University of Beirut (AUB), Beirut, Lebanon, in 1996 and 1998, respectively, and the M.S. degree in mathematics and the Ph.D. degree in electrical engineering from the University of Illinois at Urbana–Champaign (UIUC), Champaign, IL, USA, in 2002 and 2003, respectively.
			
He was a Visiting Researcher at Qualcomm, San Jose, CA, USA, in summer of 2016, where he worked on baseband receiver architectures for the IEEE 802.11ax standard. He was a Visiting Researcher at Broadcom, Sunnyvale, CA, USA, from 2012 to 2014, where he worked on the physical layer SoC architecture and algorithm development for LTE-Advanced baseband receivers. He was on research leave with Qualcomm Flarion Technologies in Bridgewater, NJ, USA, from 2006 to 2008, where he worked on modem design and implementation for 3GPP-LTE, 3GPP2-UMB, and peer-to-peer wireless networking physical layer SoC architecture and algorithm development. He was a Research Assistant at the Coordinated Science Laboratory (CSL), UIUC, from 1998 to 2003. He worked at National Semiconductor Corporation, San Francisco, CA, with the Wireless Research group in 2000. He was a Research Assistant with the Department of Electrical and Computer Engineering, AUB, in 1997, and a Teaching Assistant in 1996. He joined as a faculty member with the Department of Electrical and Computer Engineering, AUB, in 2003, where he is currently a tenured full-Professor and Chairperson. He leads the COMNETIC research group whose focus is on fundamental research spanning inter-related core areas in information processing and machine learning, wireless communications, and VLSI systems. His research interests are in the area of energy-efficient and high-performance VLSI circuits, architectures, algorithms, and systems for communications, signal processing, and computing applications.
			
Prof. Mansour is a member of the Design and Implementation of Signal Processing Systems (DISPS) Technical Committee Advisory Board of the IEEE Signal Processing Society. He served as a member of the DISPS Technical Committee from 2006 to 2013. He served as an Associate Editor for IEEE TRANSACTIONS ON CIRCUITS AND SYSTEMS II (TCAS-II) from 2008 to 2013, as an Associate Editor for the IEEE SIGNAL PROCESSING LETTERS from 2012 to 2016, and as an Associate Editor of the IEEE TRANSACTIONS ON VLSI SYSTEMS from 2011 to 2016. He served as the Technical Co-Chair of the IEEE Workshop on Signal Processing Systems in 2011, and as a member of the Technical Program Committee of various international conferences and workshops. He was the recipient of the PHI Kappa PHI Honor Society Award twice in 2000 and 2001, and the recipient of the Hewlett Foundation Fellowship Award in 2006. He has seven issued U.S. patents.
\end{IEEEbiography}


%
\beginfirstsupplement

\section{Proof of Lemma~\ref{lem:trace_formula}}
\label{supplement:proof_of_trace_formula}
First, we can assume without any loss of generality that both $\U$ and $\V$ are lower-triangular matrices with real and positive diagonal entries.
Otherwise, let $\U\!=\!\Qu\Lu$ be the QL decomposition of $\U$ and $\V\Vh \!=\! \Lv\Lvh$ be the Cholesky factorization of $\V\Vh$, where $\Lu$ and $\Lv$ are lower-triangular matrices with real and positive diagonal entries. Then
\begin{align*}
    f(\U,\V)
    &\!=\!
    \ln\det(\U\Uh) \!-\! \Trr{(\U\V)(\U\V)^{\dag}}
    \\
    &\!=\!
    \ln\det(\!\Qu\Lu\Luh\Quh\!) \!-\! \Trr{\!(\!\Qu\Lu\!)(\Lv\Lvh\!)(\Luh\Quh\!)\!}
    \\
    &\!=\!
    \ln\det(\Lu\Luh) \!-\! \Trr{(\Lu\Lv)(\Lu\Lv)^{\dag}}
    \\
    &\!=\!
    f(\Lu,\Lv).
\end{align*}
Henceforth, we assume that both $\U\!=\![u_{kj}]$ and $\V\!=\![v_{kj}]$ are lower-triangular matrices with real and positive diagonal entries. Let $\tilde{\mbf{u}}_{k}^{}\!=\![u_{k1}^{}\,u_{k2}^{}\,\cdots\,u_{k,k-1}^{}]$ denote the row vector consisting of the first $k-1$ elements of the $\nth{k}$ row of $\U$, and $\mbf{u}_{k}^{}\!=\![\tilde{\mbf{u}}_{k}^{}\,u_{kk}^{}]$. Let $\U_k^{}$ denote the leading principal matrix of $\U$ of order $k$, and let $\tilde{\U}_k^{} \!=\! \U_{k-1}^{}$. The vectors $\tilde{\mbf{v}}_{k}^{}$, $\mbf{v}_{k}^{}$, and matrices $\V_k^{}$, $\tilde{\V}_k^{}$ are similarly defined for $\V$. Let $g(\U)\!\triangleq\!\ln\det(\U\Uh)$ and $h(\U,\V) \!\triangleq\! \Trr{(\U\V)(\U\V)^{\dag}}$.

To determine $\Uopt$, we compute $\diffp*{f(\U,\V)}{\U}$ and set it to 0. We start by computing the trace $\Trr{(\U\V)(\U\V)^{\dag}}$ first,
\begin{align*}
    h
    &=
    \sum_{k=1}^{N}
        \begin{bNiceMatrix}
            \tilde{\mbf{u}}_k & u_{kk}
        \end{bNiceMatrix}
        \begin{bNiceMatrix}
            \tilde{\mbf{V}}_k & \\
            \tilde{\mbf{v}}_k & v_{kk}
        \end{bNiceMatrix}
        \begin{bNiceMatrix}
            \tilde{\mbf{V}}_k^{\dag} & \tilde{\mbf{v}}_k^{\dag} \\
                                     & v_{kk}
        \end{bNiceMatrix}
        \begin{bNiceMatrix}
            \tilde{\mbf{u}}_k^{\dag} \\
            u_{kk}
        \end{bNiceMatrix}
        \\
    &=
    \sum_{k=1}^{N}\!\!
        \left\{\!
            \tilde{\mbf{u}}_k \tilde{\mbf{V}}_k
            \tilde{\mbf{V}}_k^{\dag}\tilde{\mbf{u}}_k^{\dag}
            \!+\!
            2u_{kk}^{}\Re{\!\tilde{\mbf{v}}_k
            \tilde{\mbf{V}}_k^{\dag}\tilde{\mbf{u}}_k^{\dag}\!}
            \!+\!
            u_{kk}^2(v_{kk}^2 \!+\! \tilde{\mbf{v}}_k\tilde{\mbf{v}}_k^{\dag})
            \!
        \right\}
\end{align*}
The problem then boils down to determining the unknowns $\tilde{\mbf{u}}_k$ and $u_{kk}$ that satisfy the required derivative condition. Since $\ln\det (\U\Uh) = \sum_{k=1}^{N}\ln u_{kk}^2$ involves the diagonal elements $u_{kk}$ only, we can start by determining $\tilde{\mbf{u}}_k$ by setting the derivative of the trace term only $\diffp{h}{\tilde{\mbf{u}}_k}$ to 0. We obtain
\begin{align*}
    \tilde{\mbf{V}}_k\tilde{\mbf{V}}_k^{\dag}\tilde{\mbf{u}}_k^{\dag}
    +
    u_{kk}\tilde{\mbf{V}}_k\tilde{\mbf{v}}_k^{\dag}
    = 0,
\end{align*}
and hence
\begin{align*}
    \tilde{\mbf{u}}_k^{\opt}
    &=
    -u_{kk}
    \tilde{\mbf{v}}_k
    \tilde{\mbf{V}}_k^{-1}.
\end{align*}
We next determine $u_{kk}$. Substituting back in the trace equation, we get
\begin{align*}
    h|_{\tilde{\mbf{u}}_k=\tilde{\mbf{u}}_k^{\opt}}
    &=
    \sum_{k=1}^{N} u_{kk}^2 v_{kk}^2.
\end{align*}
Now taking derivative with respect to $u_{kk}$, including the $\ln\det$ term, we have
\begin{align*}
    \diffp{f
    }
    {u_{kk}}
    &\!=\!
    \diffp*{
        \left\{
        \ln(u_{kk}^2)
        \!-\!
        u_{kk}^2 v_{kk}^2
        \right\}
    }{u_{kk}}
    \!=\!
    \tfrac{2}{u_{kk}} \!-\! 2u_{kk}v_{kk}^2
    \!=\!
    0,
\end{align*}
implying that $u_{kk}^{\opt} \!=\! \frac{1}{v_{kk}}$. Therefore,
\begin{align*}
    \mbf{u}_k^{\opt}
    &=
    [\tilde{\mbf{u}}_k^{\opt} \, u_{kk}^{\opt}]
    =
    [
    -v_{kk}^{-1}
    \tilde{\mbf{v}}_k
    \tilde{\mbf{V}}_k^{-1}
    ~~
    v_{kk}^{-1}
    ].
\end{align*}
Next note that if we multiply $\mbf{u}_k^{\opt}$ by $\V_k^{}$ for any $k$ we obtain
\begin{align*}
    \mbf{u}_k^{\opt} \V_k^{}
    &\!=\!
    \begin{bNiceMatrix}
        -v_{kk}^{-1}
        \tilde{\mbf{v}}_k
        \tilde{\mbf{V}}_k^{-1}
        &
        v_{kk}^{-1}
    \end{bNiceMatrix}
    \!\!
    \begin{bNiceMatrix}
            \tilde{\mbf{V}}_k & \\
            \tilde{\mbf{v}}_k & v_{kk}
    \end{bNiceMatrix}
    \!=\!
    \begin{bNiceMatrix}
        \mbf{0}_{1\times (k-1)} & 1
    \end{bNiceMatrix},
\end{align*}
which implies that $\U^{\opt}\V\!=\!\I$. Hence the optimal $\U$ is the inverse of $\V$, $\U^{\opt} \!=\! \V^{-1}$, and $f(\U^{\opt},\V) \!=\! -\sum_{k=1}^{N}\ln v_{kk}^2 - N$.\IEEEQEDright 
\newpage

%
\beginnewsupplement

\begin{landscape}
\newcommand*{\thead}[1]{\multicolumn{1}{c}{\bfseries #1}}
\begin{table}
\centering
\caption{Summary of decomposition and puncturing algorithms}
\setlength{\tabcolsep}{2pt}
\def\arraystretch{1.1}
\begin{tabular}{|l|l|l|l|}
  \hline
  \thead{Algorithm} & \thead{Scheme} & \thead{Functionality} & \thead{Properties}
  \\\hline\hline
  \algnameabbr~\ref{algo:qldy} &
  \hyperref[algo:qldy]{$\mathsf{QL}$} &
  Decompose $\H$ as $\H=\Q\L$; generate $\yt=\Qh\y$ &
  $\Qh\Q\!=\!\I$; $\L$ lower-triangular, $\L(k,k)\in\mathcal{R}^+$
  \\\hline
  \multirow{2}{*}{\algnameabbr~\ref{algo:qdly}} &
  \multirow{2}{*}{\hyperref[algo:qldy]{$\mathsf{QDL}$}} &
  \multirow{2}{*}{Decompose $\H$ as $\H=\Qtilde\mbf{D}\Ltilde$; generate $\tsup{\y}=\Qtildeh\y$} &
  Square-root free QL decomposition
  \\
  &
  &
  &
  $\Qtildeh\Qtilde=\supsc{\mbf{D}}{\,-1}$; $\Ltilde$ unit lower-triangular
  \\\hline
  \multirow{3}{*}{\algnameabbr~\ref{algo:generalized_wld}} &
  \multirow{3}{*}{\hyperref[algo:qldy]{$\mathsf{WL}$}} &
  \multirow{3}{*}{Puncture $\H$ using $\W$ as $\Wh\H\!=\!\Lp$; generate $\yp\!=\!\Wh\y$} &
  One-sided puncturing algorithm
  \\
  &
  &
  &
  $\W$ non-unitary but $\diagg{\Wh\W}\!=\!\I$
  \\
  &
  &
  &
  $\Lp$ punctured lower-triangular, $\Lp(k,k)\in\mathcal{R}^+$
  \\\hline
  \multirow{3}{*}{\algnameabbr~\ref{algo:generalized_wdl}} &
  \multirow{3}{*}{\hyperref[algo:generalized_wdl]{$\mathsf{WDL}$}} &
  \multirow{3}{*}{Puncture $\H$ using $\Wtilde$ as $\Wtildeh\H\!=\!\Lptilde$; generate $\yptilde\!=\!\Wtildeh\y$} &
  One-sided square-root free puncturing algorithm
  \\
  &
  &
  &
  $\Wtilde$ non-unitary but $\diagg{\Wtildeh\Wtilde}\!=\!\supsc{\mbf{D}}{\,-1}$
  \\
  &
  &
  &
  $\Lptilde$ punctured unit lower-triangular
  \\\hline
  \multirow{3}{*}{\algnameabbr~\ref{algo:qlzd}} &
  \multirow{3}{*}{\hyperref[algo:qlzd]{$\mathsf{QLZy}$}} &
  \multirow{3}{*}{Decompose $\H$ as $\H=\Q\L\supsc{\mbf{Z}}{\,-1}$; generate $\yt=\Qh\y$} &
  QL decomposition with right reduction
  \\
  &
  &
  &
  $\Qh\Q\!=\!\I$; $\L$ lower-triangular, $\L(k,k)\in\mathcal{R}^+$
  \\
  &
  &
  &
  $\mbf{Z}$ unimodular with $\det{\mbf{Z}}=1$
  \\\hline
  \multirow{4}{*}{\algnameabbr~\ref{algo:wlzdopt}} &
  \multirow{4}{*}{\hyperref[algo:wlzdopt]{$\mathsf{WLZ}$}} &
  \multirow{4}{*}{Puncture $\H$ using $\W,\mbf{Z}$ as $\Wh\H\mbf{Z}\!=\!\Lp$; generate $\yp\!=\!\Wh\y$}
  &
  Two-sided puncturing algorithm
  \\
  &
  &
  &
  $\W$ non-unitary but $\diagg{\Wh\W}\!=\!\I$
  \\
  &
  &
  &
  $\Lp$ punctured lower-triangular, $\Lp(k,k)\in\mathcal{R}^+$
  \\
  &
  &
  &
  $\mbf{Z}$ unimodular with $\det{\mbf{Z}}=1$
  \\\hline
\end{tabular}
\label{tab:summary_decomposition_algorithms}
\end{table}
\end{landscape}

%
\beginnewsupplement

\section{QL Decomposition Algorithm}
\begin{algorithm}[hbtp]
\small
\caption{Optimized thin QL decomposition algorithm}\label{algo:qldy}
\begin{algorithmic}[0]
\LeftComment{Decompose $\H$ as $\H=\Q\L$ and generate $\yt=\Qh\y$}
\LeftComment{$\H\!:$ Complex $\Nr\!\times\!\Nt$ matrix, $\Nr\!\geq\!\Nt$}
\LeftComment{$\y\!:$ Complex $\Nr\!\times\!1$ column vector}
\LeftComment{$\Q\!:\MbyN$ matrix with orthonormal columns; $\Qh\Q=\I_{\Nt}$}
\LeftComment{$\L\!:\NbyN$ lower-triangular matrix s.t. $\L(k,k)\in\mathcal{R}^+$}
\LeftComment{$\yt\!:\Nt\!\times\!1$ such that $\yt = \Qh\y$}
\FunctionRV{QLy}{\H,\y}{[\Q, \L, \yt]}
\State $\Q\gets[\y ~ \H]$ \Comment{\textit{augment $\y$ to $\H$}}
\State $\L \gets \mbf{0}_{\Nt\times{(\Nt+1)}}^{}$
\For{$k\!=\!\Nt\!+\!1\!:\!-1\!:\!2$}\Comment{\textit{index of current column}}
    \State $\L(k\!-\!1,k) \gets \sqrt{\Q(:,k)^{\dag}\Q(:,k)}$\Comment{\textit{diagonal element}}
    \State $\Q(:,k) \!\gets\! \Q(:,k)/\L(k\!-\!1,k)$\Comment{\textit{normalize}}
    \For{$j\!=\!k\!-\!1\!:\!-1\!:\!1$}\Comment{\textit{all other cols to its left}}
        \State $\L(k\!-\!1,j) \!\gets\! \Q(:,k)^{\dag}\Q(:,j)$
        \State $\Q(:,j) \!\gets\! \Q(:,j) \!-\! \L(k\!-\!1,j)\Q(:,k)$
    \EndFor\label{QLDyendforj}
\EndFor\label{QLDyendfori}
\State $\Q \!\gets\! \Q(:,2\!:\!\Nt\!+\!1)$\Comment{last $\Nt$ cols of augmented $\Q$}
\State $\yt \!\gets\! \L(:,1)$\Comment{first col of augmented $\L$}
\State $\L \!\gets\! \L(:,2\!:\!\Nt\!+\!1)$\Comment{last $\Nt$ cols of augmented $\L$}
\EndFunctionRV
\end{algorithmic}

\end{algorithm}

\beginnewsupplement
\section{QDL Decomposition Algorithm}
\begin{algorithm}[hbtp]
\small
\caption{Optimized QDL decomposition algorithm}\label{algo:qdly}
%
\begin{algorithmic}[0]
\LeftComment{Decompose $\H$ as $\H=\Qtilde\mbf{D}\Ltilde$ and generate $\tsup{\y}=\Qtildeh\y$.} 
\LeftComment{If $[\Q, \L, \yt]\!=\!\mathsf{QLy}{(\H,\y)}$, then} \LeftComment{\makebox[\linewidth][c]{$\Qtilde=\Q\supsc{\mbf{D}}{\,-1/2}$,~ $\Ltilde=\supsc{\mbf{D}}{\,-1/2}\L$,~ $\tsup{\y}=\supsc{\mbf{D}}{\,-1/2}\yt$.}}
\LeftComment{$\H\!:$ Complex $\Nr\!\times\!\Nt$ matrix, $\Nr\!\geq\!\Nt$}
\LeftComment{$\y\!:$ Complex $\Nr\!\times\!1$ column vector}
\LeftComment{$\Qtilde\!:\MbyN$ matrix with orthogonal columns s.t. $\Qtildeh\Qtilde=\supsc{\mbf{D}}{\,-1}$}
\LeftComment{$\mbf{D}\!: \NbyN$ diagonal matrix with real positive entries
 such that}
\LeftComment{\quad~ $\mbf{D} = \supsc{\diagg{\L}}{\,2}$}
\LeftComment{$\Ltilde\!:\NbyN$ unit lower-triangular matrix; $\Ltilde(k,k)=1$}
\LeftComment{$\tsup{\y}\!:\Nt\!\times\!1$ such that $\tsup{\y} = \Qtildeh\y = \supsc{\mbf{D}}{\,-1/2}\Qh\y = \supsc{\mbf{D}}{\,-1/2}\yt$}
\FunctionRV{QDLy}{\H,\y}{[\Qtilde, \D, \Ltilde, \tsup{\y}]}
\State $\Qtilde \gets [\y ~ \H]$
\Comment{\textit{augment $\y$ to $\H$}}
\State $\D \gets \mbf{0}_{\Nt\times{\Nt}}^{}$
\Comment{\textit{normalizer diagonal matrix}}
\State $\Ltilde \gets [\mbf{0}_{\Nt\times{1}}^{} ~ \I_{\Nt}^{}]$
\Comment{\textit{normalized augmented matrix}}
\For{$k\!=\!\Nt\!+\!1\!:\!-1\!:\!2$}\Comment{\textit{index of current col}}
    \State $\D(k\!-\!1,k\!-\!1) \gets \supsc{\fnormm{\Qtilde(:,k)}}{\,2}$
    \Comment{\textit{diagonal element}}
    \For{$j\!=\!k\!-\!1\!:\!-1\!:\!1$}\Comment{\textit{all other cols to its left}}
        \State $\Ltilde(k\!-\!1,j) \!\gets\! \Qtilde(:,k)^{\dag}\Qtilde(:,j)/\D(k\!-\!1,k\!-\!1)$
        \State $\Qtilde(:,j) \!\gets\! \Qtilde(:,j) \!-\! \Ltilde(k\!-\!1,j)\Qtilde(:,k)$
    \EndFor\label{QLDy_end_for_j}
    \State $\Qtilde(:,k) \!\gets\! \Qtilde(:,k)/\D(k\!-\!1,k\!-\!1)$
\EndFor\label{QLDy_end_for_k}
\State $\Qtilde \!\gets\! \Qtilde(:,2\!:\!\Nt\!+\!1)$
\Comment{last $\Nt$ cols of augmented $\Qtilde$}
\State $\tsup{\y} \!\gets\! \Ltilde(:,1)$
\Comment{first col of augmented $\Ltilde$}
\State $\Ltilde \!\gets\! \Ltilde(:,2\!:\!\Nt\!+\!1)$
\Comment{last $\Nt$ cols of augmented $\Ltilde$}
\EndFunctionRV
\end{algorithmic}

\end{algorithm}

%
\beginnewsupplement

\section{WL Decomposition Algorithm}
\begin{algorithm}[hbtp]
\small
\caption{Optimized WL decomposition algorithm}\label{algo:generalized_wld}
%
%
%
\begin{algorithmic}[0]
\LeftComment{Generate $\W$ s.t. $\Wh\H\!=\!\Lp$, $\Wh\y\!=\!\yp$, $\diagg{\Wh\W}\!=\!\I_{\Nt}$}
\LeftComment{$\H\!:$ Complex $\Nr\!\times\!\Nt$ matrix, $\Nr\!\geq\!\Nt$}
\LeftComment{$\y\!:$ Complex $\Nr\!\times\!1$ column vector}
\LeftComment{$\nu\!:\text{puncturing order}$}
\LeftComment{$\Lp\!:\NbyN$ punctured lower-triangular matrix; $\Lp(k,k)\in\mathcal{R}^+$}
\LeftComment{$\yp\!:\Nt\!\times\!1$ such that $\yp = \Wh\y$}
\LeftComment{$\W\!:$ $\MbyN$ puncturing matrix such that $\diagg{\Wh\W} = \I_{\Nt}$}
\LeftComment{----------------------------------------------------------------------------------------}
\LeftComment{Note: $\Wh$ punctures $\H$; in manuscript, $\Wp$ punctures $\L$.}
\LeftComment{The two schemes are related as follows:}
\LeftComment{
\makebox[\linewidth][c]{$\Wh(\y\!-\!\H\x) \!=\! \Wp\Qh(\y\!-\!\H\x) \!=\! \Wp(\ytilde\!-\!\L\x)$}}
\LeftComment{
\makebox[\linewidth][c]{$\Wh \!=\! \Wp\Qh$, $\Wp\!=\!\Wh\Q$}}
\LeftComment{Also, $\Wh\Q\Qh\!=\!\Wh$ even though $\Q\Qh\!\neq\!\I$ for $\Nr\!>\!\Nt$. This is because the rows of $\Qh$ and the cols of $(\Q\Qh\!-\!\I)$ are orthogonal so that $\Qh(\Q\Qh\!-\!\I)\!=\!\mbf{0}$. Hence any matrix right-multiplied by $\Qh$ would have rows orthogonal to $(\Q\Qh\!-\!\I)$. Thus $\Wh\Q\Qh\!-\!\Wh\!=\!\Wh(\Q\Qh\!-\!\I)\!=\!\Wp\Qh(\Q\Qh\!-\!\I)\!=\!\mbf{0}$.
}
\LeftComment{----------------------------------------------------------------------------------------}
\FunctionRV{WL}{\H,\y,\nu}{[\Lp, \yp, \W]}
\State $[\Q, \L, \yt] \gets \hyperref[algo:qldy]{\mathsf{QLy}}(\H,\y)$
\Comment{\textit{QL dec.; here $\yt\!=\!\Qh\y,\Qh\!\Q\!=\!\I_{\Nt}$}}
\State $\W \gets \Q,~\Lp \gets [\yt~ \L]$
\Comment{Augment $\yt$ to $\L$}
\For{$k\!=\!\nu\!+\!2\!:\!\Nt$}\Comment{\textit{Gaussian elimination}}
    \For{$j\!=\!\nu\!+\!1\!:\!k\!-\!1$}
    \Comment{\textit{col index to puncture}}
        \State{$\alpha \!\gets\! \Lp(k,j\!+\!1)/\Lp(j,j\!+\!1)$}
        \State $\W(:,k) \!\gets\! \W(:,k) \!-\! \alpha^{\dag}\W(:,j)$
        \State $\Lp(k,1\!:\!j\!+\!1) \!\gets\! \Lp(k,1\!:\!j\!+\!1) \!-\! \alpha \Lp(j,1\!:\!j\!+\!1)$
    \EndFor\label{wld_end_for_j}
    \State $\Lp(k,1\!:\!k\!+\!1) \!\gets\! \Lp(k,1\!:\!k\!+\!1)/\fnormm{\W(:,k)}$
    \State $\W(:,k) \!\gets\! \W(:,k) / \fnormm{\W(:,k)}$
\EndFor\label{wld_end_for_k}
\State $\yp \!\gets\! \Lp(:,1)$
\Comment{first col of augmented $\Lp$}
\State $\Lp \!\gets\! \Lp(:,2\!:\!\Nt\!+\!1)$
\Comment{last $\Nt$ cols of augmented $\Lp$}
\EndFunctionRV\Comment{$\Wp\!=\!\Wh\Q$}
\end{algorithmic}

\end{algorithm}

\beginnewsupplement
\section{WDL Decomposition Algorithm}
\begin{algorithm}[hbtp]
\small
\caption{Square-root-free WDL decomposition algorithm}\label{algo:generalized_wdl}
\begin{algorithmic}[0]
\LeftComment{Square-root free version of $\mathsf{WL}()$ in \algnameabbr~\ref{algo:generalized_wld}}
\LeftComment{Generate $\mbf{D}$ and $\Wtilde$ such that $\Wtildeh\H\!=\!\Lptilde$,
$\Wtildeh\y\!=\!\yptilde$, and}
\LeftComment{\quad~ $\diagg{\Wtildeh\Wtilde}\!=\!\supsc{\mbf{D}}{\,-1}$.}

\LeftComment{$\H\!:$ Complex $\Nr\!\times\!\Nt$ matrix, $\Nr\!\geq\!\Nt$}
\LeftComment{$\y\!:$ Complex $\Nr\!\times\!1$ column vector}
\LeftComment{$\nu\!:\text{puncturing order}$}
\LeftComment{$\Lptilde\!:\NbyN$ punctured unit lower-triangular matrix; $\Lptilde(k,k)\!=\!1$}
\LeftComment{$\yptilde\!:\Nt\!\times\!1$ such that $\yptilde = \Wtildeh\y$}
\LeftComment{$\mbf{D}\!: \NbyN$ diagonal matrix with real positive entries
 such that}
\LeftComment{\quad~ $\mbf{D} = \supsc{\diagg{\Lp}}{\,2}$}
\LeftComment{$\Wtilde\!:\Nr\!\times\!\Nt$ puncturing matrix such that $\diagg{\Wtildeh\Wtilde}\!=\!\supsc{\mbf{D}}{\,-1}$}

\LeftComment{----------------------------------------------------------------------------------------}
\LeftComment{Note: $\Wtildeh$ punctures $\H$ to form $\Lptilde\!=\!\Wtildeh\H$. In manuscript, $\Wptilde$ punctures $\D\Ltilde\!=\!\supsc{\mbf{D}}{\,1/2}\L$ to form $\Lp\!=\!\Wptilde\D\Ltilde$. These quantities are related as follows:}

\LeftComment{If $[\Qtilde, \D, \Ltilde, \tsup{\y}]\!=\!\hyperref[algo:qdly]{\mathsf{QDLy}}{(\H,\y)}$, then:}
\LeftComment{\makebox[\linewidth][c]{$\Qtilde=\Q\supsc{\mbf{D}}{\,-1/2}$,~ $\Ltilde=\supsc{\mbf{D}}{\,-1/2}\L$,~ $\tsup{\y}=\supsc{\mbf{D}}{\,-1/2}\yt$}}

\LeftComment{If $[\Lp, \yp, \W]\!=\!\hyperref[algo:generalized_wld]{\mathsf{WL}}(\H,\y,\nu)$, then:}

\LeftComment{\makebox[\linewidth][c]{$\Lp\!=\!\Wh\H$, $\yp\!=\!\Wh\y$}}
\LeftComment{\makebox[\linewidth][c]{$\Wtilde\!=\!\W\supsc{\mbf{D}}{\,-1/2}$,
$\Lptilde\!=\!\supsc{\mbf{D}}{\,-1/2}\Lp$, $\mbf{D}\!=\!\supsc{\diagg{\Lp}}{\,2}$,
$\yptilde\!=\!\supsc{\mbf{D}}{\,-1/2}\yp$}}

\LeftComment{\makebox[\linewidth][c]{$\Wh\!=\!\Wp\Qh\!=\!\Wptilde\D\Qtildeh$, $\Wtildeh\!=\!\supsc{\mbf{D}}{\,-1/2}\Wptilde\D\Qtildeh$}}

\LeftComment{----------------------------------------------------------------------------------------}
\FunctionRV{WDL}{\H,\y,\nu}{[\Lptilde, \yptilde, \D, \Wtilde]}
\State $[\Qtilde, \D, \Ltilde, \tsup{\y}] \gets \hyperref[algo:qdly]{\mathsf{QDLy}}(\H,\y)$
\Comment{\textit{QDL dec.; here $\tsup{\y}\!=\!\Qtildeh\y$}}
\State $\Wtilde \gets \Qtilde$\Comment{copy in case $\Qtilde$ is needed}
\State $\Lptilde \gets [\tsup{\y}~ \Ltilde]$\Comment{Augment $\tsup{\y}$ to $\Ltilde$}
\For{$k\!=\!\nu\!+\!2\!:\!\Nt$}\Comment{\textit{Gaussian elimination}}
    \For{$j\!=\!\nu\!+\!1\!:\!k\!-\!1$}
    \Comment{\textit{col index to puncture}}
        \State{$\alpha \!\gets\! \Lptilde(k,j\!+\!1)$}
        \State $\Wtilde(:,k) \!\gets\! \Wtilde(:,k) \!-\! \alpha^{\dag}\Wtilde(:,j)$
        \State $\Lptilde(k,1\!:\!j\!+\!1) \!\gets\! \Lptilde(k,1\!:\!j\!+\!1) \!-\! \alpha \Lptilde(j,1\!:\!j\!+\!1)$
    \EndFor\label{wdl_end_for_j}
    \State $\D(k,k) \!\gets\! 1 / \supsc{\fnormm{\Wtilde(:,k)}}{\,2}$
\EndFor\label{wdl_end_for_k}
\State $\yptilde \!\gets\! \Lptilde(:,1)$
\Comment{first col of augmented $\Lptilde$}
\State $\Lptilde \!\gets\! \Lptilde(:,2\!:\!\Nt\!+\!1)$
\Comment{last $\Nt$ cols of augmented $\Lptilde$}
\EndFunctionRV
\end{algorithmic}

\end{algorithm}

%
\beginnewsupplement

\section{QLZ Decomposition Algorithm}
\begin{algorithm}[hbtp]
\small
\caption{QLZ decomposition algorithm with right reduction}\label{algo:qlzd}
%
%
\begin{algorithmic}[0]
\LeftComment{Decompose $\H$ as $\H\mbf{Z}=\Q\L$ or $\H=\Q\L\supsc{\mbf{Z}}{\,-1}$}
\LeftComment{$\H\!:$ Complex $\Nr\!\times\!\Nt$ matrix, $\Nr\!\geq\!\Nt$}
\LeftComment{$\y\!:$ Complex $\Nr\!\times\!1$ column vector}
\LeftComment{$c\!:\text{reduction control parameter}$}
\LeftComment{$\Q\!:\MbyN$ matrix with orthonormal columns; $\Qh\Q=\I_{\Nt}$}
\LeftComment{$\L$: $\NbyN$ lower-triangular matrix satisfying reduction conditions}
\LeftComment{\quad$1)~\abs{\Re{L(k,j)}} \leq 2^{-(c+1/2)} L(k,k)$ for all $j,k:j < k$}
\LeftComment{\quad$2)~\abs{\Im{L(k,j)}} \leq 2^{-(c+1/2)} L(k,k)$ for all $j,k:j < k$}
\LeftComment{\quad Note: $\L(k,k)\in\mathcal{R}^+$}
\LeftComment{$\yt = \Qh\y$}
\LeftComment{$\mbf{Z}\!:$ $\NbyN$ unimodular matrix with $\det{\mbf{Z}}=1$}
\LeftComment{$\supsc{\mbf{Z}}{\,-1}\!:$ inverse of $\mbf{Z}$; $\NbyN$ unimodular matrix with $\det{\supsc{\mbf{Z}}{\,-1}}=1$}
\FunctionRV{QLZy}{\H,\y,c}{[\Q, \L, \yt, \mbf{Z}, \supsc{\mbf{Z}}{\,-1}]}
\State $\mbf{Z} \gets \I_{\Nt}^{},~ \supsc{\mbf{Z}}{\,-1} \gets \I_{\Nt}^{}$
\Comment{Gauss matrix and its inverse}
\State $[\Q, \L, \yt] \gets \hyperref[algo:qldy]{\mathsf{QLy}}(\H,\y)$
\Comment{QL-decompose}
\For{$k\!=\!2\!:\!\Nt$}\Comment{\textit{row index}}
    \For{$j\!=\!1\!:\!k\!-\!1$}\Comment{\textit{col index}}
        \State $\zeta \gets 2^{-c} \bigl\lfloor{2^{c}\tfrac{L(k,j)}{L(k,k)}}\bigr\rceil$
        \Comment{Reduction factor}
        \If{$\zeta \neq 0$}
            \State $\L(k\!:\!\Nt,j) \gets \L(k\!:\!\Nt,j) \!-\! \zeta\cdot \L(k\!:\!\Nt,k)$
            \State $\mbf{Z}(k\!:\!\Nt,j) \gets \mbf{Z}(k\!:\!\Nt,j) \!-\! \zeta\cdot \mbf{Z}(k\!:\!\Nt,k)$
            \State $\supsc{\mbf{Z}}{\,-1}(k,1\!:\!j) \gets \supsc{\mbf{Z}}{\,-1}(k,1\!:\!j) \!+\! \zeta\cdot \supsc{\mbf{Z}}{\,-1}(j,1\!:\!j)$
        \EndIf
    \EndFor\label{QLZDyendforj}
\EndFor\label{QLZDyendfori}
\EndFunctionRV
\end{algorithmic}

\end{algorithm}

\beginnewsupplement
\section{Optimized Two-Sided WLZ Decomposition Algorithm}
\begin{algorithm}[hbtp]
\small
\caption{Two-sided WLZ decomposition algorithm}\label{algo:wlzdopt}
\begin{algorithmic}[0]
\LeftComment{Generate $\W,\mbf{Z}$ such that $\Lp=\Wh\H\mbf{Z}$}
\LeftComment{If $\H\!=\!\Q\L$, then $\L_{\mathrm{z}}\!\triangleq\!\L\mbf{Z}$ satisfies the reduction conditions}
\LeftComment{\quad$1)~\abs{\Re{\L_{\mathrm{z}}(k,j)}} \leq 2^{-(c+1/2)} \L_{\mathrm{z}}(k,k)$ for all $j,k: \nu \!<\! j \!<\! k$}
\LeftComment{\quad$2)~\abs{\Im{\L_{\mathrm{z}}(k,j)}} \leq 2^{-(c+1/2)} \L_{\mathrm{z}}(k,k)$ for all $j,k: \nu \!<\! j \!<\! k$}
\LeftComment{$\H\!:$ Complex $\Nr\!\times\!\Nt$ matrix, $\Nr\!\geq\!\Nt$}
\LeftComment{$\y\!:$ Complex $\Nr\!\times\!1$ column vector}
\LeftComment{$\nu\!:\text{puncturing order}$}
\LeftComment{$c\!:\text{reduction control parameter}$}
\LeftComment{$\Lp$: $\NbyN$ punctured lower-triangular matrix; $\Lp(k,k)\!\in\!\mathcal{R}^+$}
\LeftComment{$\yp = \Wh\y$}
\LeftComment{$\W$: $\MbyN$ matrix such that $\diagg{\Wh\W} = \I_{\Nt}$}
\LeftComment{$\mbf{Z}\!:$ $\NbyN$ unimodular matrix with $\det{\mbf{Z}}=1$}
\LeftComment{$\supsc{\mbf{Z}}{\,-1}\!:$ inverse of $\mbf{Z}$; $\NbyN$ unimodular matrix with $\det{\supsc{\mbf{Z}}{\,-1}}=1$}
\LeftComment{----------------------------------------------------------------------------------------}
\LeftComment{Note: $\Wh$ punctures $\H$; in manuscript, $\Wp$ in~\eqref{eq:Z_Wp} punctures $\L$. The two matrices are related as $\Wh\!=\!\Wp\Qh$.}
\LeftComment{----------------------------------------------------------------------------------------}
\FunctionRV{WLZ}{\H,\y,\nu,c}{[\Lp, \yp, \W, \mbf{Z}, \supsc{\mbf{Z}}{\,-1}]}
\State $[\W, \Lp, \yp] \gets \hyperref[algo:qldy]{\mathsf{QLy}}(\H,\y)$
\Comment{QL-decompose; $\yp\!=\!\Wh\y$}
\For{$k\!=\!\nu\!+\!2\!:\!\Nt$}\Comment{\textit{Reduction-elimination loop}}
    \For{$j\!=\!\nu\!+\!1\!:\!k\!-\!1$}\Comment{\textit{col index}}
        \FirstLeftComment{Reduction step}
        \State $\zeta \gets 2^{-c} \bigl\lfloor{2^{c}\tfrac{L(k,j)}{L(k,k)}}\bigr\rceil$
        \Comment{Reduction factor}
        \If{$\zeta \neq 0$}
            \State $\L(k\!:\!\Nt,j) \gets \L(k\!:\!\Nt,j) \!-\! \zeta\cdot \L(k\!:\!\Nt,k)$
            \State $\mbf{Z}(k\!:\!\Nt,j) \gets \mbf{Z}(k\!:\!\Nt,j) \!-\! \zeta\cdot \mbf{Z}(k\!:\!\Nt,k)$
            \State $\supsc{\mbf{Z}}{\,-1}(k,1\!:\!j) \gets \supsc{\mbf{Z}}{\,-1}(k,1\!:\!j) \!+\! \zeta\cdot \supsc{\mbf{Z}}{\,-1}(j,1\!:\!j)$
        \EndIf
        \LeftComment{Elimination step}
        \State{$\omega \gets \Lp(k,j)/\Lp(j,j)$}
        \State $\W(:,k) \gets \W(:,k) - \omega^{\dag}\cdot\W(:,j)$
        \State $\Lp(k,1\!:\!j) \gets \Lp(k,1\!:\!j) - \omega \cdot \Lp(j,1\!:\!j)$
        \State $\yp(k) \gets \yp(k) - \omega \cdot \yp(j) $ \Comment{update $\yp$}
    \EndFor\label{wldz_end_for_j}
    \State $\Lp(k,1\!:\!k) \gets \Lp(k,1\!:\!k)/\fnormm{\W(:,k)}$
    \Comment{normalize}
    \State $\yp(k) \gets \yp(k)/\fnormm{\W(:,k)}$
    \State $\W(:,k) \gets \W(:,k) / \fnormm{\W(:,k)}$
\EndFor\label{wldz_end_for_k}
\EndFunctionRV
\end{algorithmic}

\end{algorithm}

%
\beginnewsupplement
\begin{landscape}
\addtolength{\oddsidemargin}{-0.5in}
\addtolength{\evensidemargin}{-2in}
\addtolength{\textwidth}{2in}
\addtolength{\topmargin}{0.5in}
\addtolength{\textheight}{-2in}
\newcommand*{\thead}[1]{\multicolumn{1}{c}{\bfseries #1}}
\begin{table}
\fontsize{10}{12}\selectfont
\centering
\caption{Summary of detection algorithms}
\setlength{\tabcolsep}{2pt}
\def\arraystretch{1.1}
\begin{tabular}{|c|l|l|l|l|l|l|}
  \hline
  \thead{Algorithm} &
  \thead{Decomp. Scheme} &
  \thead{Channel} &
  \thead{Metric}&
  \thead{Marginalization} &
  \thead{Tree}&
  \thead{Metric Update}
  \\\hline
  \algnameabbr~\ref{algo:wld_Lv_mul_loc_detector_algorithm} &
  1-sided \hyperref[algo:generalized_wld]{$\mathsf{WL}$} &
  \multirow{2}{*}{left-punctured} &
  \multirow{2}{*}{$\mupnoarg \!=\! -\tfrac{1}{N_0}\fnormmsq{\yp \!-\! \Lp \x}$} &
  \multirow{2}{*}{LS on leaves} &
  multi-tree: $\Nt/\nu$ trees &
  local within tree only
  \\
  \hyperref[algo:wld_Lv_mul_loc_detector_algorithm]{$\mathsf{WLdetector}$} &
  $\nu$ parents &
  &
  &
  &
  all child nodes are leaves &
  metrics not preserved with col permutations
  \\\hline
  \algnameabbr~\ref{algo:wlzd_Lv_mul_glob_detector_algorithm} &
  2-sided \hyperref[algo:wlzdopt]{$\mathsf{WLZ}$} &
  left-punctured &
  \multirow{3}{*}{$\mupznoarg \!=\! -\tfrac{1}{N_0}\fnormmsq{\yp \!-\! \Lp\Zi \x}$} &
  \multirow{3}{*}{LS on leaves} &
  multi-tree: $\Nt/\nu$ trees &
  global across all trees
  \\
  \hyperref[algo:wlzd_Lv_mul_glob_detector_algorithm]{$\mathsf{WLZdetector}$} &
  $\nu$ parents &
  right-reduced &
  &
  &
  all child nodes are leaves &
  metrics \emph{almost} preserved with col
  \\
  &
  reduction param. $c$ &
  &
  &
  &
  &
  ~~permutations as $c$ increases
  \\\hline
  \algnameabbr~\ref{algo:awdl_detector_algorithm} &
  1-sided \hyperref[algo:generalized_wdl]{$\mathsf{WDL}$} &
  augmented &
  \multirow{2}{*}{$\muapnoarg\!=\! \tfrac{1}{\Es}\fnormmsq{\x} \!-\!\subpsc{\fnormm{\yaptilde\!-\!\Laptilde\x}}{\D}{\,2}$} &
  \multirow{2}{*}{LS on leaves} &
  multi-tree: $\Nt/\nu$ trees &
  local within tree only
  \\
  \hyperref[algo:awdl_detector_algorithm]{$\mathsf{AWDLdetector}$} &
  $\nu$ parents &
  left-punctured &
  &
  &
  all child nodes are leaves &
  metrics not preserved with col permutations
  \\\hline
  \algnameabbr~\ref{algo:awdl_box_detector_algorithm} &
  1-sided \hyperref[algo:generalized_wdl]{$\mathsf{WDL}$} &
  augmented &
  \multirow{3}{*}{$\muapnoarg\!=\! \tfrac{1}{\Es}\fnormmsq{\x} \!-\!\subpsc{\fnormm{\yaptilde\!-\!\Laptilde\x}}{\D}{\,2}$} &
  enumerate on parent 1 &
  multi-tree: $2\!\times\!\Nt/2$ trees &
  global between parent tree pairs
  \\
  \hyperref[algo:awdl_box_detector_algorithm]{$\mathsf{AWDLXdetector}$} &
  fixed $\nu\!=\!2$ parents &
  left-punctured &
  &
  ZF+window on parent 2 &
  2 trees per parent pair &
  metrics not preserved across tree pairs
  \\
  &
  window size $\eta$ &
  &
  &
  ZF-DF on leaves &
  all child nodes are leaves &
  \\\hline
  \algnameabbr~\ref{algo:lord_Lv_mul_loc_detector_algorithm} &
  1-sided \hyperref[algo:qldy]{$\mathsf{QLy}$} &
  \multirow{2}{*}{true} &
  \multirow{2}{*}{$\mu \!=\! -\tfrac{1}{N_0}\fnormmsq{\ytilde \!-\! \L \x}$} &
  \multirow{2}{*}{ZF-DF on child nodes} &
  \multirow{2}{*}{multi-tree: $\Nt/\nu$ trees} &
  \multirow{2}{*}{local within tree only}
  \\
  \hyperref[algo:lord_Lv_mul_loc_detector_algorithm]{$\mathsf{LORDdetector}$} &
  $\nu$ parents &
  &
  &
  &
  &
  \\\hline
  \algnameabbr~\ref{algo:lord_Lv_mul_glob_detector_algorithm} &
  1-sided \hyperref[algo:qldy]{$\mathsf{QLy}$} &
  \multirow{2}{*}{true} &
  \multirow{2}{*}{$\mu \!=\! -\tfrac{1}{N_0}\fnormmsq{\ytilde \!-\! \L \x}$} &
  \multirow{2}{*}{ZF-DF on child nodes} &
  \multirow{2}{*}{multi-tree: $\Nt/\nu$ trees} &
  global across trees
  \\
  \hyperref[algo:lord_Lv_mul_glob_detector_algorithm]{$\mathsf{LORDXdetector}$} &
  $\nu$ parents &
  &
  &
  &
  &
  metrics preserved with col permutations
  \\\hline
\end{tabular}
\label{tab:summary_detection_algorithms}
\end{table}
\end{landscape}

%
\beginnewsupplement

\section{WLD-Based MIMO Detection Algorithm}
\begin{algorithm}[hbtp]
\small
\caption{One-sided WLD MIMO detection algorithm}\label{algo:wld_Lv_mul_loc_detector_algorithm}
\begin{algorithmic}[1]
\LeftComment{Perform soft-output MIMO detection by puncturing $\H$ using 1-sided $\mathsf{WL}()$ decomposition scheme of \algnameabbr~\ref{algo:generalized_wld}. Process $\nu$ parent layers at a time. In each run, layers are permuted so that a new group of $\nu$ symbols are chosen as parent symbols. $\Nt/\nu$ independent runs are performed. Metrics of \textbf{parent layer symbols \underline{only}} are updated in each run. This is because, for every layer ordering of $\H$, $\Wph$ changes and is not unitary. Hence Euclidean distance metrics of the form $\fnormm{\Wph(\y\!-\!\H\x)}\!=\!\fnormm{\yp \!-\! \Lp \x}$ are not preserved when the columns of $\H$ are permuted.}
\LeftComment{$\H\!:$ Complex $\Nr\!\times\!\Nt$ matrix, $\Nr\!\geq\!\Nt$}
\LeftComment{$\y\!:$ Complex $\Nr\!\times\!1$ column vector}
\LeftComment{$N_0\!:$ noise variance}
\LeftComment{$\mathcal{X}\!:$ set of $Q$ modulation constellation symbols; $\abs{\mathcal{X}}\!=\!Q\!=\!2^q$}
\LeftComment{$\nu\!:$ puncturing order (assume $\Nt$ is a multiple of $\nu$)}
\LeftComment{$\Lambda\!:q\Nt\times1$ bit LLR vector}
\LeftComment{Note: Distance computation on line~\ref{eq_algo:wld_Lv_mul_loc_detector_algorithm_distance} is expressed in this form for brevity. It can be simplified since $\Lp$ is punctured and sparse.}
\FunctionRV{WLdetector}{\H,\y,N_0,\X,\nu}{\Lambda}
\State $Q\gets \abs{\X}, q \gets \log_2 Q$
\State $\mbf{X}
        \gets
        \text{all $\nu\!\times\!1$ vectors in $\supsc{\mathcal{\X}}{\,\nu}$}$
\Comment{$\nu\!\times\!\supsc{Q}{\,\nu}~\textit{symbol matrix}$}
\State $\mbf{x}\gets\mbf{0}_{\Nt\times1}$
\Comment{$\Nt\!\times\!1$ column symbol vector}
\State $\subsc{\mu}{1}, \subsc{\mu}{0} \gets -\subsc{\bm{\infty}}{q\Nt\!\times\! 1}$
\Comment{$q\Nt\!\times\! 1$ metric vec. initialized to $-\infty$}
\For{$t \!=\! 1 \!:\! \Nt/\nu$}
\Comment{\textit{process $\nu$ parent layers at a time}}
    \State $\pi \gets [\nu(t\!-\!1)\!+\!1\!:\!\Nt,1\!:\!\nu(t\!-\!1)]$
    \Comment{\textit{col permutation}}
    \State $[\Lp, \yp, \smallsim] \!\gets\! \hyperref[algo:generalized_wld]{\mathsf{WL}}(\H(:,\pi),\y,\nu)$
    \Comment{\textit{permuted cols}}
    \For{$j \!=\! 1\!:\!\supsc{Q}{\,\nu}$}
    \Comment{\textit{loop over all} $\nu\!\times\!1$ vectors in $\supsc{\mathcal{\X}}{\,\nu}$}
        \State $\x(1\!:\!\nu) \!\gets\! \mbf{X}(1\!:\!\nu,j)$
        \Comment{$\nu$ parent layer symbols}
        \For{$i \!=\! \nu\!+\!1\!:\!\Nt$}
        \Comment{\textit{$\Nt\!-\!\nu$ child layer symbols}}
            \State $\x(i) \!\gets\!
                    \slice{
                        \frac{
                            \yp(i) - \Lp(i,1:\nu)\x(1:\nu)
                        }
                        {
                            \Lp(i,i)
                        }
                    }
                    $
            \Comment{slice}
        \EndFor\label{wld_detector_for_i}
        \State \magenta{$\mu
                \!\gets\!
                -\supsc{\fnormm{\yp \!-\! \Lp \x}}{\,2}
                $}\label{eq_algo:wld_Lv_mul_loc_detector_algorithm_distance}
        \Comment{\textit{metric using punctured $\Lp$}}
        \State  $\mbf{b} \!\gets \mathsf{binary}(\x(1\!:\!\nu))$
        \Comment{$q\nu\!\times\!1$ \textit{binary representation}}
        \For{$k \!=\! 1 \!:\!q\nu$}
        \Comment{\textit{metrics for $q\nu$ parent symbol bits}}
            \If{$\mbf{b}(k)=1$}
                \State $\!\subsc{\mu}{1}(q\nu(t\!-\!1)\!+\!k) \!\gets\! \max\{\subsc{\mu}{1}(q\nu(t\!-\!1)\!+\!k),\mu\}$
            \Else
            \State $\!\subsc{\mu}{0}(q\nu(t\!-\!1)\!+\!k) \!\gets \! \max\{\subsc{\mu}{0}(q\nu(t\!-\!1)\!+\!k),\mu\}$
            \EndIf
        \EndFor\label{wld_detector_for_k}\Comment{$k$ \textit{loop}}
    \EndFor\label{wld_detector_for_j}\Comment{$j$ \textit{loop}}
\EndFor\label{wld_detector_for_t}\Comment{$t$ \textit{loop}}
\State $\Lambda \!\gets\! (\subsc{\mu}{1} - \subsc{\mu}{0})/N_0$
\Comment{$q\Nt\!\times 1$ \textit{vector of LLRs}}
\EndFunctionRV
\end{algorithmic}


\end{algorithm}

\beginnewsupplement
\section{WLZ-Based MIMO Detection Algorithm}
\begin{algorithm}[hbtp]
\small
\caption{Two-sided WLZ MIMO detection algorithm}\label{algo:wlzd_Lv_mul_glob_detector_algorithm}
\begin{algorithmic}[1]
\LeftComment{Perform soft-output MIMO detection by puncturing $\H$ using 2-sided $\mathsf{WLZ}()$ decomposition scheme of \algnameabbr~\ref{algo:wlzdopt}. Process $\nu$ parent layers at a time. Each run detects a new group of $\nu$ symbols chosen as parent symbols. $\Nt/\nu$ runs are performed. Metrics of \textbf{all layer symbols} are updated in each run. This approximation is possible in this case because of the right reduction step by $\mbf{Z}$. For large $c$, $\Wap\Waph\approx\I$ (i.e., almost unitary), and hence distance metrics of the form $\fnormmsq{\Waph(\ya \!-\! \Ha\x)}\propto\fnormmsq{\Waph\La(\Wmmse\y \!-\! \x)}$ are almost preserved when the columns of $\H$ are permuted.}

\LeftComment{$\H\!:$ Complex $\Nr\!\times\!\Nt$ matrix, $\Nr\!\geq\!\Nt$}
\LeftComment{$\y\!:$ Complex $\Nr\!\times\!1$ column vector}
\LeftComment{$N_0\!:$ noise variance}
\LeftComment{$\mathcal{X}\!:$ set of $Q$ modulation constellation symbols; $\abs{\mathcal{X}}\!=\!Q\!=\!2^q$}
\LeftComment{$\nu\!:$ puncturing order (assume $\Nt$ is a multiple of $\nu$)}
\LeftComment{$c\!:\text{reduction control parameter}$}
\LeftComment{$\Lambda\!:q\Nt\times1$ bit LLR vector}
\LeftComment{Note: Operation $\Lp\Zi$ on line~\ref{eq_algo:wlzd_Lv_mul_glob_detector_algorithm_LpZiop} is simply integer addition and scaling operations by powers-of-2. Also, distance computation on line~\ref{eq_algo:wlzd_Lv_mul_glob_detector_algorithm_distance} is expressed in this form for brevity. It can be simplified since $\mbf{L}_{\mathrm{z}}$ is punctured and sparse.}
\FunctionRV{WLZdetector}{\H,\y,N_0,\X,\nu,c}{\Lambda}
\State $Q\gets \abs{\X}, q \gets \log_2 Q$
\State $\mbf{X}
        \gets
        \text{all $\nu\!\times\!1$ vectors in $\supsc{\mathcal{\X}}{\,\nu}$}$
\Comment{$\nu\!\times\!\supsc{Q}{\,\nu}~\textit{symbol matrix}$}
\State $\mbf{x}\gets\mbf{0}_{\Nt\times1}$
\Comment{$\Nt\!\times\!1$ column symbol vector}
\State $\subsc{\mu}{1}, \subsc{\mu}{0} \gets -\subsc{\bm{\infty}}{q\Nt\!\times\! 1}$
\Comment{$q\Nt\!\times\! 1$ metric vec. initialized to $-\infty$}
\For{$t \!=\! 1 \!:\! \Nt/\nu$}
\Comment{\textit{process $\nu$ parent layers at a time}}
    \State $\pi \gets [\nu(t\!-\!1)\!+\!1\!:\!\Nt,1\!:\!\nu(t\!-\!1)]$
    \Comment{\textit{col permutation}}
    \State $[\Lp, \yp, \smallsim, \smallsim, \Zi] \!\gets\! \hyperref[algo:wlzdopt]{\mathsf{WLZ}}(\H(:,\pi),\y,\nu,c)$
    \State \magenta{$\mbf{L}_{\mathrm{z}} \gets \Lp \Zi$}\label{eq_algo:wlzd_Lv_mul_glob_detector_algorithm_LpZiop}
    \Comment{Integer addition/scaling operations}
    \For{$j \!=\! 1\!:\!\supsc{Q}{\,\nu}$}
    \Comment{\textit{loop over all} $\nu\!\times\!1$ vectors in $\supsc{\mathcal{\X}}{\,\nu}$}
        \State $\x(1\!:\!\nu) \!\gets\! \mbf{X}(1\!:\!\nu,j)$
        \Comment{$\nu$ parent layer symbols}
        \For{$i \!=\! \nu\!+\!1\!:\!\Nt$}
        \Comment{\textit{$\Nt\!-\!\nu$ child layer symbols}}
            \State $\x(i) \!\gets\!
                    \slice{
                        \frac{
                            \yp(i) - \mbf{L}_{\mathrm{z}}(i,1:\nu)\x(1:\nu)
                        }
                        {
                            \mbf{L}_{\mathrm{z}}(i,i)
                        }
                    }
                    $
            \Comment{slice}
        \EndFor\label{wlzd_detector_for_i}
        \State \magenta{$\mu
                \!\gets\!
                -\supsc{\fnormm{\yp \!-\! \mbf{L}_{\mathrm{z}} \x}}{\,2}
                $}\label{eq_algo:wlzd_Lv_mul_glob_detector_algorithm_distance}
        \Comment{\textit{metric using punctured $\mbf{L}_{\mathrm{z}}$}}
        \State  $\mbf{b} \!\gets \mathsf{binary}(\x)$
        \Comment{$q\Nt\!\times\!1$ \textit{binary rep. of \underline{all}} $\x$}
        \For{$k \!=\! 1 \!:\!q\Nt$}
        \Comment{\textit{update metrics for all symbol bits}}
            \If{$\mbf{b}(k)=1$}
                \State $\subsc{\mu}{1}(k) \!\gets\! \max\{\subsc{\mu}{1}(k),\mu\}$
            \Else
            \State $\subsc{\mu}{0}(k) \!\gets \! \max\{\subsc{\mu}{0}(k),\mu\}$
            \EndIf
        \EndFor\label{wlzd_detector_for_k}\Comment{$k$ \textit{loop}}
    \EndFor\label{wlzd_detector_for_j}\Comment{$j$ \textit{loop}}
\EndFor\Comment{$t$ loop}\label{wlzd_detector_for_t}
\State $\Lambda \!\gets\! (\subsc{\mu}{1} - \subsc{\mu}{0})/N_0$
\Comment{$q\Nt\!\times 1$ \textit{vector of LLRs}}
\EndFunctionRV
\end{algorithmic}

\end{algorithm}

%
\beginnewsupplement

\section{AWDL MIMO Detection Algorithms}
\begin{algorithm}[hbtp]
\small
\caption{AWDL MIMO detection algorithm}\label{algo:awdl_detector_algorithm}
\begin{algorithmic}[1]
\LeftComment{Perform soft-output MIMO detection by puncturing the augmented matrix $\Ha$ using 1-sided square-root-free \hyperref[algo:generalized_wdl]{$\mathsf{WDL}()$} decomposition scheme of \algnameabbr~\ref{algo:generalized_wdl}. Process $\nu$ parent layers at a time. In each run, layers are permuted so that a new group of $\nu$ symbols are chosen as parent symbols. $\Nt/\nu$ independent runs are performed. Metrics of \textbf{parent layer symbols \underline{only}} are updated in each run. This is because, for every layer ordering of $\Ha$, the puncturing matrix changes and is not unitary. Hence the required metrics are not preserved when the columns of $\H$ are permuted.}
\LeftComment{$\H\!:$ Complex $\Nr\!\times\!\Nt$ matrix, $\Nr\!\geq\!\Nt$}
\LeftComment{$\y\!:$ Complex $\Nr\!\times\!1$ column vector}
\LeftComment{$N_0\!:$ noise variance}
\LeftComment{$\mathcal{X}\!:$ set of $Q$ modulation constellation symbols; $\abs{\mathcal{X}}\!=\!Q\!=\!2^q$}
\LeftComment{$\nu\!:$ puncturing order (assume $\Nt$ is a multiple of $\nu$)}
\LeftComment{$\Lambda\!:q\Nt\times1$ bit LLR vector}
\LeftComment{Note: Metric computation on line~\ref{eq_algo:awdl_detector_algorithm_metric} is expressed in this form for brevity. It can be simplified since $\Laptilde$ is punctured and sparse.}
\FunctionRV{AWDLdetector}{\H,\y,N_0,\X,\nu}{\Lambda}
\State $Q\gets \abs{\X}, ~q \gets \log_2 Q$
\State $\Es\gets \tfrac{1}{Q}\sum_{x\in\mathcal{X}}\supsc{\abs{x}}{\,2}$
\Comment{Avg. symbol energy}
\State $\mbf{X}
        \gets
        \text{all $\nu\!\times\!1$ vectors in $\supsc{\mathcal{\X}}{\,\nu}$}
        $
\Comment{$\nu\!\times\!\supsc{Q}{\,\nu}~\textit{matrix of symbols}$}
\State $\mbf{x}\gets\mbf{0}_{\Nt\times1}$
\Comment{$\Nt\!\times\!1$ column symbol vector}
\State $\subsc{\mu}{1}, \subsc{\mu}{0} \gets -\subsc{\bm{\infty}}{q\Nt \!\times\! 1}$
\Comment{$q\Nt\!\times\! 1$ metric vec. initialized to $-\infty$}
\State  $\Ha \!\gets \!
            \begin{bNiceMatrix}
                \tfrac{1}{\sqrt{N_0}}\H  \\
                \tfrac{1}{\sqrt{\Es}}\subsc{\I}{\Nt}
            \end{bNiceMatrix}
        $,
        ~
        $\ya \!\gets\! \tfrac{1}{\sqrt{N_0}}\!
            \begin{bNiceMatrix}
                \y  \\
                \subsc{\mbf{0}}{\Nt\times 1}
            \end{bNiceMatrix}
        $
\Comment{augmented $\Ha,\ya$}
\For{$t \!=\! 1 \!:\! \Nt/\nu$}
\Comment{\textit{process $\nu$ parent layers at a time}}
    \State $\pi \gets [\nu(t\!-\!1)\!+\!1\!:\!\Nt,1\!:\!\nu(t\!-\!1)]$
    \Comment{\textit{column permutation}}
    \State $[\Laptilde, \yaptilde, \D] \!\gets\! \hyperref[algo:generalized_wdl]{\mathsf{WDL}}(\Ha(:,\pi),\ya,\nu)$
    \Comment{\textit{permuted cols}}
    \For{$j \!=\! 1\!:\!\supsc{Q}{\,\nu}$}
    \Comment{\textit{loop over all} $\nu\!\times\!1$ vectors in $\supsc{\mathcal{\X}}{\,\nu}$}
        \State $\x(1\!:\!\nu) \!\gets\! \mbf{X}(1\!:\!\nu,j)$
        \Comment{$\nu$ parent layer symbols}
        \For{$i \!=\! \nu\!+\!1\!:\!\Nt$}
        \Comment{\textit{$\Nt\!-\!\nu$ child layer symbols}}
            \State $\x(i) \!\gets\!
                    \slice{
                        \frac{
                            \yaptilde(i) - \Laptilde(i,1:\nu)\x(1:\nu)
                        }
                        {
                            1 - 1/(\Es\D(i,i))
                        }
                    }
                    $
            \Comment{slice}
        \EndFor\label{awdl_for_i}
        \State \magenta{$\mu
                \!\gets\!
                \tfrac{1}{\Es} \supsc{\normm{\x}}{\,2}
                \!-\!
                \supsc{(\yaptilde \!-\! \Laptilde \x)}{\dag}\D(\yaptilde \!-\! \Laptilde \x)
                $}\label{eq_algo:awdl_detector_algorithm_metric}
        \Comment{\textit{metric}}
        \State  $\mbf{b} \!\gets \mathsf{binary}(\x(1\!:\!\nu))$
        \Comment{$q\nu\!\times\!1$ \textit{binary representation}}
        \For{$k \!=\! 1 \!:\!q\nu$}
        \Comment{\textit{metrics for $q\nu$ parent symbol bits}}
            \If{$\mbf{b}(k)=1$}
                \State $\subsc{\mu}{1}(q\nu(t\!-\!1)\!+\!k) \!\gets\! \max\{\subsc{\mu}{1}(q\nu(t\!-\!1)\!+\!k),\mu\}$
            \Else
            \State $\subsc{\mu}{0}(q\nu(t\!-\!1)\!+\!k) \!\gets \! \max\{\subsc{\mu}{0}(q\nu(t\!-\!1)\!+\!k),\mu\}$
            \EndIf
        \EndFor\label{awdl_for_k}\Comment{$k$ \textit{loop}}
    \EndFor\label{awdl_for_j}\Comment{$j$ \textit{loop}}
\EndFor\label{awdl_for_t}\Comment{$t$ \textit{loop}}
\State $\Lambda \!\gets\! \subsc{\mu}{1} - \subsc{\mu}{0}$
\Comment{$q\Nt\!\times 1$ \textit{vector of LLRs}}
\EndFunctionRV
\end{algorithmic}


\end{algorithm}

\beginnewsupplement
\section{AWDL-BOX MIMO Detection Algorithms}\vspace{0in}
\begin{algorithm}[hbtp]
\fontsize{10}{9.65}\selectfont
\caption{AWDL-BOX MIMO detection algorithm}\label{algo:awdl_box_detector_algorithm}
\begin{algorithmic}[1]
\LeftComment{Optimized version of \hyperref[algo:awdl_detector_algorithm]{$\mathsf{AWDLdetector}$} for $\nu\!=\!2$. Process $2$ parent layers at a time, by enumerating over parent 1 and doing ZF-DF for parent 2. The search for parent 2 is expanded to a window of size $\eta$ around the ZF solution. The parents are switched and the process is repeated for a second run. In each pair of runs, a new pair symbols is chosen as parents. $\Nt/2$ pairs of runs are performed. Metrics of \textbf{all symbols} are updated in each pair of runs. This is because metrics are preserved if parent layers are permuted and child layers are permuted independently, but metrics are not preserved for col arbitrary permutations.}
\LeftComment{$\H\!:$ Complex $\Nr\!\times\!\Nt$ matrix, $\Nr\!\geq\!\Nt$}
\LeftComment{$\y\!:$ Complex $\Nr\!\times\!1$ column vector}
\LeftComment{$N_0\!:$ noise variance}
\LeftComment{$\mathcal{X}\!:$ set of $Q$ modulation constellation symbols; $\abs{\mathcal{X}}\!=\!Q\!=\!2^q$}
\LeftComment{$\eta\!:$ window size around ZF solution for parent 2}
\LeftComment{$\Lambda\!:q\Nt\times1$ bit LLR vector}
\LeftComment{Note: WDL decomposition on line~\ref{eq_algo:awdl_box_detector_algorithm_WDL} can be optimized for each pair of runs since right-most $\Nt\!-\!2$ cols of $\Laptilde$ do not change.}
\LeftComment{Note: Metric computation on line~\ref{eq_algo:awdl_box_detector_algorithm_metric} is expressed in this form for brevity. It can be simplified since $\Laptilde$ is punctured and sparse.}
\FunctionRV{AWDLXdetector}{\H,\y,N_0,\X,\blue{\eta}}{\Lambda}
\State $Q\gets \abs{\X}, ~q \gets \log_2 Q, ~\Es\gets \tfrac{1}{Q}\sum_{x\in\mathcal{X}}\supsc{\abs{x}}{\,2}$
\State $\mbf{x}\gets\mbf{0}_{\Nt\times1}$
\Comment{$\Nt\!\times\!1$ column symbol vector}
\State $\subsc{\mu}{1}, \subsc{\mu}{0} \gets -\subsc{\bm{\infty}}{q\Nt\!\times\! 1}$
\Comment{$q\Nt\!\times\! 1$ metric vec. initialized to $-\infty$}
\State  $\Ha \!\gets \!
            \begin{bNiceMatrix}
                \tfrac{1}{\sqrt{N_0}}\H  \\
                \tfrac{1}{\sqrt{\Es}}\subsc{\I}{\Nt}
            \end{bNiceMatrix}
        $,
        ~
        $\ya \!\gets\! \tfrac{1}{\sqrt{N_0}}\!\!
            \begin{bNiceMatrix}
                \y  \\
                \subsc{\mbf{0}}{\Nt\times 1}
            \end{bNiceMatrix}
        $
\For{$t \!=\! 1 \!:\! \Nt/2$}
\Comment{\textit{process $2$ parent layers at a time}}
    \For{$p \!=\! 1\!:\!2$}
    \Comment{parent layers order: $[1,2]$ or $[2,1]$}
        \State $\pi \gets [2t\!-\!2\!+\!p, 2t\!-\!p\!+\!1, 2t\!+\!1 \!:\! \Nt, 1\!:\!2t\!-\!2]$
        \State $[\Laptilde, \yaptilde, \D] \!\gets\! \hyperref[algo:generalized_wdl]{\mathsf{WDL}}(\Ha(:,\pi),\ya,2)$
        \label{eq_algo:awdl_box_detector_algorithm_WDL}
        \Comment{$\nu\!=\!2$}
        \For{$j \!=\! 1\!:\!Q$}
        \Comment{loop over all symbols in $\mathcal{\X}$}
            \State $\x(1) \!\gets\! \mathcal{X}(j)$
            \Comment{parent layer symbol}
            \State $z \!\gets\!
                        \slice{
                            \frac{
                                \yaptilde(2) - \Laptilde(2,1)\mbf{x}(1)
                            }
                            {
                                1 - 1/(\Es\D(2,2))
                            }
                        }
                    $
            \Comment{slice layer $2$}
            \State $\mathcal{W}(z)
                \gets
                \text{$\eta$ closest symbols in $\mathcal{X}$ to $z$}
                $
            \ForAll{$\omega \!\in\! \mathcal{W}(z)$}
            \Comment{$\eta$ closest symbols to $z$}
                \State $\x(2) \!\gets\! \omega$
                \Comment{set as layer 2 symbol}
                \For{$i \!=\! 3\!:\!\Nt$}
                \Comment{\textit{$\Nt\!-\!2$ child layer symbols}}
                    \State $\x(i) \!\gets\!
                            \slice{
                                \frac{
                                    \yaptilde(i) - \Laptilde(i,1:2)\x(1:2)
                                }
                                {
                                    1 - 1/(\Es\D(i,i))
                                }
                            }
                            $
                    \Comment{slice}
                        \State
                            \magenta{$\mu
                            \!\gets\!
                            \tfrac{1}{\Es}\fnormmsq{\x} \!-\!\subpsc{\fnormm{\yaptilde\!-\!\Laptilde\x}}{\D}{\,2}
                            $}\label{eq_algo:awdl_box_detector_algorithm_metric}
                    \Comment{\textit{metric}}
                    \State  $\mbf{b} \!\gets\! \mathsf{binary}(\x(1\!:\!2,1))$
                    \Comment{binary repres.}
                    \For{$k \!=\! 1 \!:\!2q$}
                    \Comment{parent bits metrics}
                        \State $r \!\gets\! (k\!-\!1\!+\!(p\!-\!1)q)\%(2q) \!+\! 1$
                        \Comment{index}
                        \If{$\mbf{b}(k)=1$}
                            \State $\subsc{\mu}{1}(2q(t\!-\!1)\!+\!r) \!\gets\!$
                            \State $\qquad\max\{\subsc{\mu}{1}(2q(t\!-\!1)\!+\!r),\mu\}$
                        \Else
                        \State $\subsc{\mu}{0}(2q(t\!-\!1)\!+\!r) \!\gets \!$
                        \State $\qquad\max\{\subsc{\mu}{0}(2q(t\!-\!1)\!+\!r),\mu\}$
                        \EndIf
                    \EndFor\label{awdlx_for_k}\Comment{$k$ \textit{loop}}
                \EndFor\label{awdlx_for_i}\Comment{$i$ \textit{loop}}
            \EndFor\label{awdlx_for_omega}\Comment{$\omega$ \textit{loop}}
        \EndFor\label{awdlx_for_j}\Comment{$j$ \textit{loop}}
    \EndFor\label{awdlx_for_p}\Comment{$p$ \textit{loop}}
\EndFor\label{awdlx_for_t}\Comment{$t$ \textit{loop}}
\State $\Lambda \!\gets\! \subsc{\mu}{1} - \subsc{\mu}{0}$
\Comment{$q\Nt\!\times 1$ \textit{vector of LLRs}}
\EndFunctionRV
\end{algorithmic}


\end{algorithm}

%
\beginnewsupplement

\section{LORD MIMO Detection Algorithm}
\begin{algorithm}[hbtp]
\small
\caption{LORD MIMO detection algorithm}\label{algo:lord_Lv_mul_loc_detector_algorithm}
\begin{algorithmic}[1]
\LeftComment{LORD soft-output MIMO detection using $\mathsf{QLy}()$ decomposition scheme of \algnameabbr~\ref{algo:qldy}. Process $\nu$ parent layers at a time. In each run, layers are permuted so that a new group of $\nu$ symbols are chosen as parent symbols. $\Nt/\nu$ independent runs are performed. Metrics of \textbf{parent layer symbols only} are updated in each run.}
\LeftComment{$\H\!:$ Complex $\Nr\!\times\!\Nt$ matrix, $\Nr\!\geq\!\Nt$}
\LeftComment{$\y\!:$ Complex $\Nr\!\times\!1$ column vector}
\LeftComment{$N_0\!:$ noise variance}
\LeftComment{$\mathcal{X}\!:$ set of $Q$ modulation constellation symbols; $\abs{\mathcal{X}}\!=\!Q\!=\!2^q$}
\LeftComment{$\nu\!:$ puncturing order (assume $\Nt$ is a multiple of $\nu$)}
\LeftComment{$\Lambda\!:q\Nt\times1$ bit LLR vector}
\LeftComment{Note: Distance computation on line~\ref{eq_algo:lord_Lv_mul_loc_detector_algorithm_distance} is expressed in this form for brevity. It can be simplified since $\L$ is lower-triangular.}
\FunctionRV{LORDdetector}{\H,\y,N_0,\X,\nu}{\Lambda}
\State $Q\gets \abs{\X}, q \gets \log_2 Q$
\State $\mbf{X}
        \gets
        \text{all $\nu\!\times\!1$ vectors in $\supsc{\mathcal{\X}}{\,\nu}$}
        $
\Comment{$\nu\!\times\!\supsc{Q}{\,\nu}~\textit{matrix of symbols}$}
\State $\mbf{x}\gets\mbf{0}_{\Nt\times1}$
\Comment{$\Nt\!\times\!1$ column symbol vector}
\State $\subsc{\mu}{1}, \subsc{\mu}{0} \gets -\subsc{\bm{\infty}}{q\Nt\!\times\! 1}$
\Comment{$q\Nt\!\times\! 1$ metric vec. initialized to $-\infty$}
\For{$t \!=\! 1 \!:\! \Nt/\nu$}
\Comment{\textit{process $\nu$ parent layers at a time}}
    \State $\pi \gets [\nu(t\!-\!1)\!+\!1\!:\!\Nt,1\!:\!\nu(t\!-\!1)]$
    \Comment{\textit{column permutation}}
    \State $[\smallsim, \L, \ytilde] \!\gets\! \hyperref[algo:qldy]{\mathsf{QLy}}(\H(:,\pi),\y,\nu)$
    \Comment{\textit{permuted cols}}
    \For{$j \!=\! 1\!:\!\supsc{Q}{\,\nu}$}
    \Comment{\textit{loop over all} $\nu\!\times\!1$ vectors in $\supsc{\mathcal{\X}}{\,\nu}$}
        \State $\x(1\!:\!\nu) \!\gets\! \mbf{X}(1\!:\!\nu,j)$
        \Comment{$\nu$ parent layer symbols}
        \For{$i \!=\! \nu\!+\!1\!:\!\Nt$}
        \Comment{\textit{$\Nt\!-\!\nu$ child layer symbols}}
            \State $\x(i) \!\gets\!
                    \slice{
                        \frac{
                            \yt(i) - \L(i,1:\nu)\x(1:\nu)
                        }
                        {
                            \L(i,i)
                        }
                    }
                    $
            \Comment{slice}
        \EndFor\label{lord_for_i}
        \State $\mu
                \!\gets\!
                -\supsc{\fnormm{\yt \!-\! \L \x}}{\,2}
                $\label{eq_algo:lord_Lv_mul_loc_detector_algorithm_distance}
        \Comment{\textit{metric using full $\L$}}
        \State  $\mbf{b} \!\gets \mathsf{binary}(\x(1\!:\!\nu))$
        \Comment{$q\nu\!\times\!1$ \textit{binary representation}}
        \For{$k \!=\! 1 \!:\!q\nu$}
        \Comment{\textit{metrics for $q\nu$ parent symbol bits}}
            \If{$\mbf{b}(k)=1$}
                \State $\!\subsc{\mu}{1}(q\nu(t\!-\!1)\!+\!k) \!\gets\! \max\{\subsc{\mu}{1}(q\nu(t\!-\!1)\!+\!k),\mu\}$
            \Else
            \State $\!\subsc{\mu}{0}(q\nu(t\!-\!1)\!+\!k) \!\gets \! \max\{\subsc{\mu}{0}(q\nu(t\!-\!1)\!+\!k),\mu\}$
            \EndIf
        \EndFor\label{lord_for_k}\Comment{$k$ \textit{loop}}
    \EndFor\label{lord_for_j}\Comment{$j$ \textit{loop}}
\EndFor\label{lord_for_t}\Comment{$t$ \textit{loop}}
\State $\Lambda \!\gets\! (\subsc{\mu}{1} - \subsc{\mu}{0})/N_0$
\Comment{$q\Nt\!\times 1$ \textit{vector of LLRs}}
\EndFunctionRV
\end{algorithmic}


\end{algorithm}

\beginnewsupplement
\section{Optimized LORD MIMO Detection Algorithm}
\begin{algorithm}[hbtp]
\small
\caption{Optimized LORD MIMO detection algorithm}\label{algo:lord_Lv_mul_glob_detector_algorithm}
\begin{algorithmic}[1]
\LeftComment{Optimized version of $\mathsf{LORDdetector}$ in \algnameabbr~\ref{algo:lord_Lv_mul_loc_detector_algorithm} to globally update metrics in each run. Process $\nu$ parent layers at a time. In each run, layers are permuted so that a new group of $\nu$ symbols are chosen as parent symbols. $\Nt/\nu$ independent runs are performed. Metrics of \textbf{all layer symbols} are updated in each run. This is possible because Euclidean distance metrics do not change under column permutation of $\H$.}
\LeftComment{$\H\!:$ Complex $\Nr\!\times\!\Nt$ matrix, $\Nr\!\geq\!\Nt$}
\LeftComment{$\y\!:$ Complex $\Nr\!\times\!1$ column vector}
\LeftComment{$N_0\!:$ noise variance}
\LeftComment{$\mathcal{X}\!:$ set of $Q$ modulation constellation symbols; $\abs{\mathcal{X}}\!=\!Q\!=\!2^q$}
\LeftComment{$\nu\!:$ puncturing order (assume $\Nt$ is a multiple of $\nu$)}
\LeftComment{$\Lambda\!:q\Nt\times1$ bit LLR vector}
\LeftComment{Note: Distance computation on line~\ref{eq_algo:lord_Lv_mul_glob_detector_algorithm_distance} is expressed in this form for brevity. It can be simplified since $\L$ is lower-triangular.}
\FunctionRV{LORDXdetector}{\H,\y,N_0,\X,\nu}{\Lambda}
\State $Q\gets \abs{\X}, q \gets \log_2 Q$
\State $\mbf{X}
        \gets
        \text{all $\nu\!\times\!1$ vectors in $\supsc{\mathcal{\X}}{\,\nu}$}
        $
\Comment{$\nu\!\times\!\supsc{Q}{\,\nu}~\textit{matrix of symbols}$}
\State $\mbf{x}\gets\mbf{0}_{\Nt\times1}$
\Comment{$\Nt\!\times\!1$ column symbol vector}
\State $\subsc{\mu}{1}, \subsc{\mu}{0} \gets -\subsc{\bm{\infty}}{q\Nt\!\times\! 1}$
\Comment{$q\Nt\!\times\! 1$ metric vec. initialized to $-\infty$}
\For{$t \!=\! 1 \!:\! \Nt/\nu$}
\Comment{\textit{process $\nu$ parent layers at a time}}
    \State $\pi \gets [\nu(t\!-\!1)\!+\!1\!:\!\Nt,1\!:\!\nu(t\!-\!1)]$
    \Comment{\textit{column permutation}}
    \State $[\smallsim, \L, \ytilde] \!\gets\! \hyperref[algo:qldy]{\mathsf{QLy}}(\H(:,\pi),\y,\nu)$
    \Comment{\textit{permuted cols}}
    \For{$j \!=\! 1\!:\!\supsc{Q}{\,\nu}$}
    \Comment{\textit{loop over all} $\nu\!\times\!1$ vectors in $\supsc{\mathcal{\X}}{\,\nu}$}
        \State $\x(1\!:\!\nu) \!\gets\! \mbf{X}(1\!:\!\nu,j)$
        \Comment{$\nu$ parent layer symbols}
        \For{$i \!=\! \nu\!+\!1\!:\!\Nt$}
        \Comment{\textit{$\Nt\!-\!\nu$ child layer symbols}}
            \State $\x(i) \!\gets\!
                    \slice{
                        \frac{
                            \yt(i) - \L(i,1:\nu)\x(1:\nu)
                        }
                        {
                            \L(i,i)
                        }
                    }
                    $
            \Comment{slice}
        \EndFor\label{lordx_for_i}
        \State $\mu
                \!\gets\!
                -\supsc{\fnormm{\yt \!-\! \L \x}}{\,2}
                $\label{eq_algo:lord_Lv_mul_glob_detector_algorithm_distance}
        \Comment{\textit{metric using full $\L$}}
        \State  $\mbf{b} \!\gets \mathsf{binary}(\x)$
        \Comment{$q\Nt\!\times\!1$ \textit{binary rep. of \underline{all}} $\x$}
        \For{$k \!=\! 1 \!:\!q\Nt$}
        \Comment{\textit{update metrics for all symbol bits}}
            \If{$\mbf{b}(k)=1$}
                \State $\subsc{\mu}{1}(k) \!\gets\! \max\{\subsc{\mu}{1}(k),\mu\}$
            \Else
            \State $\subsc{\mu}{0}(k) \!\gets \! \max\{\subsc{\mu}{0}(k),\mu\}$
            \EndIf
        \EndFor\label{lordx_for_k}\Comment{$k$ \textit{loop}}
    \EndFor\label{lordx_for_j}\Comment{$j$ \textit{loop}}
\EndFor\Comment{$t$ loop}\label{lordx_for_t}
\State $\Lambda \!\gets\! (\subsc{\mu}{1} - \subsc{\mu}{0})/N_0$
\Comment{$q\Nt\!\times 1$ \textit{vector of LLRs}}
\EndFunctionRV
\end{algorithmic}


\end{algorithm}

%
\beginnewsupplement
\begin{landscape}
\begin{figure}
  \centering
  \vspace{-0.25in}
  \includegraphics[scale=1.6]{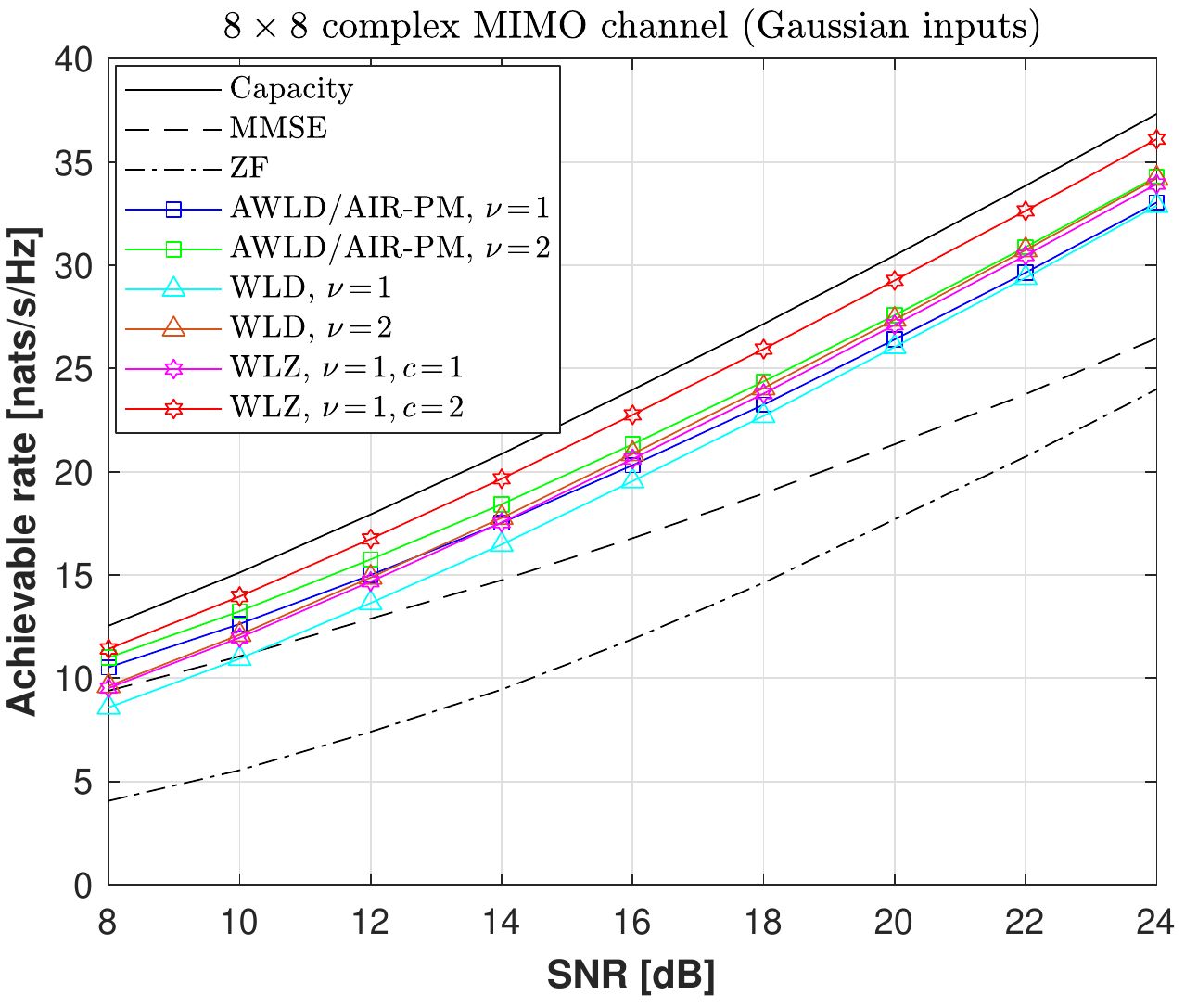}
  \vspace{-0.1in}
  \caption{\normalsize {Comparison of AIRs for $8\!\times\! 8$ MIMO channels with Gaussian inputs. For the AWLD, WLD, and WLZ algorithms, parent layers are optimally selected so as to maximize $\ILBWLD$ in~\eqref{eq:ILB_WLD_expression}.}}
  \label{figsup:achievable_rates_opt_8x8}
\end{figure}
\end{landscape}

%
\beginnewsupplement
\begin{landscape}
\begin{figure}
  \centering
  \vspace{-0.25in}
  \includegraphics[scale=1.6]{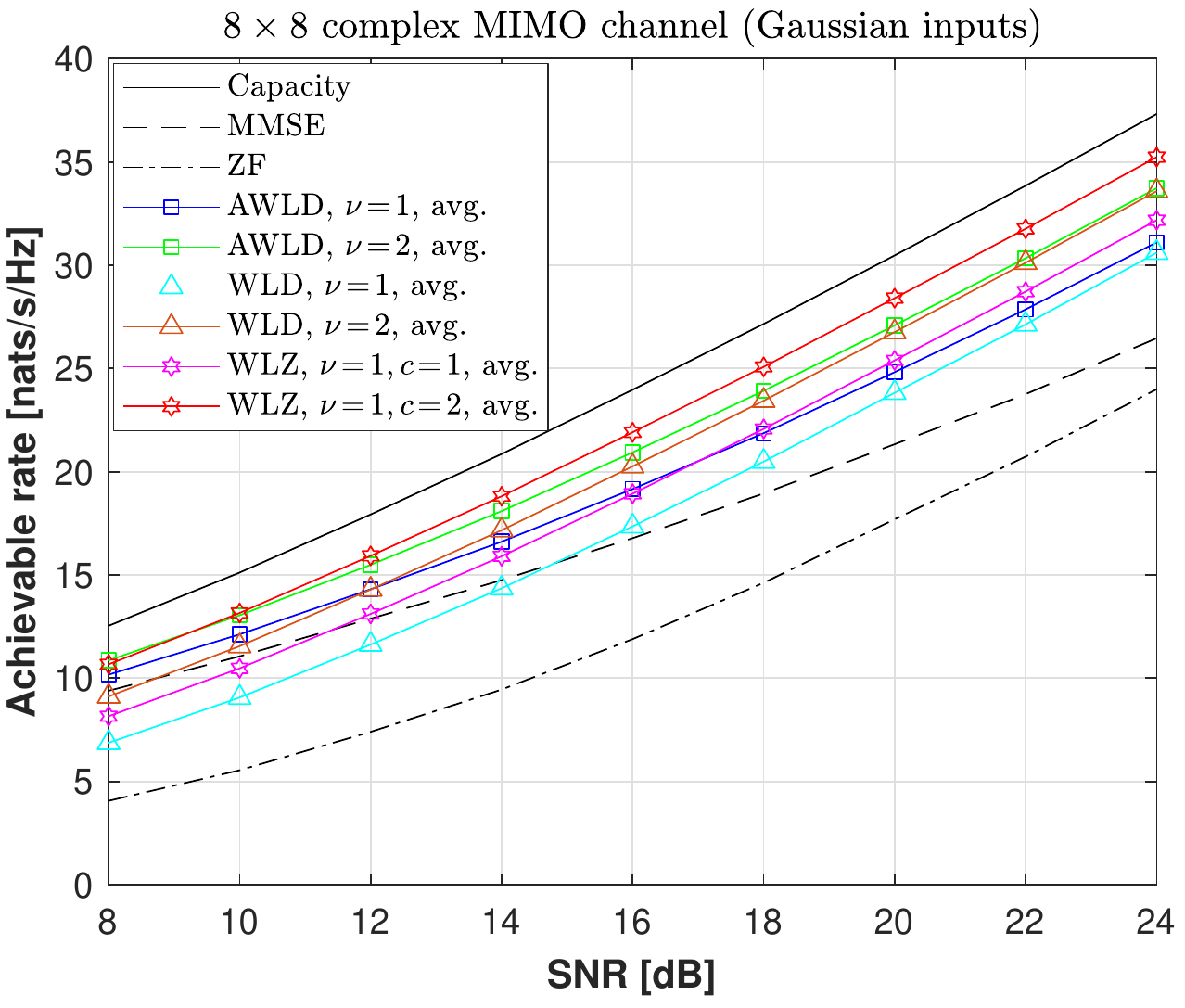}
  \vspace{-0.1in}
  \caption{\normalsize {Comparison of AIRs for $8\!\times\! 8$ MIMO channels with Gaussian inputs. The AIRs for the AWLD, WLD, and WLZ algorithms are averaged over all possible parent layer selections.}}
  \label{figsup:achievable_rates_avg_8x8}
\end{figure}
\end{landscape}

%
\beginnewsupplement
\begin{landscape}
\begin{figure}
  \centering
  \vspace{-0.25in}
  \includegraphics[scale=1.6]{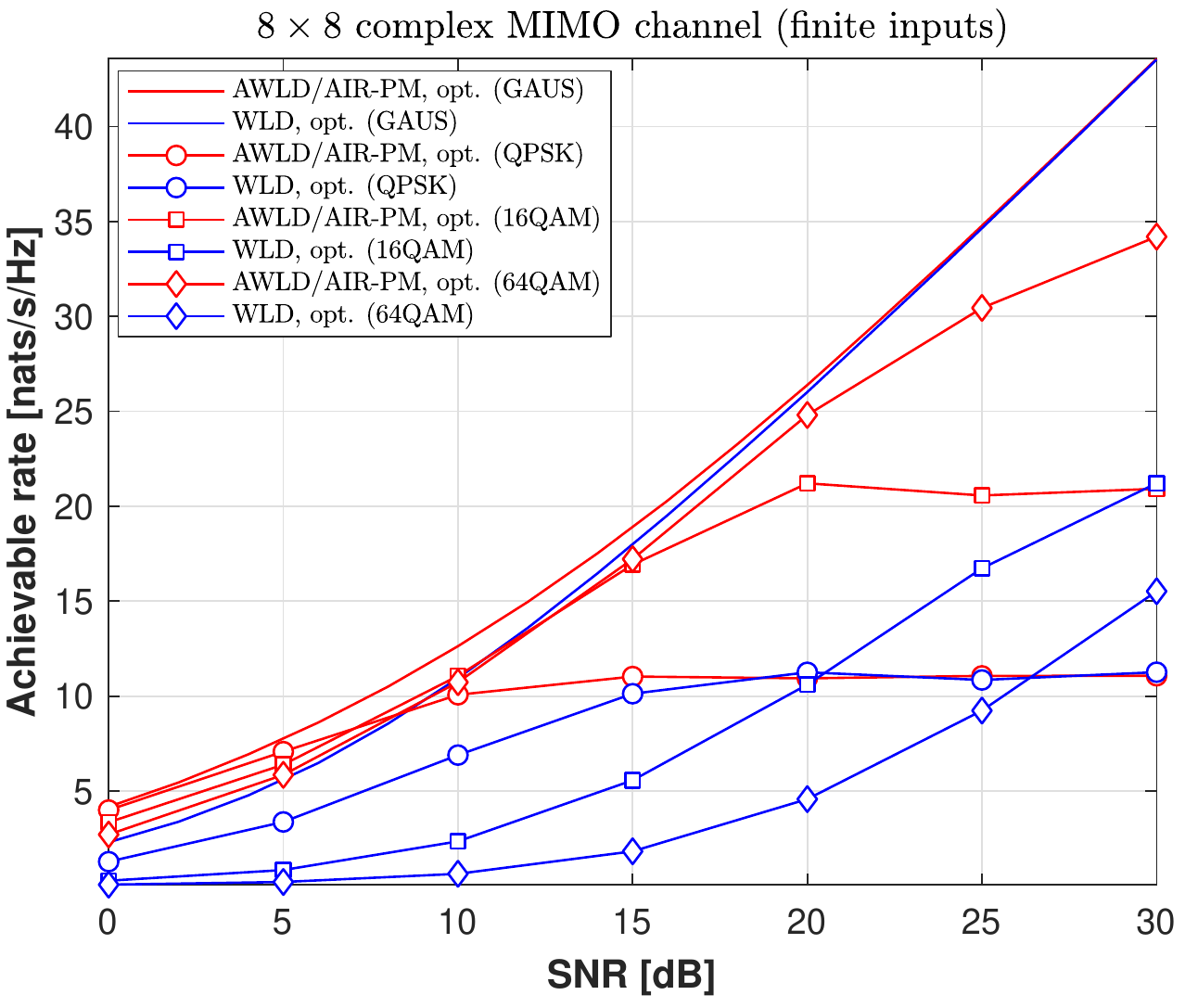}
  \vspace{-0.1in}
  \caption{\normalsize {Comparison of AIRs for $8\!\times\! 8$ MIMO channels with finite inputs. For the AWLD, WLD, and WLZ algorithms with QPSK, 16QAM, and 64QAM inputs, parent layers are selected so as to maximize $\ILBWLD$ in~\eqref{eq:ILB_WLD_expression} if Gaussian inputs were assumed.}}
  \label{figsup:finite_inputs_rates_opt_8x8}
\end{figure}
\end{landscape}

%
\beginnewsupplement
\begin{landscape}
\begin{figure}
  \centering
  \vspace{-0.25in}
  \includegraphics[scale=1.6]{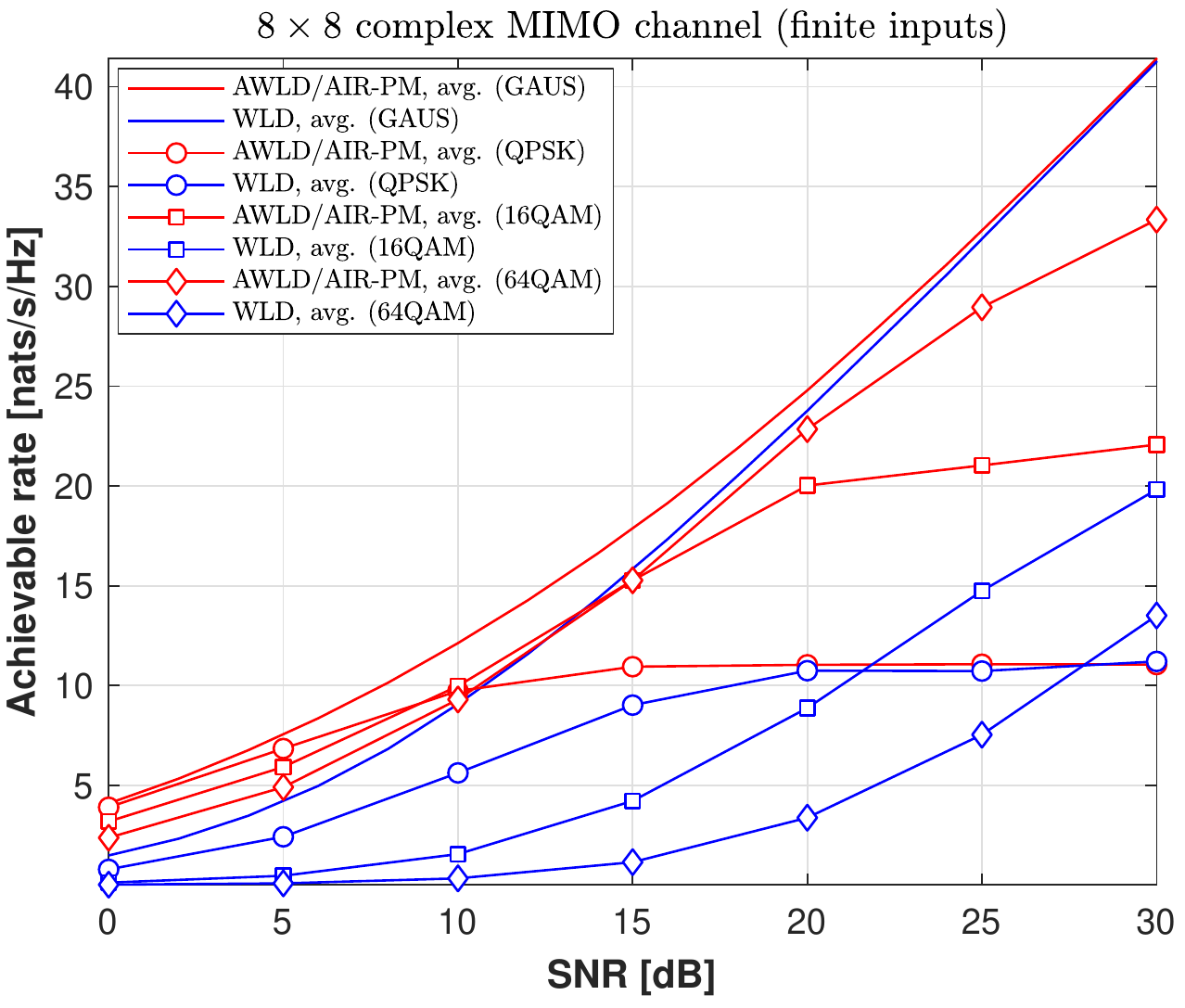}
  \vspace{-0.1in}
  \caption{\normalsize {Comparison of AIRs for $8\!\times\! 8$ MIMO channels with finite inputs. The AIRs for the AWLD, WLD, and WLZ algorithms are averaged over all possible parent layer selections.}}
  \label{figsup:finite_inputs_rates_avg_8x8}
\end{figure}
\end{landscape}

%
\beginnewsupplement
\begin{landscape}
\begin{figure}
  \centering
  \includegraphics[scale=1.6]{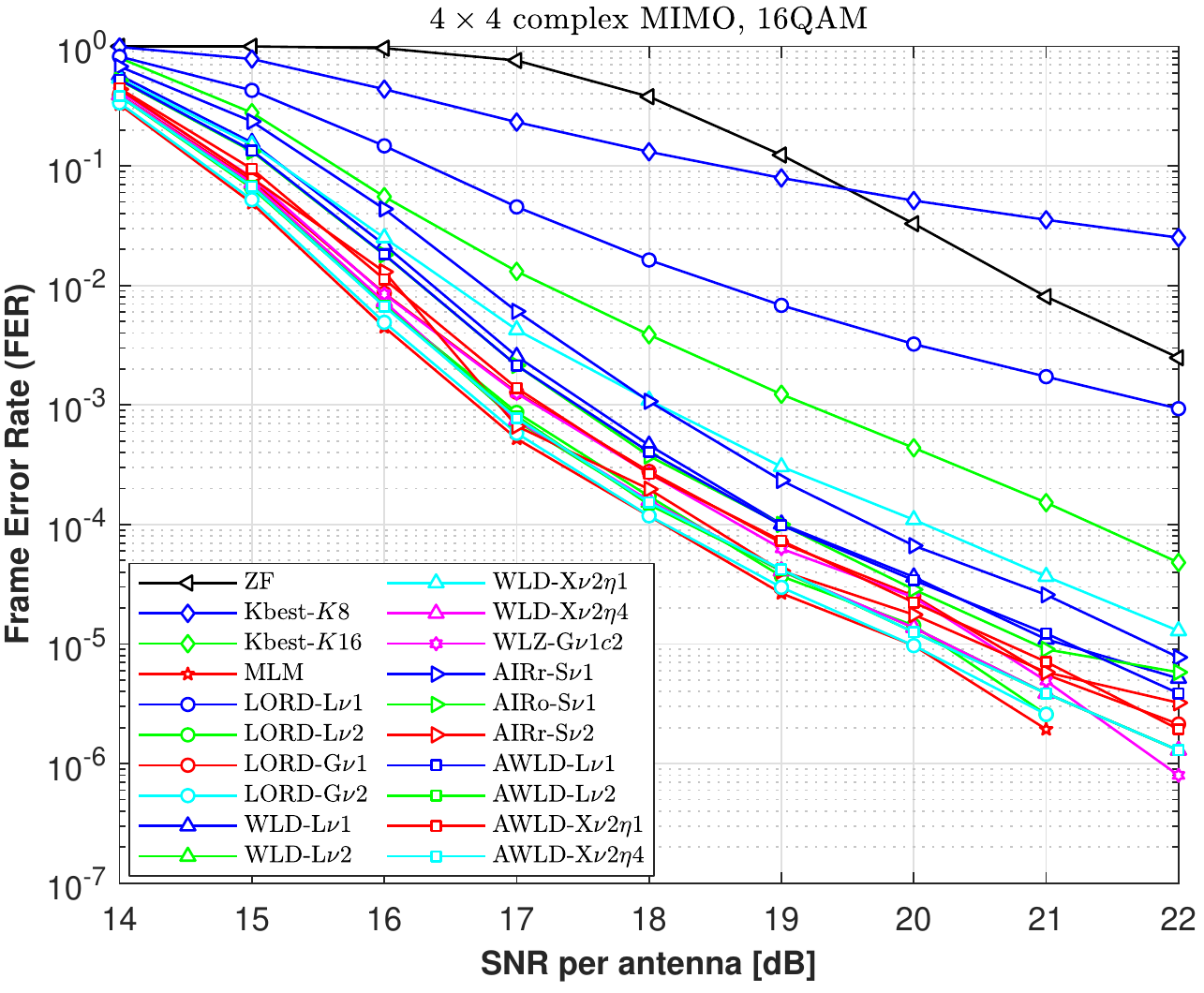}
  \caption{\normalsize Frame error-rate of $4\!\times\!4$ complex MIMO channels, 16QAM}
  \label{figsup:fer_plot_4x4_16QAM}
\end{figure}
\end{landscape}

\beginnewsupplement
\begin{landscape}
\begin{figure}[hbtp]
  \centering
  \includegraphics[scale=1.6]{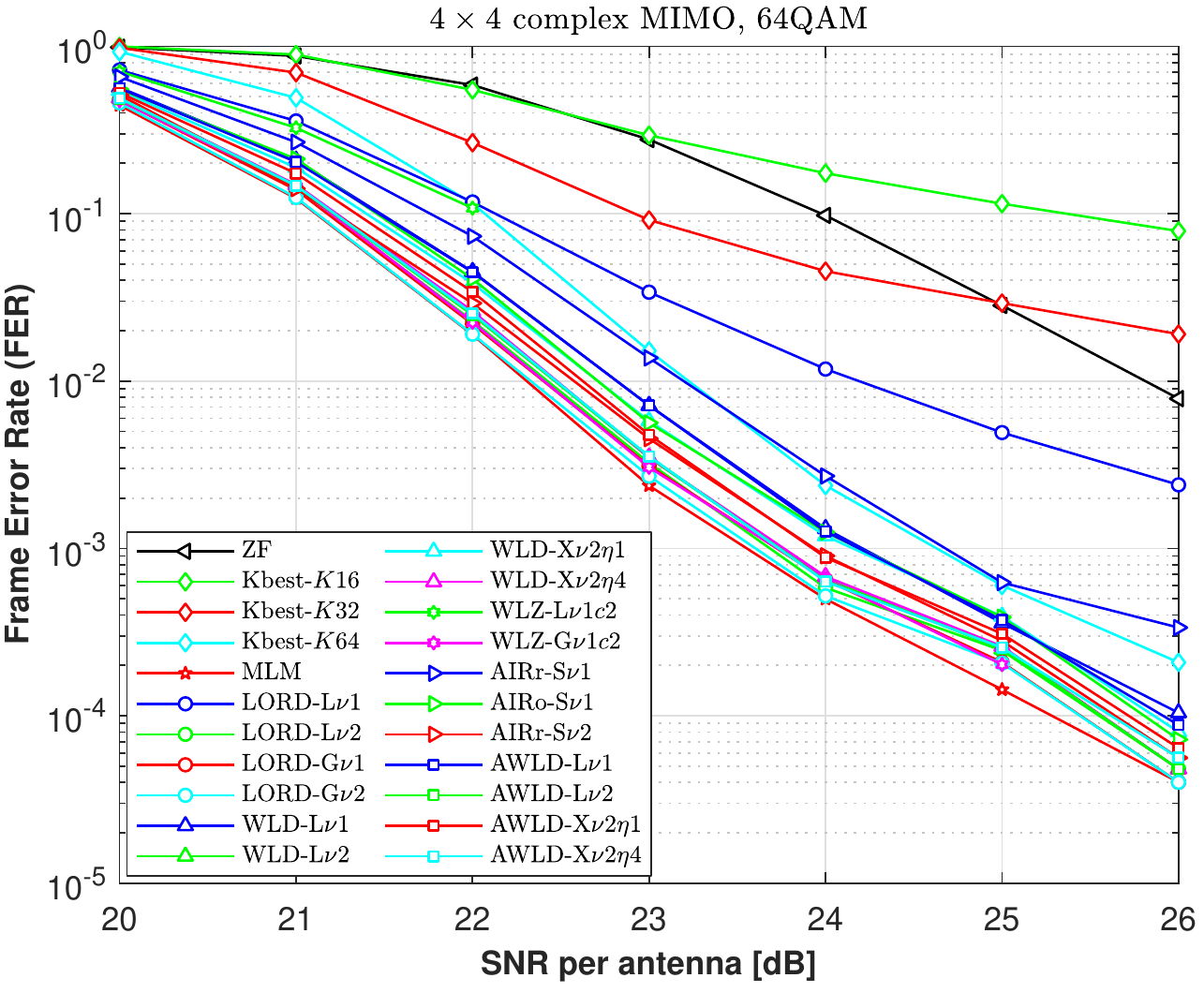}
  \caption{\normalsize Frame error-rate of $4\!\times\!4$ complex MIMO channels, 64QAM}\label{figsup:fer_plot_4x4_64QAM}
\end{figure}
\end{landscape}

\beginnewsupplement
\begin{landscape}
\begin{figure}[hbtp]
  \centering
  \includegraphics[scale=1.6]{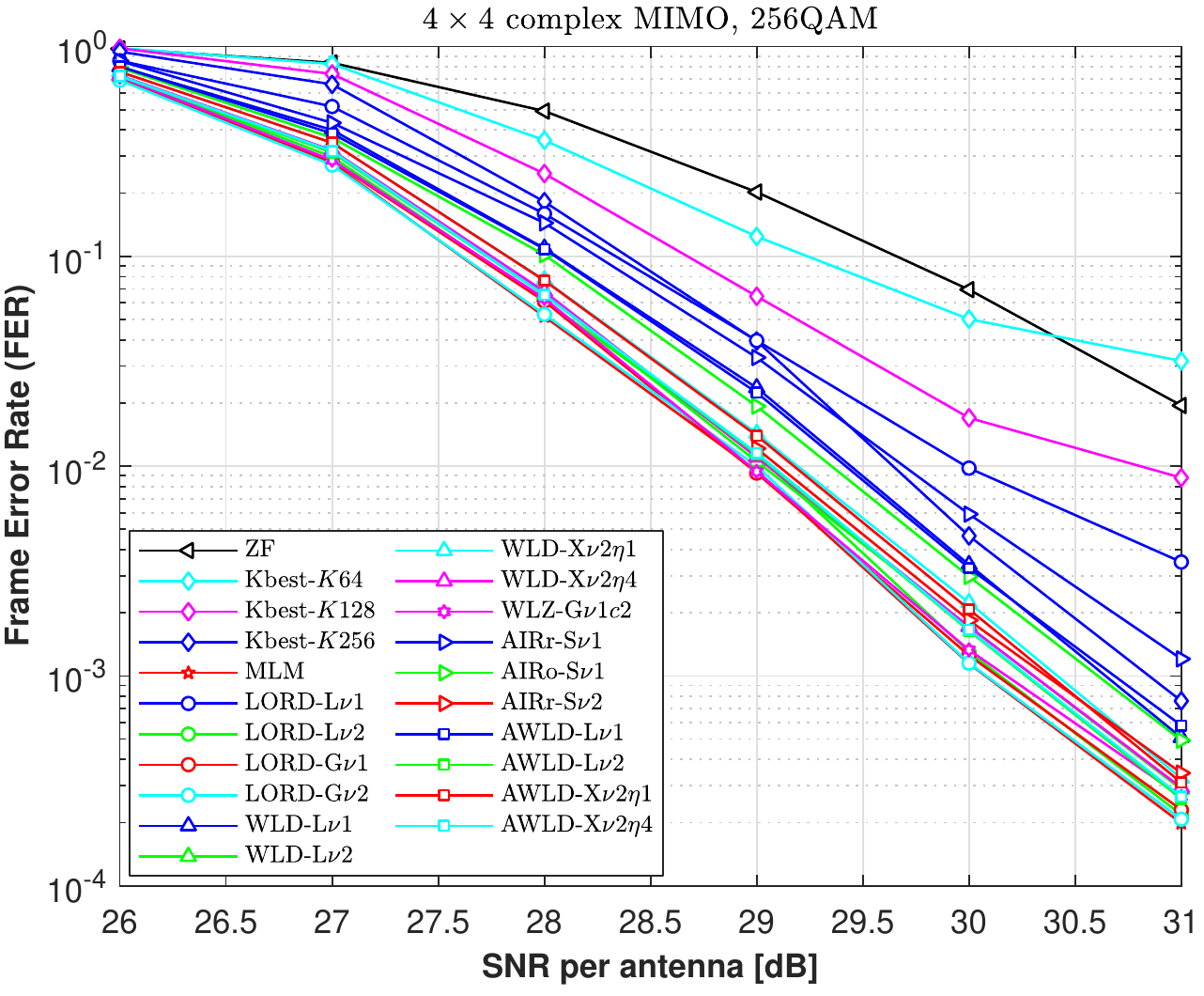}
  \caption{\normalsize Frame error-rate of $4\!\times\!4$ complex MIMO channels, 256QAM}\label{figsup:fer_plot_4x4_256QAM}
\end{figure}
\end{landscape}

\beginnewsupplement
\begin{landscape}
\begin{figure}[hbtp]
  \centering
  \includegraphics[scale=1.6]{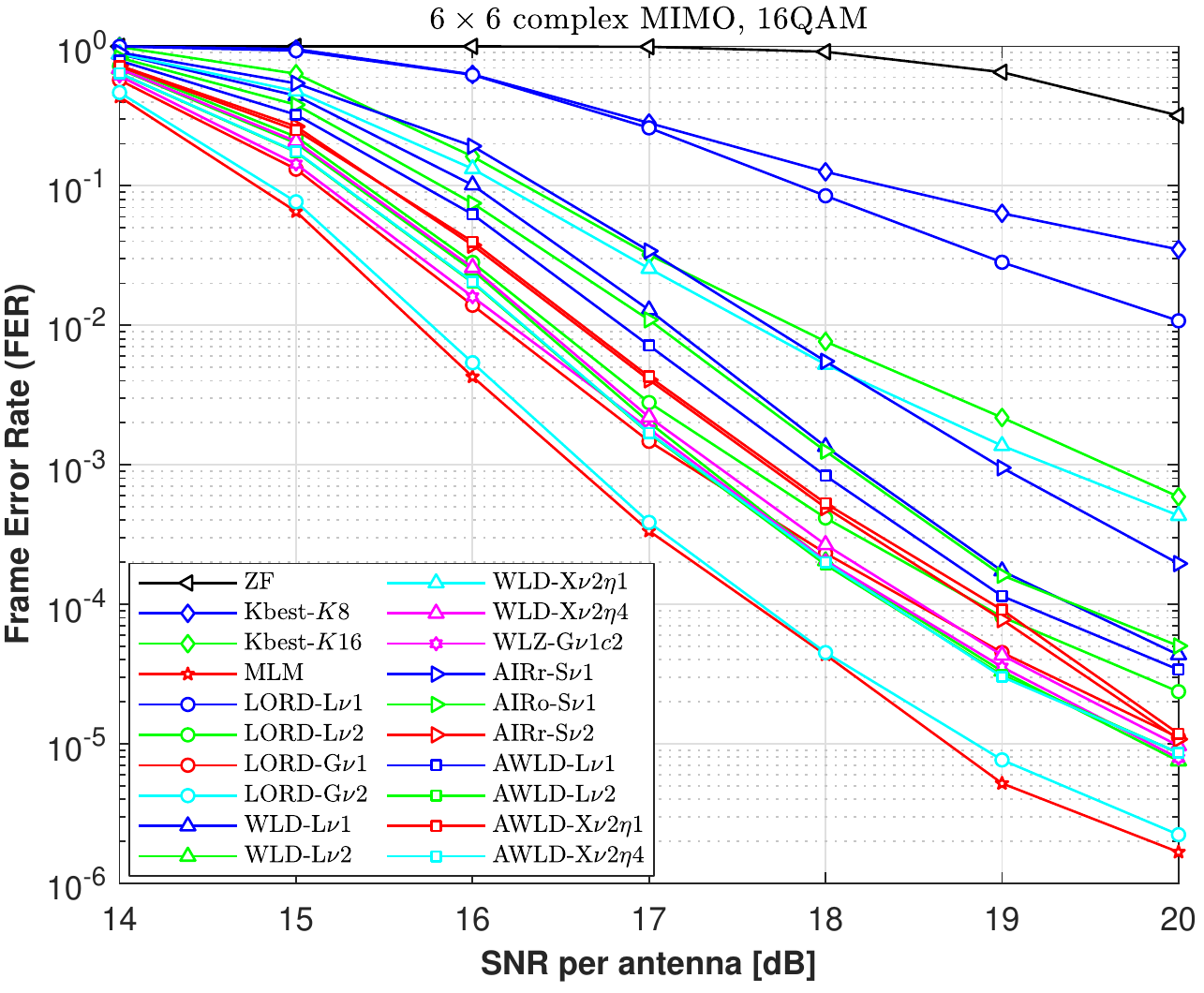}
  \caption{\normalsize Frame error-rate of $6\!\times\!6$ complex MIMO channels, 16QAM}\label{figsup:fer_plot_6x6_16QAM}
\end{figure}
\end{landscape}

\beginnewsupplement
\begin{landscape}
\begin{figure}[hbtp]
  \centering
  \includegraphics[scale=1.6]{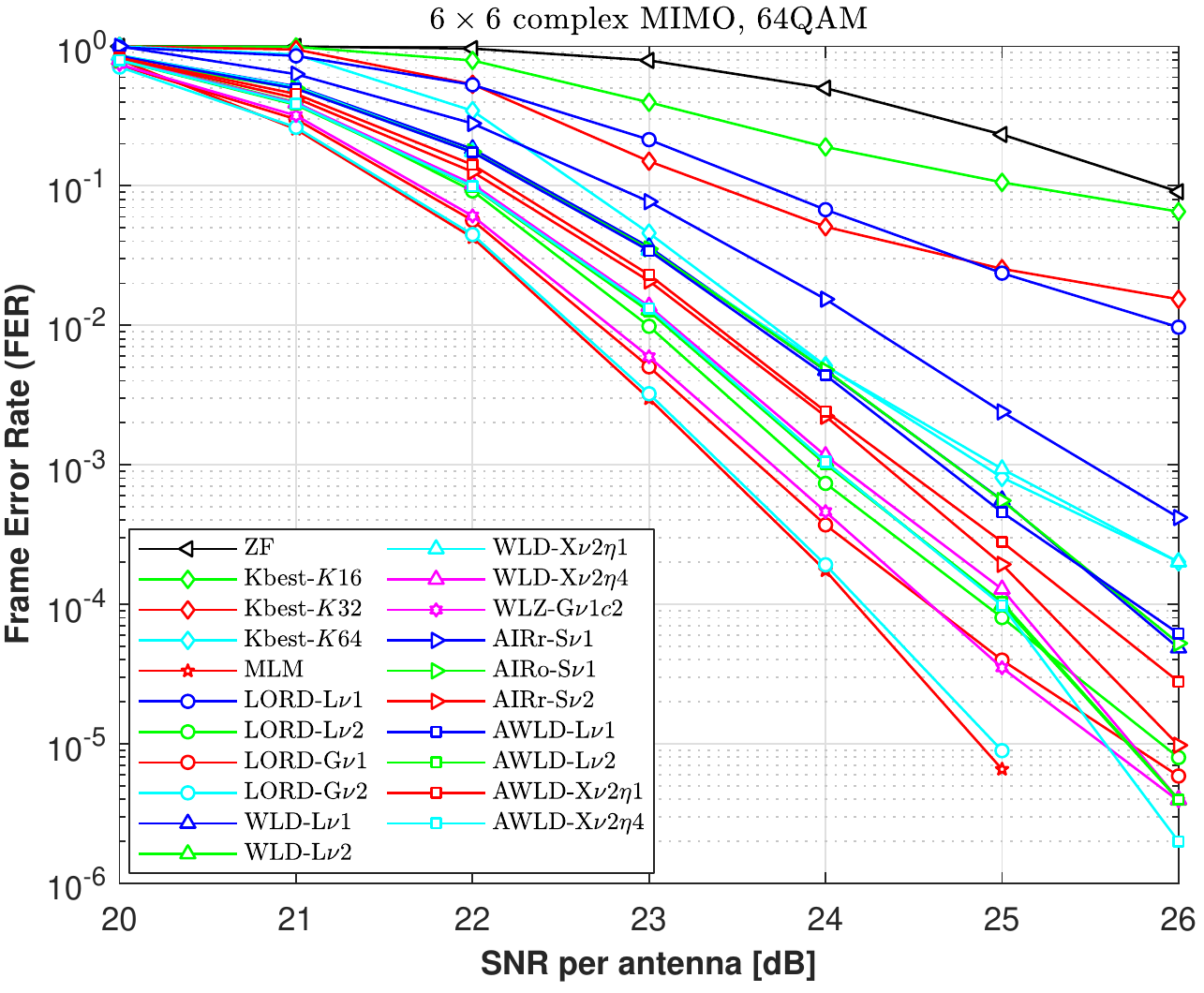}
  \caption{\normalsize Frame error-rate of $6\!\times\!6$ complex MIMO channels, 64QAM}\label{figsup:fer_plot_6x6_64QAM}
\end{figure}
\end{landscape}

\beginnewsupplement
\begin{landscape}
\begin{figure}[hbtp]
  \centering
  \includegraphics[scale=1.6]{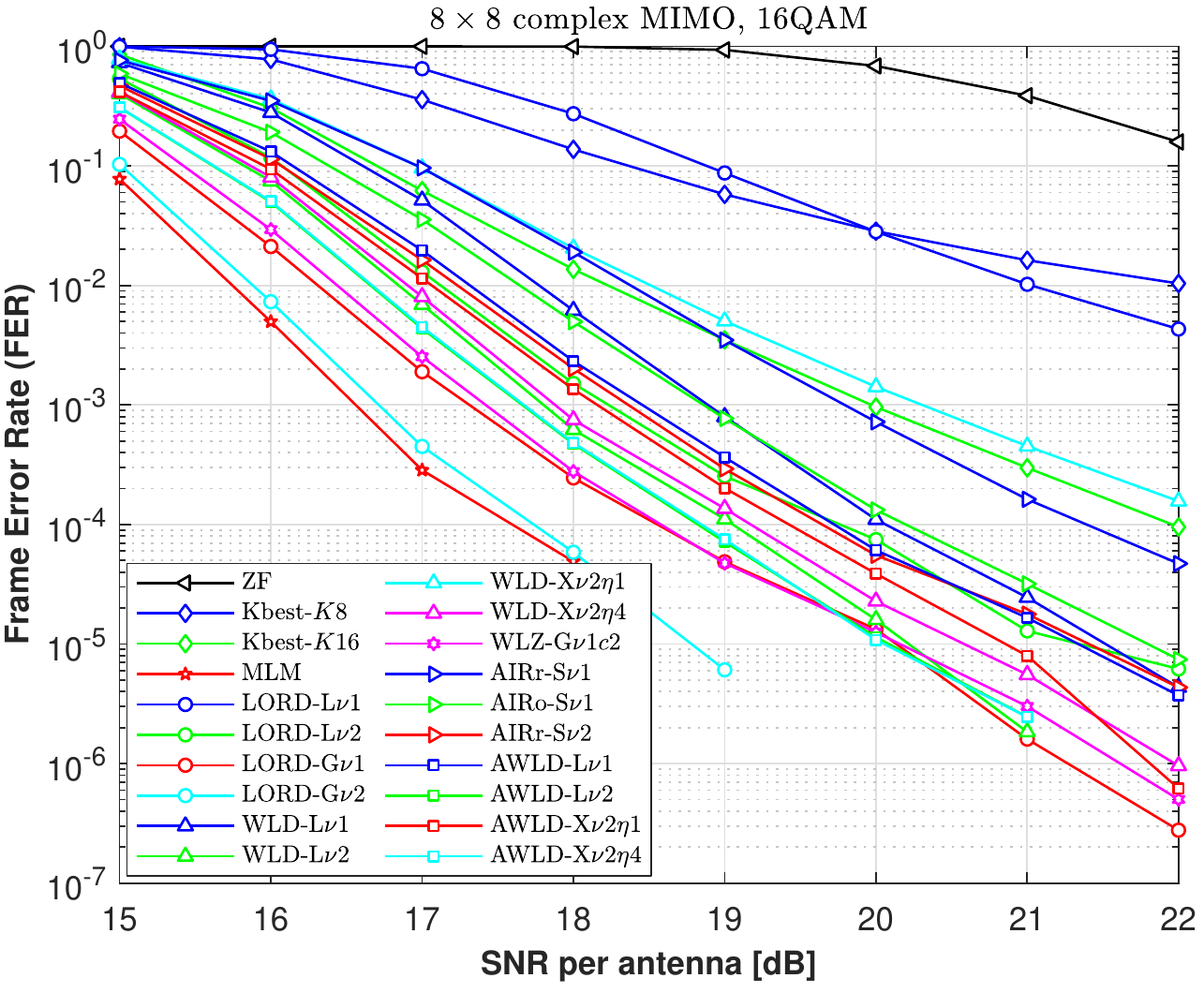}
  \caption{\normalsize Frame error-rate of $8\!\times\!8$ complex MIMO channels, 16QAM}\label{figsup:fer_plot_8x8_16QAM}
\end{figure}
\end{landscape}

\beginnewsupplement
\begin{landscape}
\begin{figure}[hbtp]
  \centering
  \includegraphics[scale=1.6]{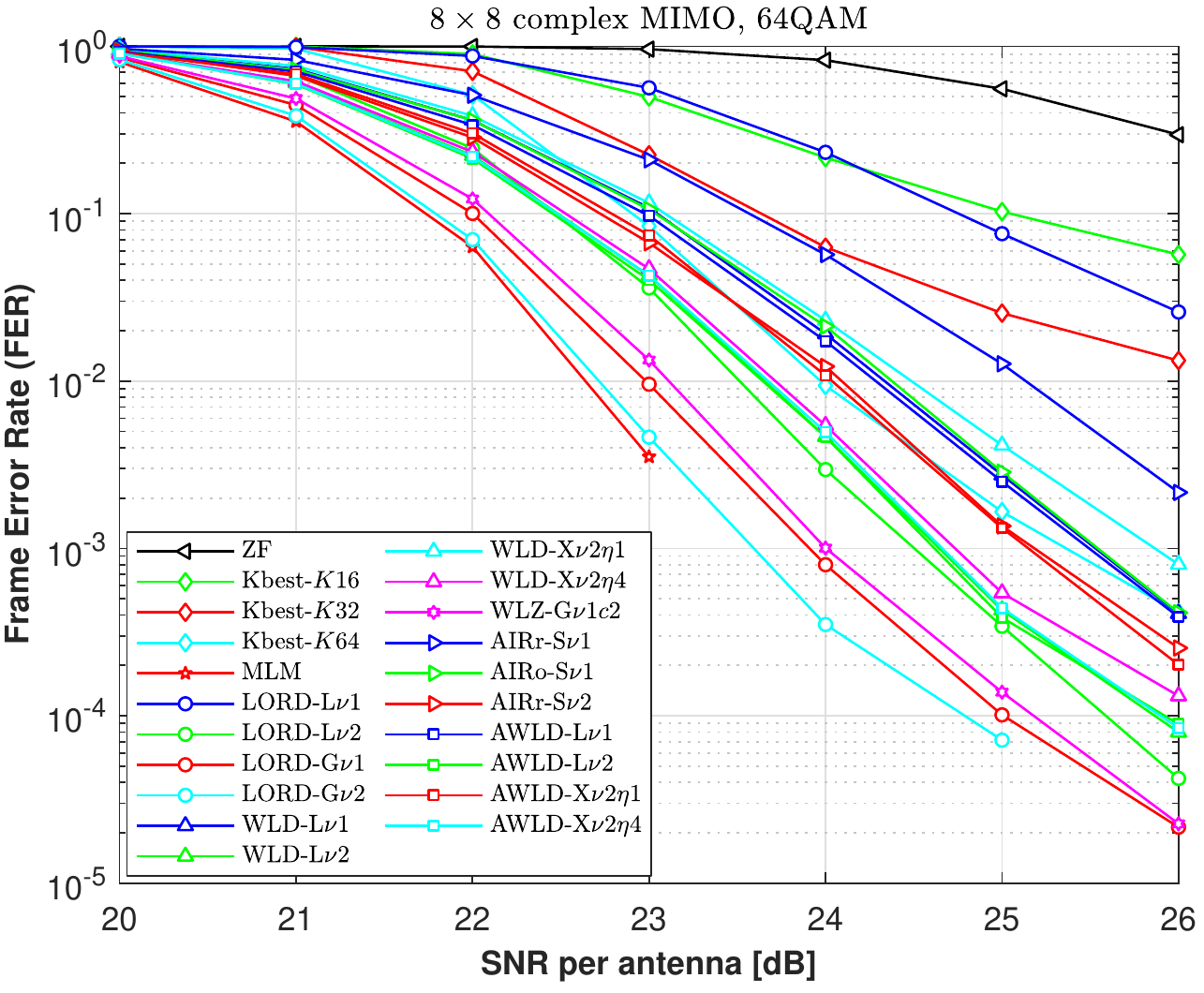}
  \caption{\normalsize Frame error-rate of $8\!\times\!8$ complex MIMO channels, 64QAM}\label{figsup:fer_plot_8x8_64QAM}
\end{figure}
\end{landscape}

\beginnewsupplement
\begin{landscape}
\begin{figure}[hbtp]
  \centering
  \includegraphics[scale=1.6]{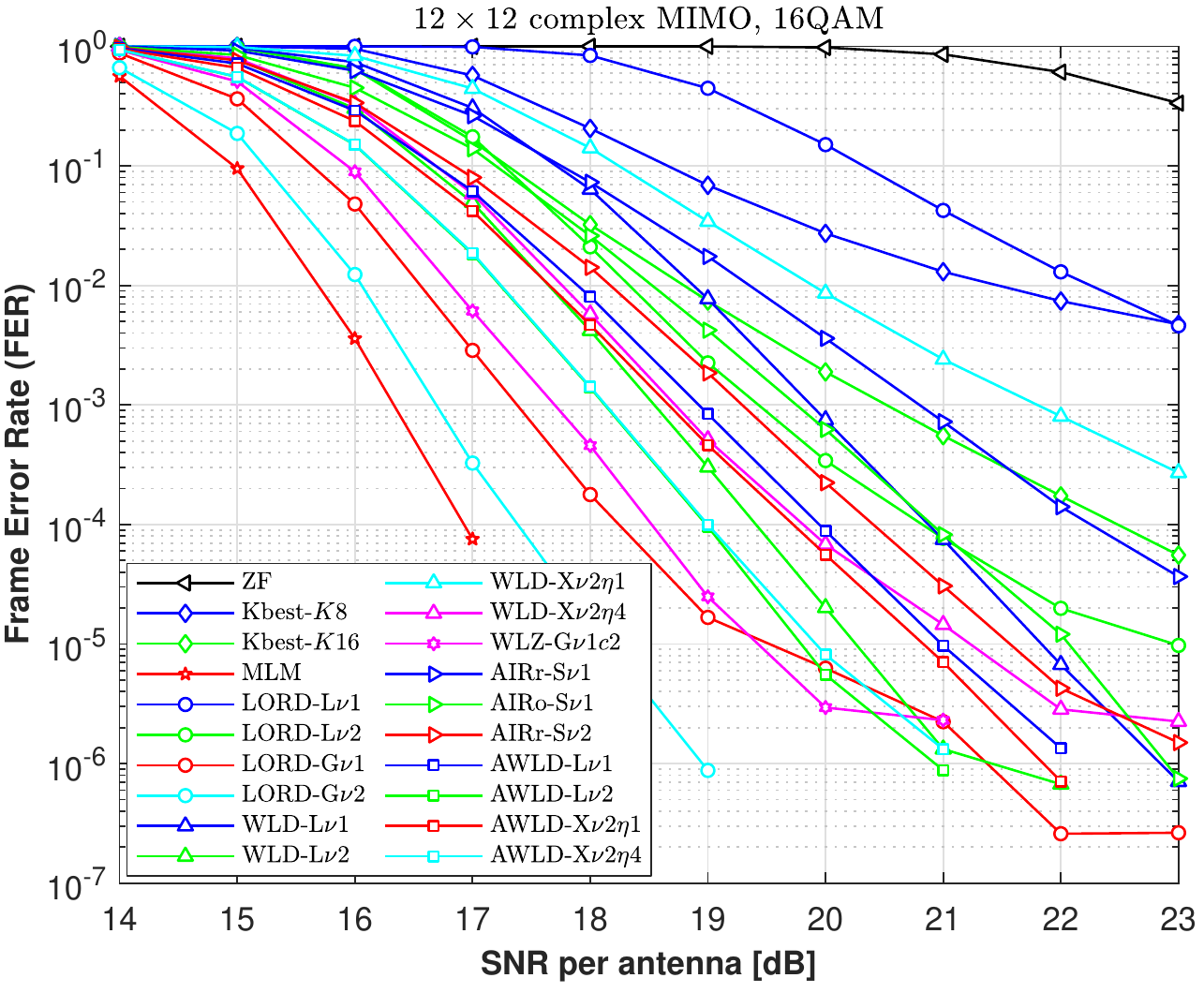}
  \caption{\normalsize Frame error-rate of $12\!\times\! 12$ complex MIMO channels, 16QAM}\label{figsup:fer_plot_12x12_16QAM}
\end{figure}
\end{landscape}

\beginnewsupplement
\begin{landscape}
\begin{figure}[hbtp]
  \centering
  \includegraphics[scale=1.6]{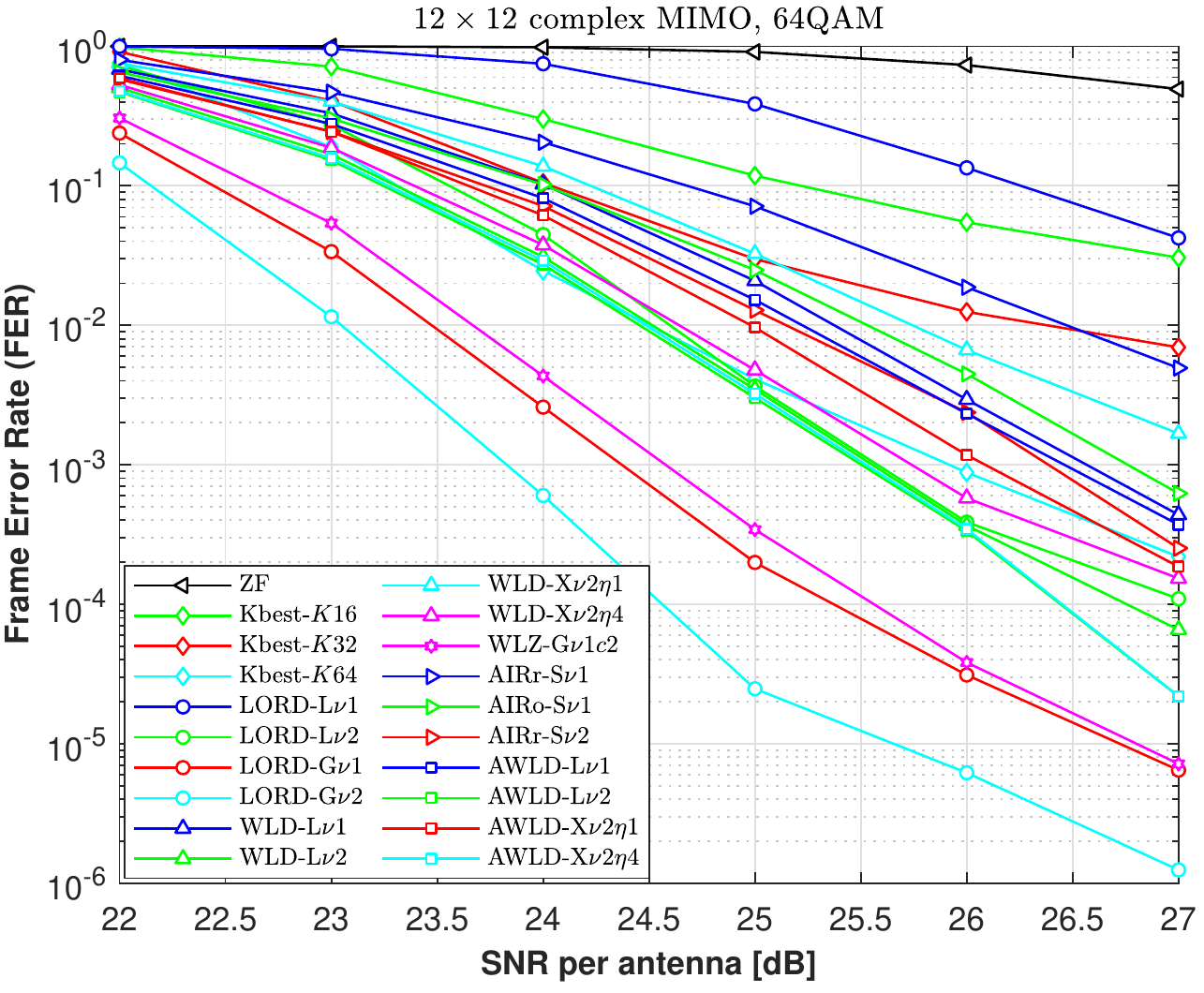}
  \caption{\normalsize Frame error-rate of $12\!\times\! 12$ complex MIMO channels, 64QAM}\label{figsup:fer_plot_12x12_64QAM}
\end{figure}
\end{landscape}

\beginnewsupplement
\begin{landscape}
\begin{figure}[hbtp]
  \centering
  \includegraphics[scale=1.6]{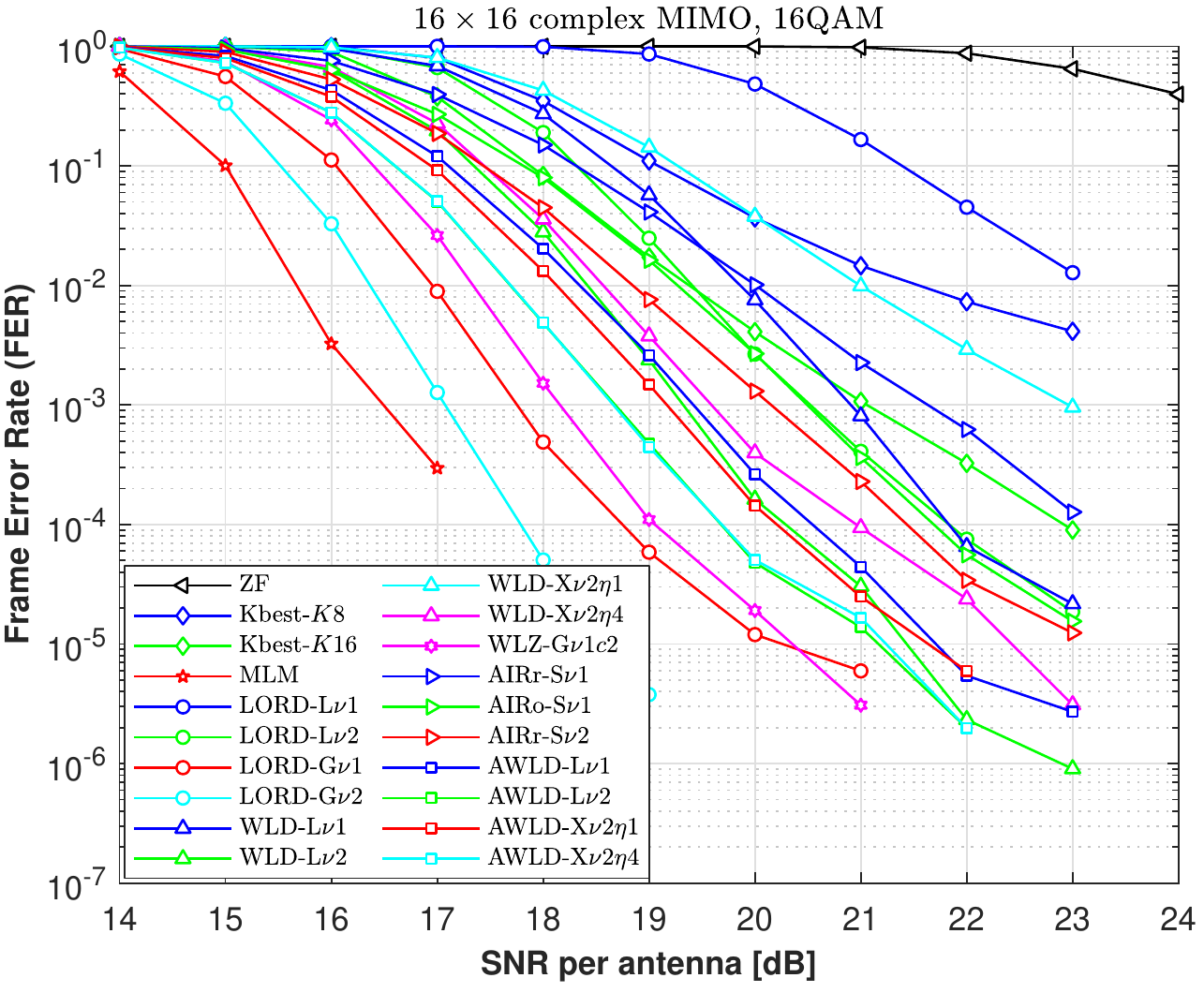}
  \caption{\normalsize Frame error-rate of $16\!\times\! 16$ complex MIMO channels, 16QAM}\label{figsup:fer_plot_16x16_16QAM}
\end{figure}
\end{landscape}

\beginnewsupplement
\begin{landscape}
\begin{figure}[hbtp]
  \centering
  \includegraphics[scale=1.6]{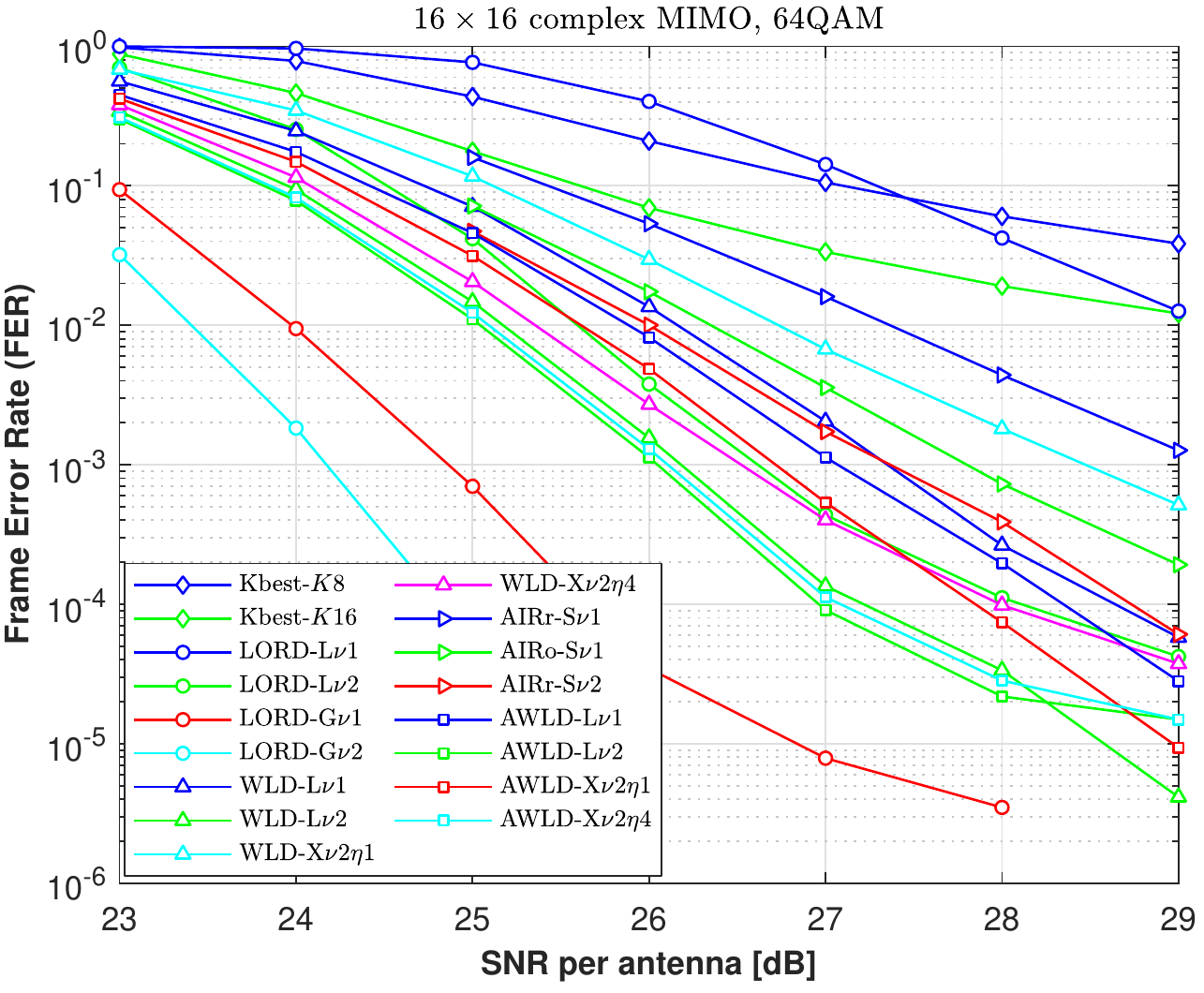}
  \caption{\normalsize Frame error-rate of $16\!\times\! 16$ complex MIMO channels, 64QAM}\label{figsup:fer_plot_16x16_64QAM}
\end{figure}
\end{landscape}

\beginnewsupplement
\begin{landscape}
\begin{figure}[hbtp]
  \centering
  \includegraphics[scale=1.6]{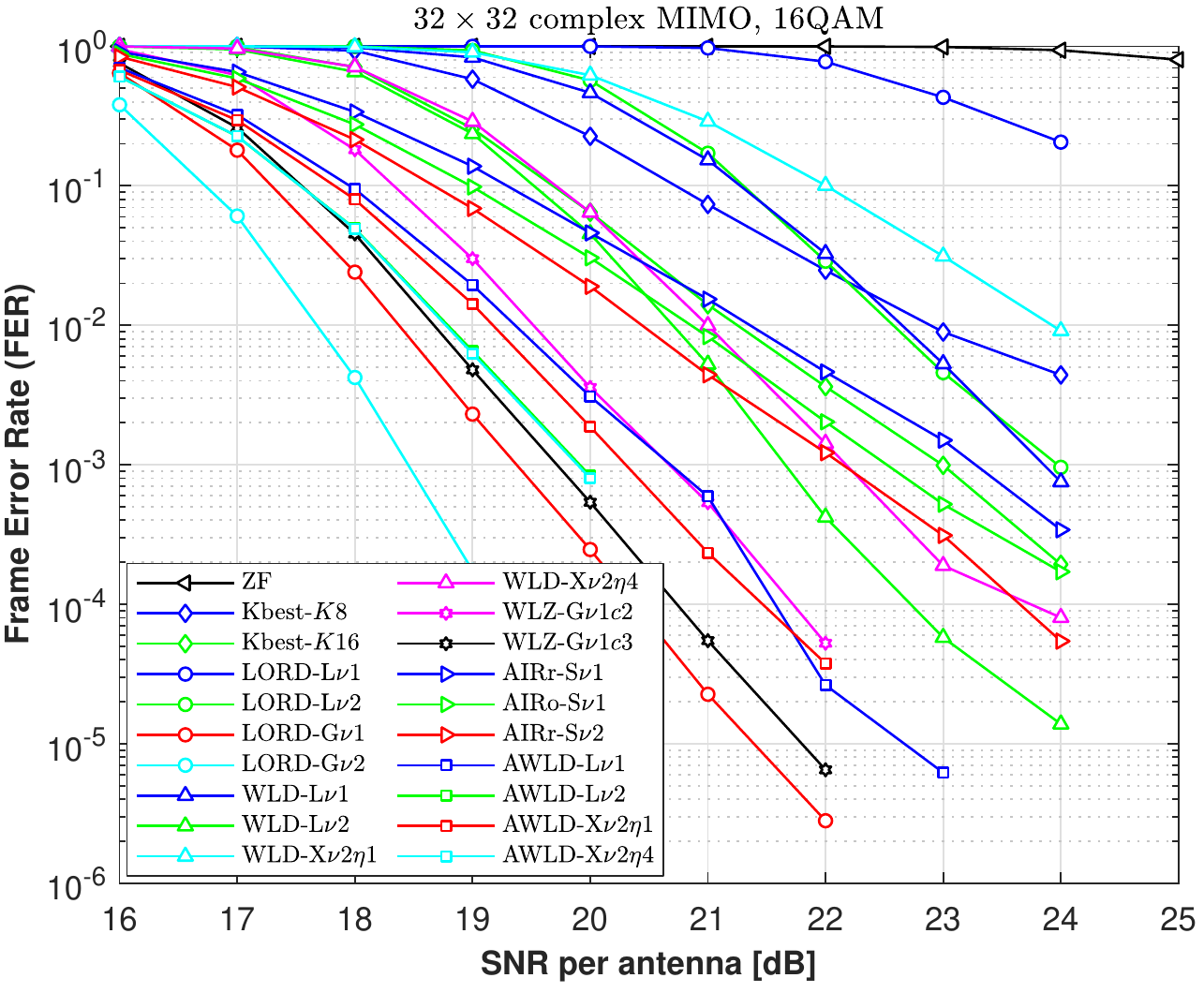}
  \caption{\normalsize Frame error-rate of $32\!\times\! 32$ complex MIMO channels, 16QAM}\label{figsup:fer_plot_32x32_16QAM}
\end{figure}
\end{landscape}

\beginnewsupplement
\begin{landscape}
\begin{figure}[hbtp]
  \centering
  \includegraphics[scale=0.8]{LLR_bit_1_dist_plot_4x4_16QAM}\hspace{-0.00in}
  \includegraphics[scale=0.8]{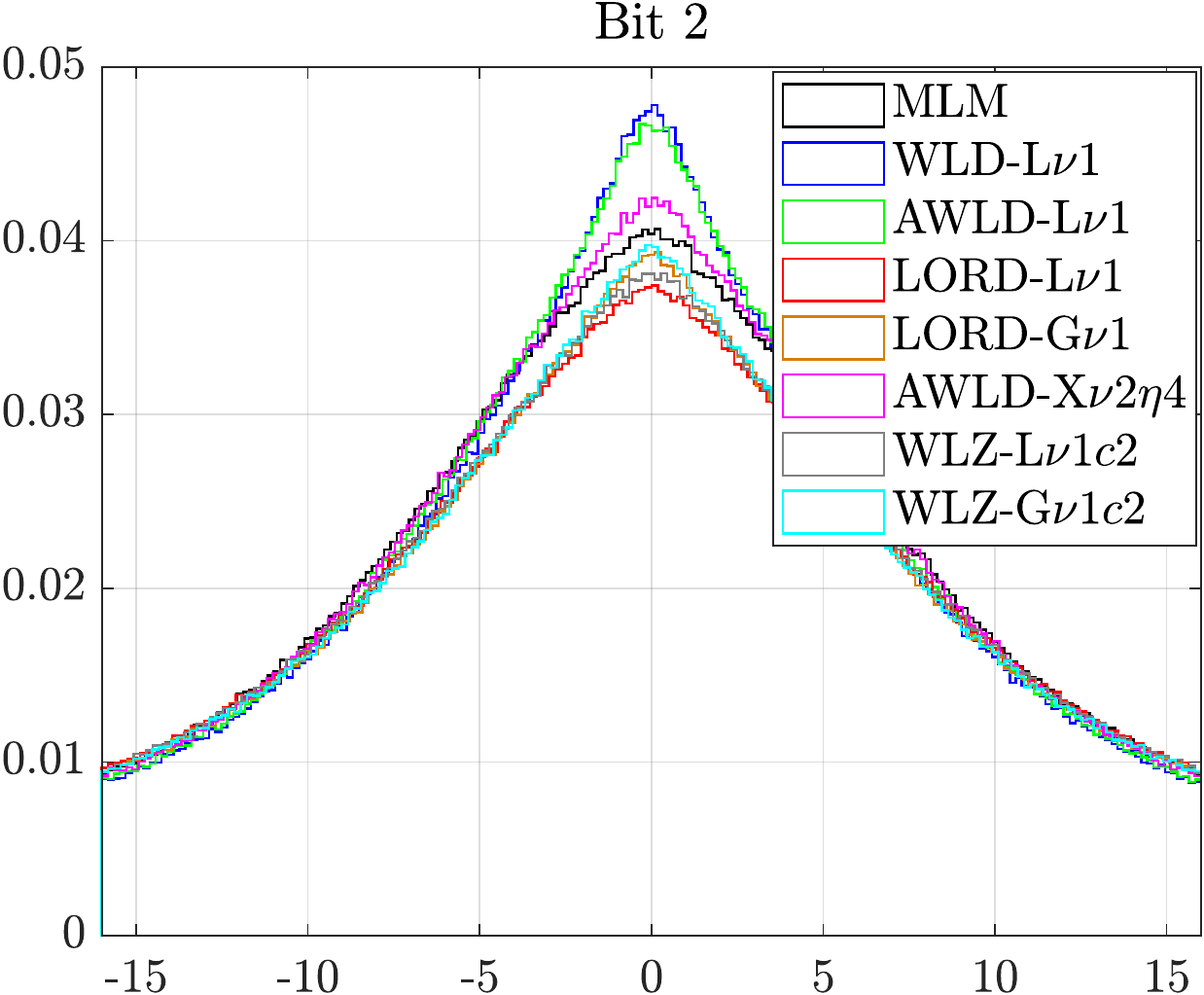}
  \\\vspace{0.025in}
  \includegraphics[scale=0.8]{LLR_bit_3_dist_plot_4x4_16QAM}\hspace{-0.00in}
  \includegraphics[scale=0.8]{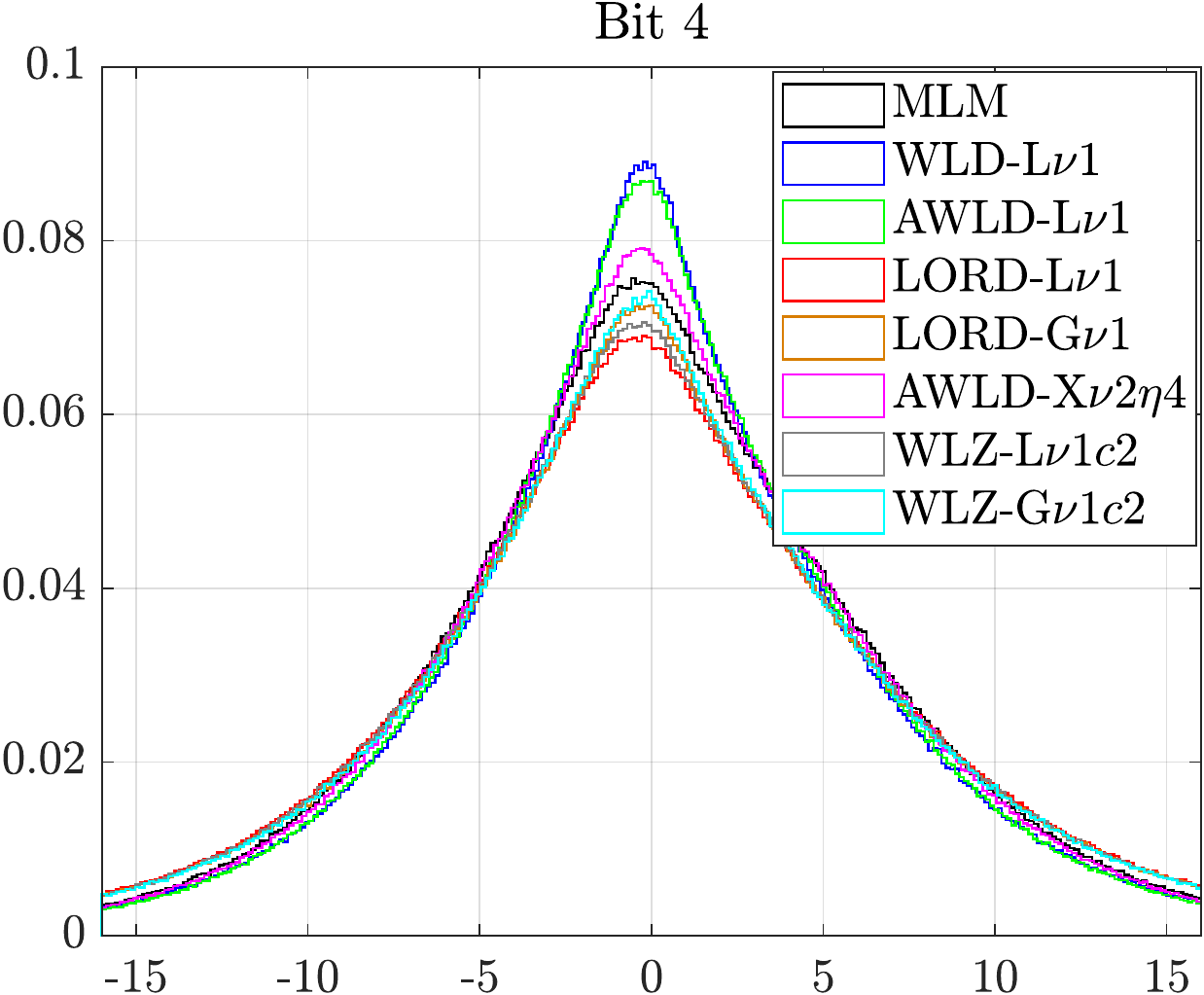}
  \vspace{-0.05in}
  \caption{\normalsize Distribution of bit LLRs of one symbol: $4\!\times\! 4$ complex MIMO channel, 16QAM, $\text{SNR}\!=\!\unit[20]{dB}$.}
  \label{figsup:LLR_bit_1_4_dist_plot_4x4_16QAM}
\end{figure}
\end{landscape}

\beginnewsupplement
\begin{landscape}
\begin{figure}
  \centering
  \includegraphics[scale=1.6]{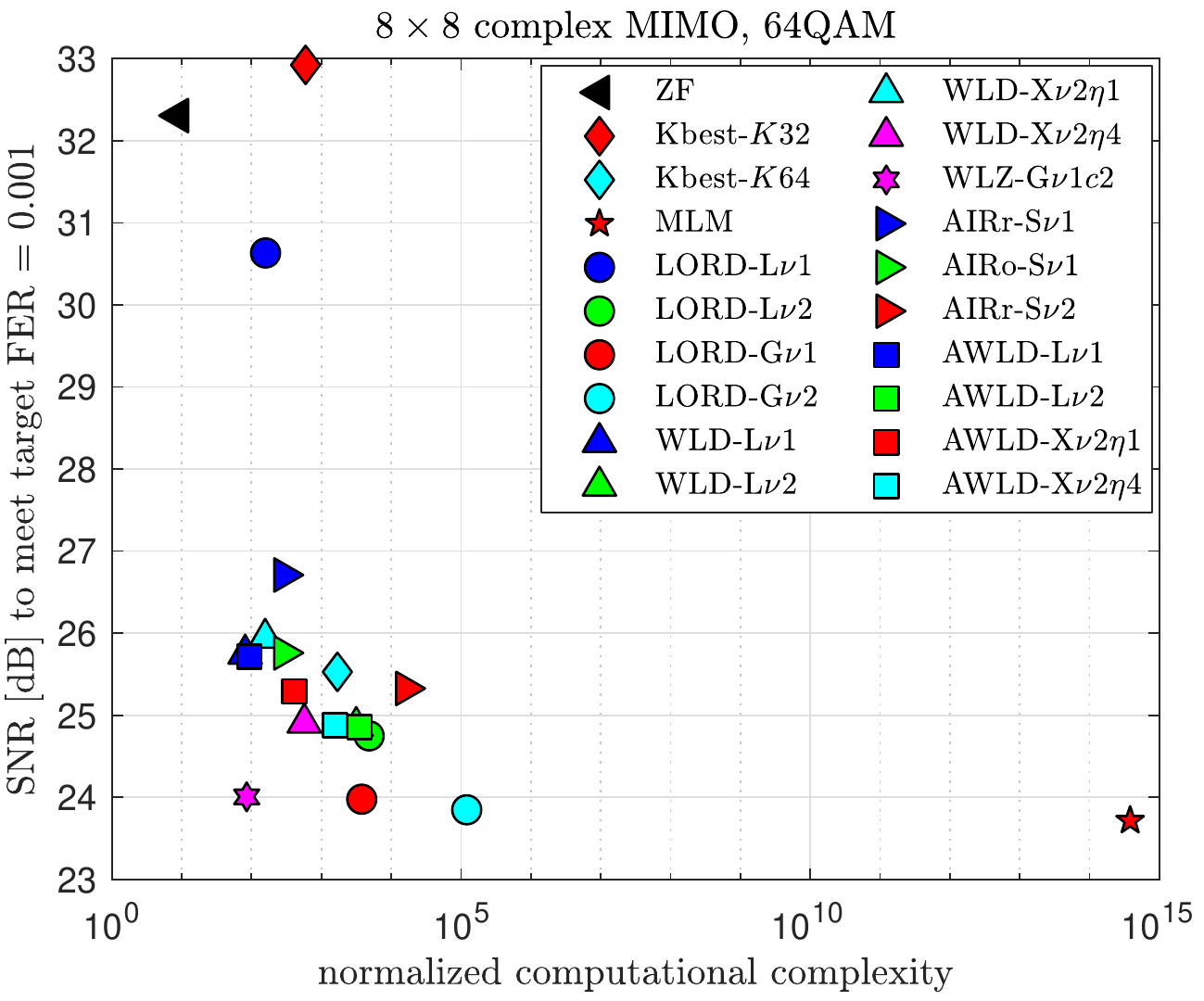}
  \vspace{-0.05in}
  \caption{\normalsize SNR to meet target FER of $0.1\%$ versus complexity.}\label{figsup:complexity_fer_8x8_64QAM}
\end{figure}
\end{landscape}

\end{document}